\documentclass{ut-thesis}

\usepackage[]{natbib}
\usepackage{graphicx}
\usepackage{tabstackengine}
\usepackage{amsmath}
\usepackage{mathrsfs}
\usepackage{amssymb}
\usepackage[encapsulated]{CJK}
\usepackage{ucs}
\usepackage[utf8x]{inputenc}
\usepackage[encapsulated]{CJK}
\usepackage{tabularx}
\usepackage{listings}
\usepackage{hyperref}
\hypersetup{
    colorlinks=true,
    linkcolor=blue,
    filecolor=blue,      
    urlcolor=blue,
    citecolor=blue,
    anchorcolor=blue
}
\usepackage{appendix}
\usepackage{chngcntr}
\usepackage{etoolbox}
\usepackage{lipsum}
\usepackage{float}
\usepackage{wrapfig}
\usepackage{listings} 
\AtBeginEnvironment{subappendices}{%
\chapter*{Chapter 2 Appendix}
\addcontentsline{toc}{chapter}{Chapter 2 Appendices}
\counterwithin{figure}{section}
\counterwithin{table}{section}
}
\usepackage{lmodern}

\newcommand{\cntext}[1]{\begin{CJK}{UTF8}{gbsn}#1\end{CJK}}

\degree{Doctor of Philosophy}
\department{Department of Astronomy and Astrophysics}
\gradyear{2018}
\author{Jielai Zhang (\cntext{张洁莱})}
\title{The Development and Scientific Application of the Dragonfly Telephoto Array}

\providecommand{\sorthelp}[1]{}

\begin{document}

\begin{preliminary}

\maketitle


\begin{abstract}
The low surface brightness visible wavelength Universe below 29 mag arcsec$^{-2}$ is teeming with unexplored astrophysical phenomena. Structures fainter than this surface brightness are extremely difficult to image due to systematic errors of sky subtraction and scattered light in the atmosphere and in the telescope. In Chapter 1, I show how The Dragonfly Telephoto Array (Dragonfly for short) addresses these systematics via a combination of hardware and software and is able to image at a level of 30 mag arcsec$^{-2}$ or fainter. In Chapter 2, I describe the Dragonfly Pipeline and how it is optimized for low surface brightness imaging, how it automatically rejects problematic exposures, and its cloud-orchestration. In Chapter 3, I present a study of the outer disk of the nearby spiral galaxy NGC 2841 using Dragonfly (in Sloan \textit{g} and \textit{r} bands) as well as archival data in UV from the Galaxy Evolution Explorer Satellite and rest frame 21 cm data using the Very Large Array. While it is commonly accepted that gas dominates over stars in galaxy outer disks, I find that in NGC 2841, this is not the case. The stellar disk extends to five times R$_{25}$, and there is more stellar than gas mass at all radii. Surprisingly there is a constant ratio of stellar to gas mass beyond 30 kpc, where the disk is also warped. I propose the most likely formation mechanism for this outer disk is co-planar satellite accretion. In Chapter 4, I present a study of thermally emitted and scattered light from dust in the optically thin regions of the Spider HI Cloud, using Dragonfly and Herschel Space Observatory data. Such a study is novel because scattered light from diffuse optically thin clouds is faint and difficult to observe. The main scientific result is a measurement of the ratio of thermally emitted to scattered light. This ratio can be used to test dust models. In closing the thesis (Chapter 5), I look forward to further improvements in the Dragonfly Pipeline, a population study of the formation mechanisms of galaxy disks and to carrying out tests of dust models.  
\end{abstract}



\null\vfill
\begin{center}
To my husband, my mum, my family, \par
thank you for wanting me to be happy and loving me.
\end{center}
\null\vfill

\clearpage

\null\vfill

\setlength\parindent{230pt} One does not discover new lands \par
\setlength\parindent{90pt} without consenting to lose sight of the shore for a very long time.\par
\setlength\parindent{320pt} --Andre Gide

\vfill\vfill
\setlength\parindent{1em}

\newpage  


\begin{acknowledgements}
\addcontentsline{toc}{chapter}{\numberline{}Acknowledgments}
No words will be able to truly show my appreciation of my PhD advisors, Roberto Abraham and Peter Martin. Let me make an attempt nevertheless. Bob, I could not have gotten to this point in my PhD without your enduring encouragement, advise and support throughout my time in Toronto. Not only are you an excellent scientific mentor (our plans always seem to work out!), your appreciation of your students as multifaceted human beings, your continuous motivation for learning how best to supervise and your infinite passion for the work we have done together was the calm I could always rely on. Peter, I have learned so much from the way you do science in the last year of my PhD. Thank you for all the weekends and late afternoons you pulled out of your schedule to meet with me, thank you for patiently answering my endless questions regarding dust and the ISM, and thank you for the mentorship you have provided on my work in the West African International Summer School for Young Astronomers during my PhD. Whenever I speak to others about my experiences during my PhD, I find myself expressing how lucky I am to be mentored by Bob and Peter. I would not be the researcher, nor would I be the human being I am today without your guidance, support and kindness. 

I am thankful to Pieter van Dokkum for his piercing insights into the world of pixels and block averaging. The way you view data and how you turn it into science never ceases to enlighten me. Allison Merritt, working on Dragonfly would not have been the same without you. I am forever thankful that you were right there with me, sometimes going around in circles, but always coming out the other end together. Over time, the Dragonfly team grew, and it has been a pleasure and a privilege to be part of such an outstanding, coherent team. Thank you Deborah Lokhorst, Shany  Danieli and Lamiya Mowla for making working on Dragonfly even more fun than before! Special thanks goes to Mike and Lynn Rice, Grady, Eugene and Richard at the New Mexico Skies Observatory. Your professionalism, dedication and friendship makes working on Dragonfly a great pleasure. 

Throughout my PhD, I have received help from a great many people. Thank you to my committee members, Chris Matzner and Ray Carlberg for helpful suggestions and feedback. Thank you to Howard Yee for speaking with me at length each time I came to you with questions. Thank you to Chris Matzner and Yanqin Wu for teaching two of the most memorable classes I have ever taken. Thank you Dr. Rhea-Silvia Remus. In your short visit to Toronto, your scientific expertise has helped me a lot in interpreting my science, and your friendship and support is very much appreciated. I would like to say a special thank you to Mike Reid. Since the very beginning of my PhD, you have mentored me in the art of expressing science in a clear and inspiring way. This has helped me explain things to myself, and I know it will continue to help me into the future. Thank you Bryan Gaensler and Renee Hlozek for your thoughtful and insightful guidance every time I came to you for advice. Thank you Margaret Meaney for your support and friendship, for pointing me to opportunities and replying in the middle of the Christmas holidays when I emailed you in panic. Thank you to Anabele Pardi for help with proof reading this thesis, you are an angel! 

A very special thank you goes to Steven Janssens and Heidi White for making my time in Toronto full of the joys of friendship, being understood, and having a home away from home. APLEJIBS (you know who you are), I will continue to think of us as the best of all time. Thank you to my fellow graduate students, who have made coming into work every day a source of joy.  

I am very appreciative of the thoughtful and constructive feedback from my thesis Final Oral Examination (FOE) external examiner, Laura Parker, and FOE committee member, Mike Reid. 
 
\end{acknowledgements}

\tableofcontents
\addcontentsline{toc}{chapter}{\numberline{}Table of Contents}


\listoffigures
\addcontentsline{toc}{chapter}{\numberline{}List of Figures}
\listoftables
\addcontentsline{toc}{chapter}{\numberline{}List of Tables}


\end{preliminary}


\chapter{Introduction}

\section{Motivation for Low Surface Brightness Observations}
Astrophysical phenomena on vastly different physical scales are predicted to exhibit low surface brightness emission (at or below 29 mag arcsec$^{-2}$) at visible wavelengths. These include dust grains in the interstellar medium, degree scale planetary rings in our Solar System, light echos of historical stellar explosions, the outer disks of galaxies, the stellar halos of galaxies, and the cosmic web. Beyond what is predicted, whenever a new parameter space is opened for exploration, new discoveries of unforeseen phenomena can occur. The Dragonfly Telephoto Array (Dragonfly for short) was built to open up the parameter space of ultra-deep visible wavelength imaging. Over the past 40 years the cutting edge of astronomical imaging has transitioned from photographic plates on ground based telescopes to CCD detectors on space telescopes. This has improved the ability to detect small and faint galaxies by a factor of about 600. Over this same period of time, there has been no improvement in our ability to detect large but faint (low surface brightness) structures in the Universe. This is because the limiting factor in this case is not photon statistics or image resolution, it is control of systematic errors, such as the scattering of light inside the telescope itself, sky subtraction, flat fielding and the wide-angle point-spread-function. These systematic errors are addressed via a combination of hardware and software for Dragonfly. 

The work contained in this thesis includes the development of the data management and reduction pipeline for data taken by Dragonfly, as well as observations of the outskirts of galaxy disks and dust in the Milky Way. 

\subsection{Outskirts of Galaxy Disks}
The sizes of galaxy disks and the extent to which they have well-defined edges remain poorly understood. Galaxy sizes are often quantified using $R_{25}$, the isophotal radius corresponding to $B=25$ mag arcsec$^{-2}$, however this is an arbitrary choice. In fact, the literature over the last three decades has produced conflicting views regarding whether there is a true physical edge to galactic stellar disks. Early studies seemed to show a truncation in the surface brightness profiles of disks at radii where star formation is no longer possible due to low gas density~\citep{vanderKruitSearle1982}. More recent investigations have found examples of galaxy disks where the visible wavelength profile is exponential all the way down to the detection threshold~\citep{Bland-Hawthorn2005,vanDokkum2014,vlajic2011}. There is considerable confusion in the literature regarding the relationship between the profile shape and the size of the disk. Most disks fall in one of three classes of surface brightness profile types~\citep{PohlenTrujillo2006,Erwin2008}: Type I (up-bending), Type II (down-bending) and Type III (purely exponential). The existence of Type II disks has been pointed to as evidence for physical truncation in disks. However, while the position of the inflection in the profile can certainly be used to define a physical scale for the disk, this scale may not have any relationship to the ultimate edge of the disk~\citep{Bland-Hawthorn2005,PohlenTrujillo2006}.

The common view in the literature is that the HI disks of galaxies are considerably larger than their stellar disks. This arises from the observation that HI emission extends much further in radius than the starlight detected in deep images~\citep{vanderKruitFreeman2011,Elmegreen2016}. A rationale for this is the possible existence of a minimum gas density threshold for star formation~\citep{FallEfstathiou1980,Kennicutt1989}, although this idea is challenged by the fact that extended UV (XUV) emission is seen in many disks at radii where the disks are known to be globally stable~\citep{Leroy2008}. Studies suggest that large scale instability is decoupled from local instability, and the latter may be all that is required to trigger star formation. For example, a study by~\cite{Dong2008} analyzed the Toomre stability of individual UV clumps in the outer disk of M83. They found that even though the outer disk is globally Toomre stable, individual UV clumps are consistent with being Toomre unstable. These authors also found that the relationship between gas density and the star-formation rate of the clumps follows a local Kennicutt-Schmidt law. In a related investigation,~\cite{Bigiel2010} carried out a combined analysis of the HI and XUV disks of 22 galaxies and found no obvious gas surface density threshold below which star formation is cut off, suggesting that the Kennicutt-Schmidt law extends to arbitrarily low gas surface densities, but with a shallower slope.

On the basis of these considerations, it is far from clear that we have established the true sizes of galactic disks at any wavelength. Absent clear evidence for a physical truncation, the `size' of a given disk depends mainly on the sensitivity of the observations. This basic fact applies to both the radio and the visible wavelength observations, and relative size comparisons which do not account for the sensitivity of the observations can be rather misleading. For example, it is commonly seen that the gas in galaxies extends much further in single dish observations than it does in interferometric observations, because single dish observations probe down to lower column densities~\citep{Koribalski2016}. At visible wavelengths, the faintest surface brightness probed by observations has been stalled at $\sim 29.5$ mag/arcsec$^2$ for several decades~\citep{Abraham2016book}, with this surface brightness `floor' set by systematic errors~\citep{Slater2009}. 

The Dragonfly Telephoto Array (Dragonfly for short) addresses some of these systematic errors and is optimized for low surface brightness observations; see~\cite{Abraham2014} for more details. Dragonfly has demonstrated the capability to  routinely reach $\sim$32 mag arcsec$^{-2}$ in azimuthally averaged profiles~\citep{vanDokkum2014,Merritt2016a}. The present thesis uses Dragonfly to study the stellar disk of spiral galaxy NGC 2841, and compare it to neutral gas mapped by The HI Nearby Galaxies Survey (THINGS)~\citep{Walter2008}, and the XUV emission mapped by The Galaxy Evolution Explorer (GALEX) satellite~\citep{Thilker2007}. Instead of just comparing sizes of disks in different wavelengths, we will compare mass surface densities of gas, stars and star formation up to the sensitivity limit of the respective data sets.

\subsection{Dust in the Milky Way}

Dust is a critical component of the interstellar medium (ISM) and plays an important role in galactic evolution. It can serve as a catalyst for the formation of molecular hydrogen, heat the ISM via the photoelectric effect in the presence of UV light, help dense regions cool, and impart radiation pressure to gas. For those who do stellar or extragalactic observations, dust in the Milky Way is a source of extinction, and so must be characterized to measure the intrinsic color and brightness for a source of interest. For those who study cosmology, dust polarizes the signal coming from the cosmic microwave background (CMB), and so must be characterized to measure the polarization of the CMB due to cosmological effects. Suffice to say, understanding the physical and radiative properties of dust is a critical element in many areas of astronomy.

There are several observable radiative processes that throw light on properties of dust in the ISM. They include extinction, polarization, scattering of light, thermal emission, luminescence and microwave emission associated with rotation of small dust grains. Observed scattering of light by dust grains is mostly associated with reflection nebulae scattering visible wavelength light from nearby bright stars, or with dust around X-ray sources scattering X-rays at small-angles~\citep{Drain}. 

Observations of the visible wavelength low surface brightness scattered light from dust away from bright stars has proved to be difficult. Dragonfly has demonstrated the ability to do low surface brightness visible wavelength imaging. Its spectacular low surface brightness performance means that light scattered by interstellar dust in low column density lines of sight, away from the Galactic plane, illuminated by the integrated light of the Milky Way itself, is not only visible, but has become the biggest source of light pollution for extragalactic studies. This offers a novel way to study the dust and the ISM. The comparison of scattered light with thermal emission from dust can reveal properties of the dust itself, a phase of the ISM known as dark molecular gas, and the column density of the ISM. 

\subsubsection{Dark Molecular Gas and Dust Emissivity}
The ISM pervades galaxies and is the medium from which stars are made, and the medium into which living and dying stars shed material. Yet, the make up of the ISM is still poorly known. This is because the ISM is diffuse, all phases of gas (molecular, atomic, and ionized) in the ISM, as well as dust, are difficult to observe and measurements of column densities are very nuanced. 

The most abundant element in the ISM is hydrogen. Neutral hydrogen is typically observed at radio wavelengths via the 21 cm hyperfine line. A photon with a wavelength of 21 cm is emitted from a hydrogen atom when the relative spins of the electron and proton in the atom flips from parallel to anti-parallel. Collisions can easily flip the alignment of the electron and proton spin to the higher energy state (of being parallel), and for gas with temperatures above 50 K, velocity dispersion above 10 km s$^{-1}$ and column density less than $9 \times 10^{21} \text{ cm}^{-2}$ this line is optically thin~\citep{KulkarniHeiles1988}.

Molecular hydrogen (H$_2$) is a critical component of the ISM, however, unlike neutral hydrogen, is not directly observable in emission at temperatures associated with the ISM. Typically molecular gas is traced via carbon monoxide (CO) emission. CO emits strongly via the 2.6 mm rotational line (J=1$\rightarrow$0). It can be assumed that where there is a CO detection, there must also be H$_2$ present. This is because Hydrogen can self shield at an A$_V$ of $\sim$0.05 mag, whereas CO starts to self shield at an A$_V$ of $\sim$0.2 mag~\citep{Li2018, planck2011-7.0}. The conversion factor to get the column density for H$_2$ given the integrated line intensity of CO emission is the so-called X-factor, often denoted as X$_{\text{CO}}$. The standard conversion equation is as follows:

\begin{equation}
\text{N}(\text{H}_2) = \text{X}_{\text{CO}} \cdot \text{W}(^{12}\text{C}^{16}\text{O, } \text{J}=1 \rightarrow 0)
\label{eqn:XCO}
\end{equation} 

\noindent
where N(H$_2$) is in cm$^{-2}$ and W(CO) is in K km s$^{-1}$. Typical values for X$_{\text{CO}}$ are $1-3 \times 10^{20} \text{ cm}^{-2} (\text{K km s}^{-1})^{-1}$ \citep{planck2011-7.0}. It is important to note that this CO line is generally optically thick, and so the emission is a measure of the temperature of the optical depth $\tau = 1$ surface. It is not a direct measure of the column density of CO. However, if the assumption is made that the molecular cloud is virialized, then the line intensity measures the mass of the virialized cloud~\citep{planck2011-7.0, Bolatto2013}. This assumption is likely not true for all but the densest regions~\citep{planck2011-7.0}. 

There are other ways to probe the amount of molecular ISM, including looking at UV absorption by H$_2$, dust emission in the infrared, $\gamma$-ray emission due to cosmic rays colliding with nucleons, and OH continuum absorption~\citep{planck2011-7.0, Li2018}. All these independent tracers of molecular gas have been used to reveal that while CO is the most commonly used tracer of molecular gas, it is not a good tracer of low density H$_2$~\citep{planck2011-7.0, Li2018, deVries1987}. There are regions where these independent tracers of molecular gas detected more gas than the total amount of atomic and molecular gas as traced by 21 cm line emission and CO total intensity. This diffuse gas component has been termed ``dark molecular gas" or DMG. It is often defined as molecular hydrogen in regions where the column density of CO is too low for detection. 

Dust is a good tracer of total mass in the ISM. However, the use of thermal dust emission is complicated by changes in dust emissivity. In other words, more dust emission can be a result of the fact that there is actually more dust, or simply because the same column density of dust is emitting more. If scattered light is correlated with neutral hydrogen, excess scattered light relative to a linear dependence can directly indicate the presence of dark gas without the complication of changes in the emissivity of dust. Furthermore, scattered light correlated with the thermal emission from dust can directly trace the changes in emissivity without the complication of needing to deal with difficulties in estimating total hydrogen column density or consideration of the phases of the ISM.

\subsubsection{Dust opacity}
\begin{figure*}[!htbp]
\centering
\includegraphics[width=0.85\textwidth]{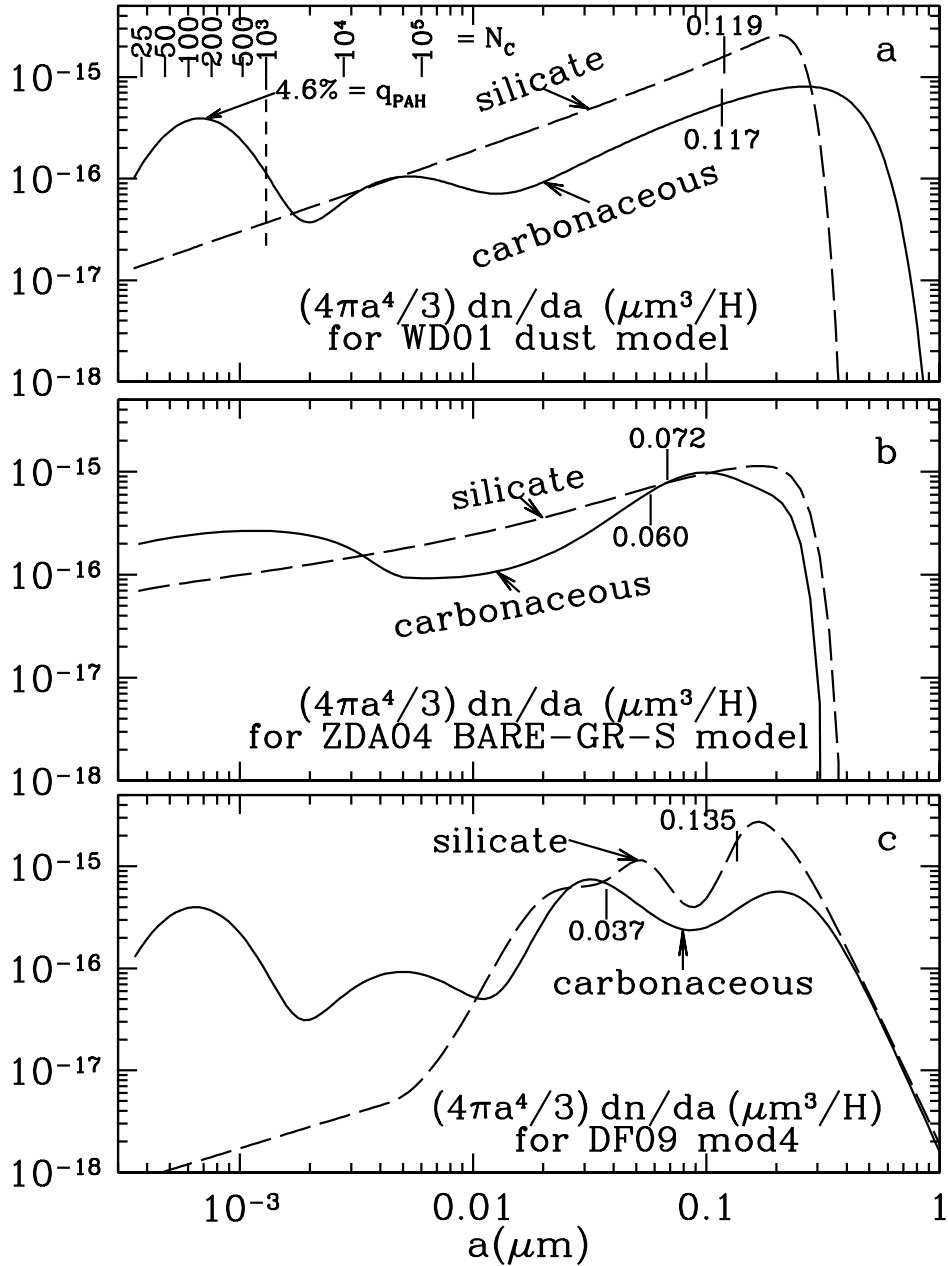}
\caption{Original figure and figure caption from~\cite{Drain}: ``Size distribution for silicate and carbonaceous grains for dust models from (a)~\cite{WeingartnerDrain2001}, (b)~\cite{Zubko2004}, and (c)~\cite{DrainFraisse2009}. The quantity plotted, (4$\pi$a$^3$)\textit{dn}/\textit{d}ln\textit{a} is the grain volume per H per logarithmic interval in \textit{a}. In each case, tick-marks indicate the ``half-mass" radii for the silicate grains and carbonaceous grains." Here \textit{a} is the grain size.}
\label{fig:grain_size_distribution}
\end{figure*}

Another important property of dust is its dust opacity, which is measured in cm$^2$ per gram. Opacity is the effective interaction area, or cross section, per gram~\citep{RybickiLightman} of dust. The opacity of dust to light at different wavelengths depends on the grain size of the dust, and in particular whether it is much larger than, similar in size to, or much smaller than the wavelength of incident light~\citep{Drain}. This means that calculating the opacity of dust in the ISM requires precise knowledge of the grain size distribution of this dust. Grain sizes span a range from least 0.01$\mu$m to 0.2$\mu$m, and different dust models have different size distributions. Some grain distributions are presented in Figure~\ref{fig:grain_size_distribution}, taken from~\cite{Drain}, with the original caption. Without knowledge of the dust grain size distribution, dust opacity is difficult to know precisely a priori. 

The total column density of dust can be found if the dust opacity is known using the following equation~\citep{Lombardi2014,planck2013-p06b}:
\begin{equation}
\Sigma_{\text{dust}} = \frac{\tau_\nu}{\kappa_\nu}
\label{eqn:taukappaSigmaD}
\end{equation} 
\noindent
In this equation, $\tau_\nu$ is the optical depth, $\kappa_\nu$ is the dust opacity and $\Sigma_{\text{dust}}$ is the mass column density of dust. Both $\tau_\nu$ and$\kappa_\nu$ are dependent upon the wavelength of incident radiation. Another expression of the same relationship from~\cite{planck2011-7.12} is:
\begin{equation}
\tau_\nu = \sigma_{e,\nu} \Sigma_{\text{HI}}
\label{eqn:tausigmaSigmaHI}
\end{equation} 
\noindent
where $\sigma_{e,\nu}$ is the emission cross section of the ISM per H atom. $\tau_\nu$ for the wavelengths associated with thermal emission from dust is optically thin even in dense molecular clouds. $\tau_\nu$, as well as the temperature of the emitting dust, can be measured by modelling the dust thermal emission with a modified black body (MBB) that accounts for the optical thinness of dust~\citep{Lombardi2014}. An expression for the MBB from Equation 2 in \cite{planck2013-p06b} is:
\begin{equation}
I_\nu = \tau_\nu B_\nu(T) 
\end{equation} 
\noindent
where $I_\nu$ is the specific intensity, $B_\nu(T)$ is the Planck function for a black body at temperature T, and $\tau_\nu$ is the dust optical depth that modifies the Planck function. However, studies such as~\cite{Ossenkopf1994,Ormel2011,Kohler2012} and~\cite{planck2013-p06b} show that dust opacity, or the cross section per H in the ISM changes as gas becomes more dense, including in the diffuse ISM, making the total column density of dust difficult to determine.

By studying scattered light from dust in the ISM using the Dragonfly Telephoto Array, we can sidestep the complications associated with dust opacity in measuring dust column densities and hence trace total mass column densities in the ISM. We can also directly measure the ratios of opacity at infrared and visible wavelengths, which can be used to inform dust models.  

\section{Limiting factors for Low Surface Brightness Observations}
Our ability to detect faint point sources has improved continuously with the design, construction and use of larger and larger telescopes. In principle (i.e. neglecting seeing) larger telescopes allow for better resolution, and larger telescopes can collect more photons and thereby reduce Poisson noise. However, resolution and Poisson noise are not the factors that limit our ability to do low surface brightness observations. Instead, the dominant factors are systematic issues. Systematics include issues such as not accounting for scattered light (caused by the atmosphere, as well as optical components of telescopes, instruments and detectors), and inaccurate sky subtraction. This section will review the literature with regard to how these systematic issues affect low surface brightness observations. 

\subsection{Scattered Light and the Wide-Angle Point-Spread-Function}
When taking an image, light from a point source is broadened by diffraction and by scattering, the product of which is described by the point-spread function (PSF). The PSF is defined by the optical properties of the telescope and by any medium through which the light has traveled. This includes the atmosphere of the Earth for ground based astronomical observations. A small fraction of the light is scattered to extremely large angles, and the exact shape of and power in this wide-angle PSF is extremely difficult to determine, because it is very faint. This light that is thrown to wide angles from every source on the image causes two important problems when it comes to low surface brightness imaging. Firstly, it can create a surface brightness floor below which any measurements of surface brightness cannot be trusted because every pixel is affected by scattered light at that level~\citep{Slater2009}. Secondly, measurements of the surface brightness of faint outskirts of galaxies can be affected by contamination of scattered light from central bright regions of the respective galaxies. 

\begin{figure*}[!htbp]
\centering
\includegraphics[width=0.85\textwidth]{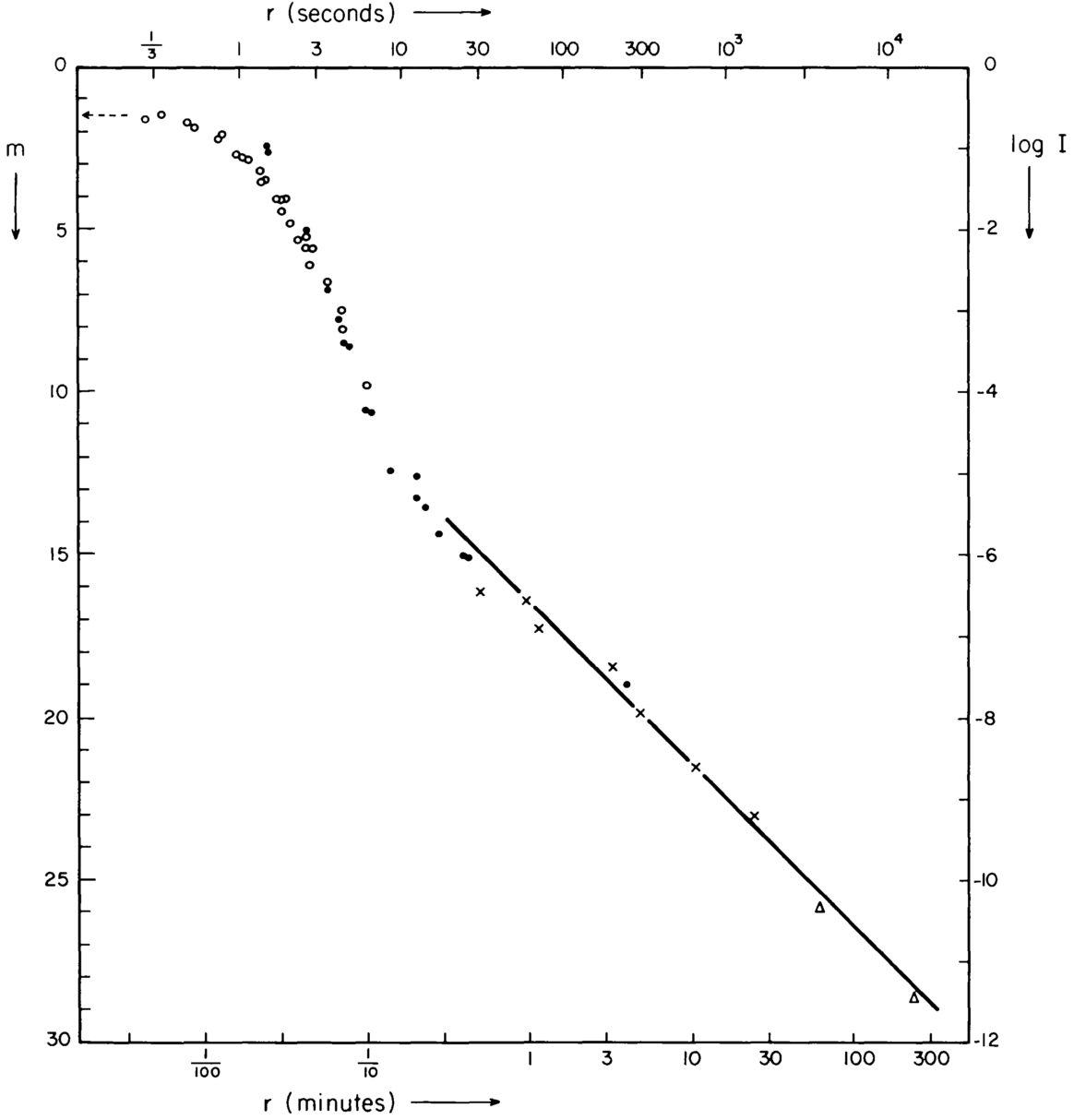}
\caption{Figure and image caption taken directly from~\cite{King1971}. Surface brightness, in magnitude per square second, in the image of a star of magnitude zero. Open circles are derived from 60-inch Cassegrain images, closed circles from diameters of NPS stars on the Palomar Observatory Sky Survey (POSS), and the crosses from other stars on POSS. Straight line is inverse-square law found by de Vaucouleurs. Triangles are from sky brightness near the sun.}
\label{fig:KingPSF}
\end{figure*}

In one of the earliest studies of the wide-angle PSF, \cite{King1971} combined stellar images taken with the Mount Wilson 60-inch reflector and the Palomar 48-inch telescope to measure a PSF out to 5 degrees and determined that it has three parts. As shown in Figure~\ref{fig:KingPSF}, the components of the PSF are: a central seeing disk, an exponential drop and an inverse-square halo. The wide-angle inverse-square halo part of the PSF has been named different things in literature. For example,~\cite{King1971} named it the aureole, while~\cite{deJong2008} called it the tail of the PSF.~\cite{King1971} points out that while the empirical PSF is a good description of the data, it is unclear what the origin of it is, and how it changes with the ``instrument, its condition, the site and the weather"\footnote{This quotes~\cite{King1971} directly.}.  

A more recent investigation of the effect of scattered light or the wide-angle PSF on measured low surface brightness features around galaxies was presented by~\cite{deJong2008}. This paper showed that claimed measurements of stellar halos around an edge-on galaxy by~\cite{ZibettiFerguson2004} in the Hubble Ultra Deep Field~\citep{Beckwith2006} could entirely be due to scattered light. It also showed that scattered light could explain 20-80\% of the light of stellar halos claimed to be detected by~\cite{Zibetti2004} in stacked Sloan Digital Sky Survey images of edge-on galaxies. \cite{Sandin2014} showed that the measured stellar halo around another edge-on disk galaxy, NGC 5907~\citep{Sackett1994}, can be completely accounted for by scattered light as well. \cite{Sandin2015} went on to show that scattered light could also be responsible for the measured stellar halos and thick disks around edge-on disk galaxies, up-bending surface brightness profiles around face on disk galaxies and halos around elliptical galaxies. 

\begin{figure*}[!htbp]
\centering
\includegraphics[width=0.85\textwidth]{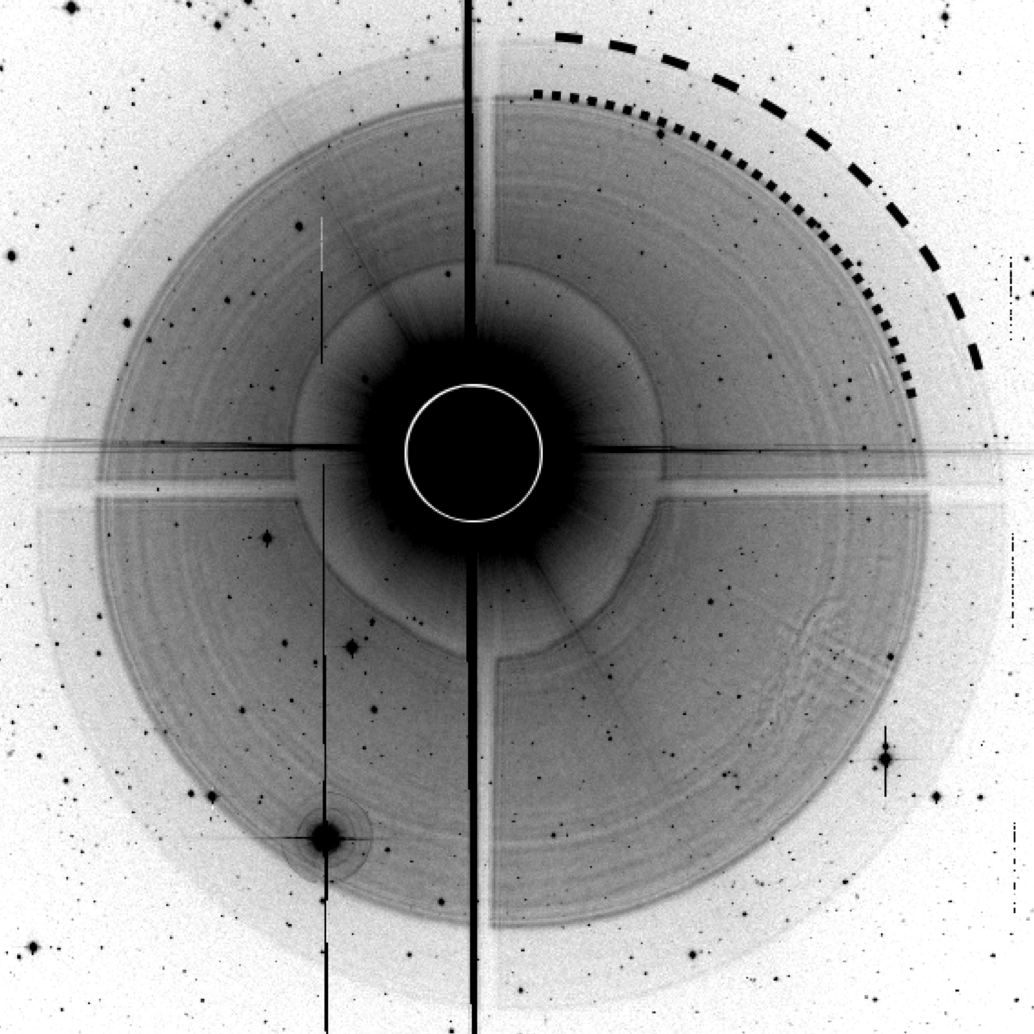}
\caption{Original figure and figure caption from~\cite{Slater2009}: Example image of Arcturus, scaled logarithmically. Black dashed arc indicates the outer edge of the reflection off the top surface of the filter, and the dotted arc indicates the edge of the bottom surface filter reflection. The white circle in the center shows the extent of the dewar window reflection, which is not visible under this scaling. The image is 41' $\times$ 41'. (Note that in the original text, it says the image is 41" by 41", but this is inconsistent with the description in the main body of the paper). }
\label{fig:SlaterArcturus}
\end{figure*}

~\cite{Slater2009} made a careful characterization of the internal reflections and wide-angle PSF of the Burrell Schmidt Telescope. They showed that for images taken by the Burrell Schmidt Telescope, every single pixel is affected by scattered light at the 29.5 mag arcsec$^{-2}$ level. They found that internal reflections off the CCD and other optical surfaces create bright rings around stars. An example of this is shown in Figure~\ref{fig:SlaterArcturus}, which shows an image of Arcturus taken by the Burrell Schmidt Telescope. These authors found that light which has made multiple reflections will contribute at the 30 mag arcsec$^{-2}$ level, but that was below their detection limits, and so was not modeled. After modeling internal reflections, a PSF was measured to a radius of one degree. Unlike~\cite{King1971},~\cite{Slater2009} found that the aureole falls off as $r^{\alpha}$, where $\alpha=-1.6$ beyond 5'. Other studies have found a range of values for $\alpha$ between -1.6 and -3~\citep{Kormendy1973,Shectman1974,Bernstein2007,deJong2008,Sandin2014}.  

There is considerable debate on the origin of the stellar aureole. While the atmosphere was considered to be a potential candidate, ~\cite{Racine1996} speculated that the stellar aureole is mainly caused by scattering from dust, microripples and microroughness on the primary mirror, combined with multiple reflections of various optical surfaces. \cite{Bernstein2007} agreed that the stellar aureole, because of its steep slope ($r^{\alpha}$, where $\alpha\approx-2$), cannot be due to the atmosphere because atmospheric scattering processes (Rayleigh and Mie scattering) would produce much shallower slopes.~\cite{Sandin2014} suggests that at least part of the stellar aureole is caused by an obstructed pupil in combination with multiple reflections, and that a power-law slope shallower than -2 might be due to the addition of dust deposits and the degeneration of coatings on optical surfaces. It is interesting to note that literature from geophysical research identifies the aureole with the atmosphere. In fact,~\cite{AtmIceCrystals2013} proposes that the monitoring of the stellar aureole component of the PSF be used to monitor the size distribution of high atmospheric ice crystals for the study of climate change. Whatever the causes of the stellar aureole, it is important to emphasize that, as noted by~\cite{King1971},~\cite{Sandin2014, Sandin2015} and~\cite{Bernstein2007}, the wide-angle PSF varies with time, telescope, and position on the CCD.  

It is clear from literature that scattered light and the wide-angle PSF pose huge issues for low surface brightness observations. The Dragonfly Telephoto Array and the Dragonfly Pipeline Software incorporate novel methods to address this source of systematic error. 

\subsection{Sky Background Subtraction}
In order to undertake high precision photometry of faint sky sources, the sky background must be subtracted to remove the flux it is contributing to the pixels that the source is occupying. For faint objects, a slight error in sky background subtraction can drastically change measurements, since faint structures can be thousands of times fainter than the night sky~\citep{Sandin2015}. 

For the detection and photometry of point sources or sources with a small angular size, only a local background need be subtracted. The local background can be the combination of all other sources, and not just the sky. For this purpose, a median value over a large enough aperture around the source is often satisfactory (e.g.~\citealt{Lupton2001, Abazajian2009}). When it comes to background subtraction for faint extended sources, the median value of a large aperture can still be affected by the faint light from the extended source itself~\citep{West2005,Hyde2009}. Furthermore, if an aperture that is too large is used, then this is not a local measure of the sky. One way to overcome this problem is to mask sources, and then subtract a fit to a smooth spatial function to the variation of the remaining sky background pixels~\citep{Blanton2011,West2005}. This method of sky subtraction was tested by inserting fake galaxies into real data. The method was able to recover galaxy light for fake galaxies with a half light radius as big as 100 arcsec~\citep{Blanton2011}.  

\section{The Dragonfly Telephoto Array} \label{sec:DragonflyHardware}

The Dragonfly Telephoto Array is designed to image extended low surface brightness structures at a level of 30 mag arcsec$^{-2}$ or fainter. The key design decisions are: (1) To avoid the use of mirrors, because one cause for the wide-angle PSF is dust, microroughness or microripples on reflective surfaces. The use of mirrors is particularly damaging because light is scattered into the optical path. (2) To avoid an obstructed pupil, because it can diffract light away from the central peak of the PSF. (3) To uses lenses with the best anti-reflective coatings available in order to reduce the amount of light being directed into the wide-angle PSF. (4) To use a fast lens that has a small focal ratio, as the number of photons received per pixel for a source with a given surface brightness varies with the inverse of the square of the focal ratio (keeping all other aspects constant). (5) To have a wide field of view to facilitate imaging of local galaxies that extend over large angular scales in the sky.

With these design decisions in mind, the Dragonfly Telephoto Array consists of 48 commercial Canon 400 mm f/2.8 IS II USM telephoto lenses. These fast telephoto lenses have a wide fields of view (about 10 degree$^2$) and the industry's best anti-reflective coatings. The use of multiple lenses together in an array, simultaneously observing the same field of view, means multiple images are taken with independent optical paths. When these images are combined, the effect of scattered light originating from internal optics of the telescope is further reduced. 

\begin{figure*}[!htbp]
\centering
\includegraphics[width=0.7\textwidth]{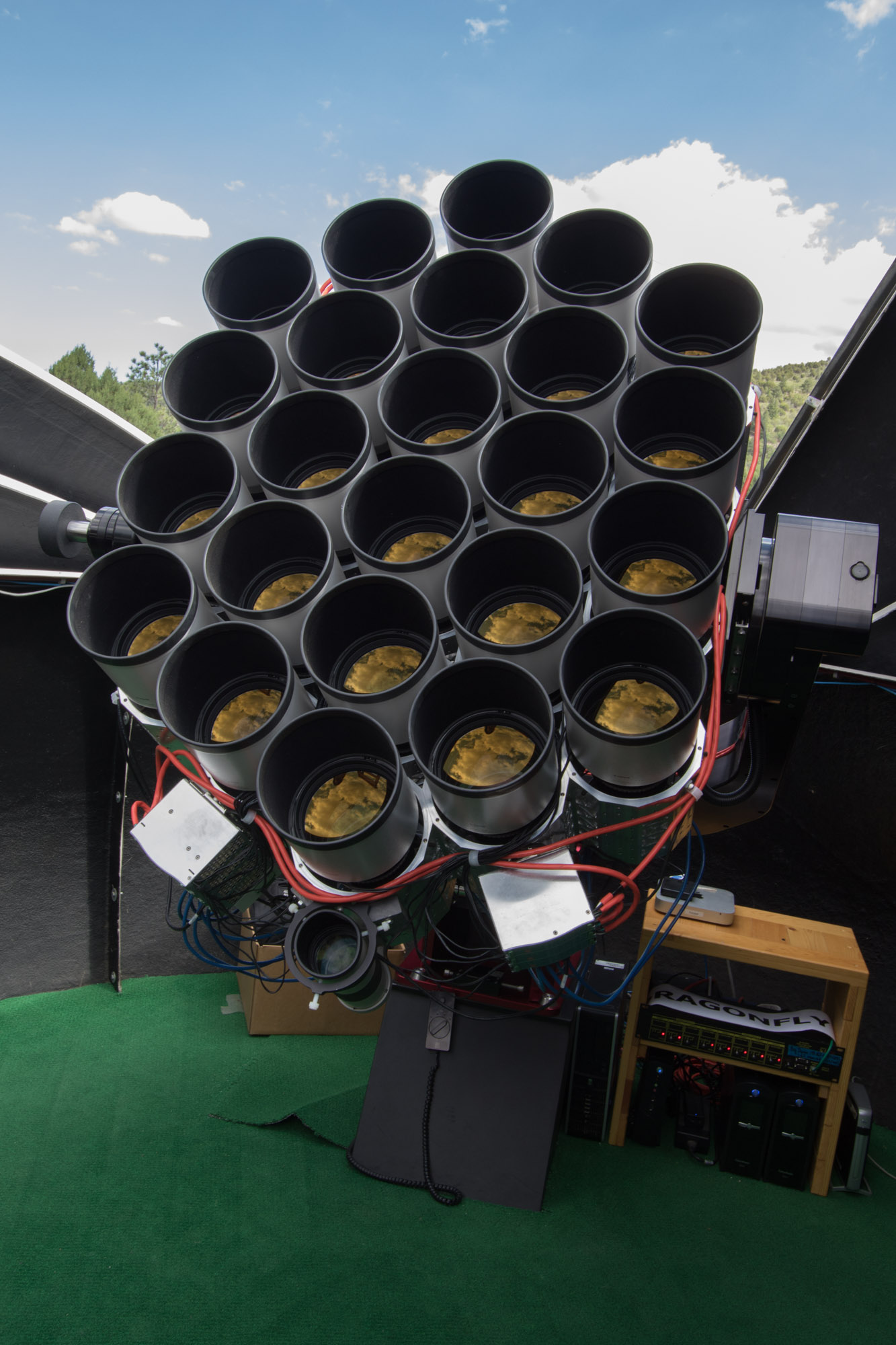} 
\caption{Image of one of two 24-lens array mounts of the Dragonfly Telephoto Array.}
\label{fig:DragonflyPhoto}
\end{figure*}

\begin{figure*}[!htbp]
\centering
\includegraphics[width=0.7\textwidth]{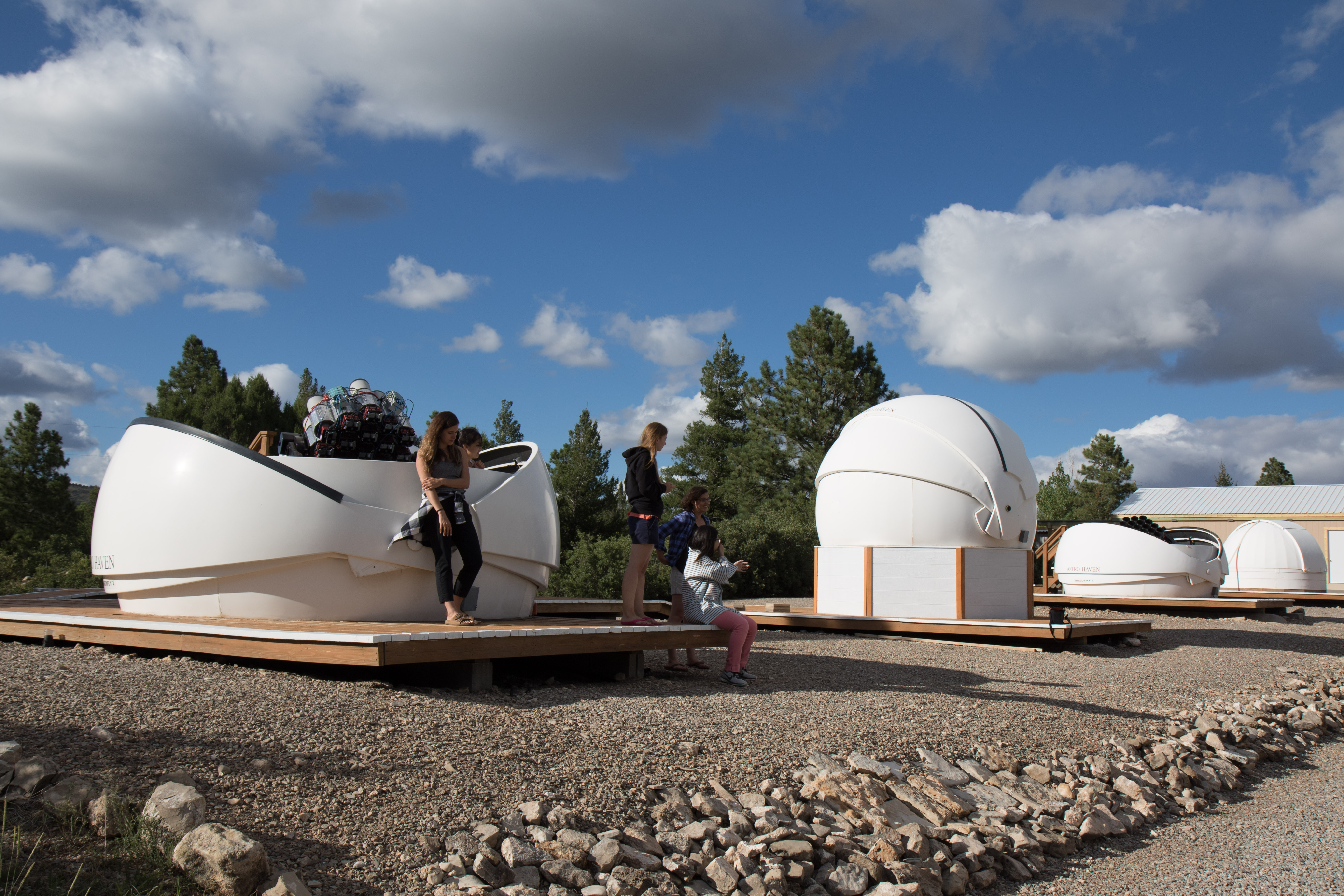}
\caption{Image of where Dragonfly is hosted at the New Mexico Skies Observatory. The two open domes are the two 24-lens arrays that make up Dragonfly.}
\label{fig:NMSPhoto}
\end{figure*}

Each lens is equipped with a Sloan-\textit{g} or \textit{r} filter, and coupled with a CCD camera manufactured by SBIG Inc. (model ST, STF or STT). The lenses are focused using a serial lens controller manufactured by Birger Engineering Inc. Each lens subsystem is controlled by a miniature PC computer (an Intel Compute Stick). The 48 subsystems are arranged on two separate Paramount Taurus mounts, manufactured by Software Bisque Inc., each holding 24 lenses~\footnote{A single mount cannot bear the weight of a 48 lens array, and so the subsystems are split over two mounts.}. Figure~\ref{fig:DragonflyPhoto} shows what one of the 24-lens array mounts of the telescope look like. The other mount looks similar. The telescope is situated at the New Mexico Skies Observatory in Cloudcroft, New Mexico, U.S.A. A photo of the telescope site is shown in Figure~\ref{fig:NMSPhoto}. Each lens is mounted to point at approximately the same field of view, such that they are not offset from one another by more than about 10'. 

\begin{figure*}[!htbp]
\centering
\includegraphics[width=0.85\textwidth]{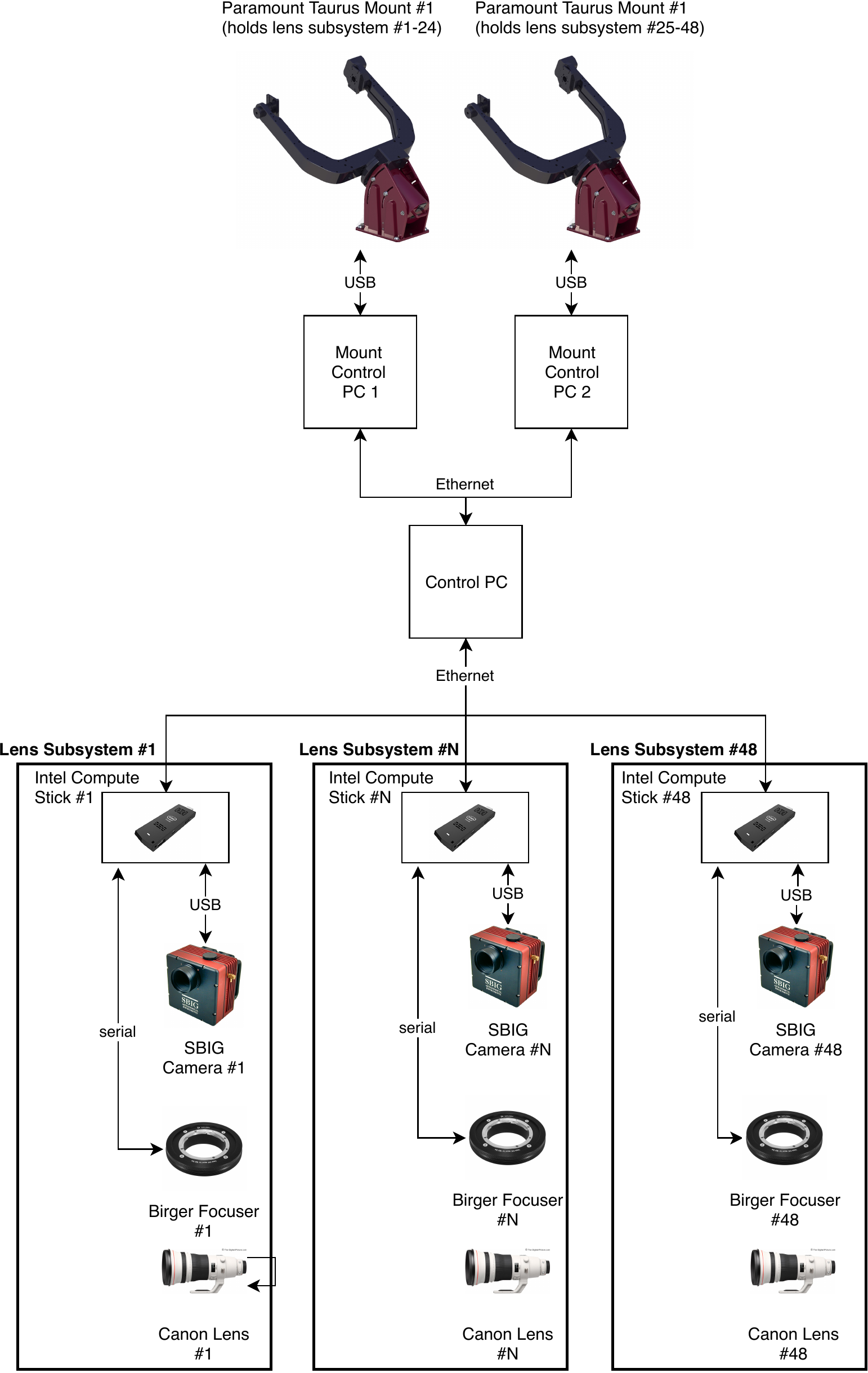}
\caption{Functional diagram of Dragonfly hardware components and how they communicate with one another.}
\label{fig:hardware}
\end{figure*}

Dragonfly is controlled via an Internet Of Things (IoT) software architecture: each lens-camera subsystem's Intel Compute Stick is connected on the local network. During observing, control sequence messages are sent from one central control PC to each subsystem's Intel Compute Stick. Each subsystem carries out independent focusing, image acquisition, image storage and pre-processing. The IoT software architecture allows subsystems not to depend on each other's functionality, so the addition of lens-camera subsystems is relatively simple because subsystems are independent and redundant. A functional diagram of the current Dragonfly hardware components and how they communicate with one another is shown in Figure~\ref{fig:hardware}.

\begin{figure*}[!htbp]
\centering
\includegraphics[width=0.85\textwidth]{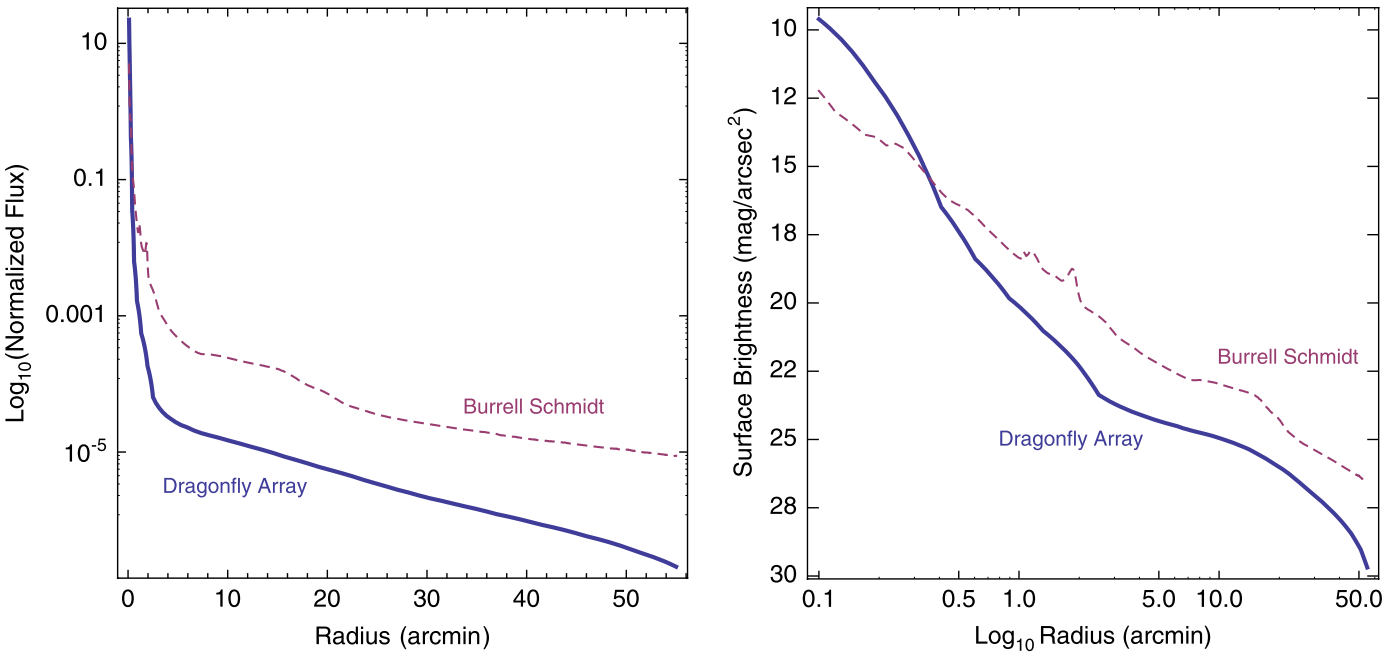}
\caption{The Dragonfly PSF. Original figure and caption from ~\cite{Abraham2014}: Comparison of the wide-angle halo point-spread function of the Canon telephoto lenses used in the Dragonfly Telephoto Array with the corresponding PSF of the 0.9 m Burrell Schmidt telescope on Kitt Peak~\citep{Slater2009}. The Burrell Schmidt is optimized for low surface brightness imaging and has produced a well-known ultra-deep image of the Virgo cluster~\citep{Mihos2005}. The minimum radius plotted is 0.1', and in both cases $\gtrsim$ 90\% of the total stellar flux is interior to this, so only a small fraction of a star’s total light is contained within the wide-angle halo (mainly due to scattering and internal reflections) shown in these plots. Left: Flux as a function of linear radius, after normalizing both profiles to contain identical total fluxes. Note that in large part thanks to nano-fabricated anti-reflection coatings on some of its elements, the Canon lenses have a factor of 5–10 less halo light at radii$>5'$. Right: The r-band AB mag arcsec$^2$ surface brightness profile of Vega as a function of logarithmic radius for the Dragonfly Array and for the Burrell Schmidt telescope.}
\label{fig:dragonflypsf}
\end{figure*}

The potential of the Dragonfly Telephoto Array for low surface brightness observations can be shown by comparing its wide-angle PSF to that obtained by other telescopes. A wide-angle PSF of a single Dragonfly lens-camera subsystem was constructed by taking several images of Vega, each of different integration time. The different images are able to probe different parts of the PSF, e.g. a short integration image probes the inner bright part of the PSF. The stitched-together PSF is shown in Figure~\ref{fig:dragonflypsf}. Also shown in this Figure, taken directly from~\cite{Abraham2014} is the PSF of the Burrell Schmidt Telescope on Kitt Peak. This telescope is well known for a deep image of the Virgo cluster, and so is a good point of comparison for how well Dragonfly does in the world of low surface brightness imaging. Note how Dragonfly's PSF is more than a factor of 6 lower than that of the Burrell Schmidt Telescope at large angular scales. 

\section{Thesis Overview}
This thesis is organized as follows. Chapter 2 details the Dragonfly Pipeline, and how it optimizes for low surface brightness data reduction. Chapters 3 and 4 describe how Dragonfly was used to study the stellar disk of NGC 2841, and dust in the Milky Way, respectively. Chapter 5 concludes with a summary of this thesis and some thoughts on the future of the ideas presented. 

\chapter{The Dragonfly Data Management and Reduction Software Package}

\section{Introduction and Motivation}
The Dragonfly Telephoto Array (Dragonfly) is built to observe extended features in the sky at levels of 30 mag arcsec$^{-2}$ or fainter. This poses a great challenge both at the hardware and software level. Hardware is required to minimize internal reflections within the telescope, but it is software that carefully accounts for the remaining systematic issues such as sky subtraction and the wide-angle point spread function (PSF)\footnote{I led the development of the Dragonfly Pipeline with inputs from the other three members of the Dragonfly team, Prof. Roberto Abraham, Prof. Pieter van Dokkum and Dr. Allison Merritt}. 

To place the challenge into context, it is interesting to compare the precision required by Dragonfly to that required by researchers working on exoplanets. Exoplanet research puts forward one of the most stringent requirements on relative photometry. Researchers in this field need to be able to measure brightness dips in the light curves of stars at 0.1\% precision or fainter. Sky features at a level of 30 mag arcsec$^{-2}$ are 0.05\% the brightness of a dark, moonless night sky (approximately 22 mag arcsec$^{-2}$ in the \textit{g}-band at New Mexico Skies\footnote{this is a representative value as measured by the Dragonfly Pipeline over the years}, where Dragonfly is hosted). While exoplanet studies require photometric precision on small scales, imaging in low surface brightness poses a challenge on larger scales (30" to degree scale). Sources that contribute to a local background on these scales, such as the wide-angle PSF and sky background, need to be properly accounted for. Addressing sky subtraction and the wide-angle PSF properly during data reduction are therefore key requirements for any data reduction process we choose to adopt. 

Appropriate treatment of images, including proper sky subtraction and management of the wide-angle PSF, during the data reduction process could be done by having an astronomer carefully inspect images, processing and stacking only the good science exposures by hand. However, this is extremely time intensive: Dragonfly observations of a single target typically have over 3,000 science exposures, with a similar number of calibration frames (dark and flat exposures). This is why a completely automated pipeline that takes in raw images and outputs stacked images, and satisfactorily deals with systematics such as sky subtraction and the wide-angle PSF, is needed. Very importantly, the needed software should be clear and standardized so that whichever collaborator reduces the data, what was done to the images is methodically documented, with a careful record kept of which images went into the final combined image, which didn't, and why. This is all managed by the Dragonfly Pipeline Software. In order to facilitate human assessment of the performance of the pipeline, the software subroutines are designed to be modular, with clearly defined inputs and outputs. The performance of each subroutine can be assessed by inspecting saved input and output frames to see if a given step in the pipeline had the desired outcome. 

The amount of time it takes to process the data for a single target is typically over a week for a medium depth ($\sim$2,000 - 3,000 science exposures) stacked final image. Quite often, based on the way observations are scheduled, several targets complete their allocated observing time on similar dates. Machines dedicated to data reduction could therefore be idle for a long time, and then suddenly be over-subscribed and a long queue of sets of data for different targets would await reduction, delaying the processing by weeks. Storage of data could also be an issue. The data reduction process creates numerous intermediate data products that can take ~20 times more disk space than the original raw data set. The intermediate products can eventually be deleted after the reduction output is confirmed to be satisfactory, but in the meantime storage requirements can be enormous. Since Dragonfly data is used by many collaborators, in order for all collaborators to use the Dragonfly Pipeline Software, they would each need dedicated machines with tremendous storage space, processing power and all required software installed, posing a logistical challenge.

As we will show below, cloud-based processing provides a solution to these problems. The Canadian Advanced Network for Astronomy Research\footnote{Acknowledgements from canfar.net, quoted directly: ``CANFAR is a consortium of Canadian university astronomers, Compute Canada, and the National Research Council Canada’s Canadian Astronomy Data Centre with support from CANARIE and the Canadian Space Agency."} (CANFAR) is a national platform for data-intensive astronomical research computing in the cloud, and it is equipped with cloud data storage facilities that interface with powerful cloud computing facilities\footnote{Acknowledgements from canfar.net, quoted directly: ``All the cloud services used by CANFAR are the Compute Canada (CC) OpenStack offerings."}. The Dragonfly Pipeline harnesses the resources of  CANFAR to operate efficiently.

The data management, data flow and backup aspects of the Dragonfly Pipeline Software should not be overlooked. It is critical that all data is backed up, easily accessible, sorted and indexed in a manner that allows it to be efficiently queried. This ensures that data is easily accessible by both collaborators and by the automated data reduction pipeline. For this reason, information on raw data frames is stored in a MySQL relational database. Customized scripts that interface with the database and data storage and retrieval software allow users and the Pipeline to easily find the raw data they are seeking and download it. It is also critical that information derived regarding each Dragonfly dark, flat and science exposure as it passes through the Dragonfly Pipeline is recorded. This ensures that bulk properties of the Dragonfly system and observing conditions can be analyzed. This information allows questions like these to be answered: what is the typical sky brightness at the New Mexico Skies observing site in \textit{g} and \textit{r}-band? What is the typical seeing? What are the typical zeropoints of the camera-lens subsystems? How do all these measures change with time? For these purposes, information derived via the Dragonfly Pipeline is also saved on the Dragonfly Database and can be easily queried. 

This chapter is organized as follows. First, an overview of the Dragonfly Pipeline Software Architecture is provided in Section~\ref{sec:pipelinearchitecture}, followed by brief explanations of each step in the pipeline in Section~\ref{sec:pipelinesteps}.  By that point, you will have a good idea of how the pipeline works overall. From there, full details of steps in the Dragonfly Pipeline that are unique to its design are presented. Section~\ref{sec:systematics} delves into sky subtraction, and the wide-angle PSF. Section~\ref{sec:rejection} describes the different algorithms for automatic rejection of problematic exposures. Details on image registration, scaling and combination is presented in Section~\ref{sec:2p6}. Explanations on the way data is accessed and the Dragonfly database are given in Section~\ref{sec:databackupandaccess}. Finally, the cloud orchestration of the Dragonfly Pipeline is presented in Section~\ref{sec:cloud-orch}.  

\section{Overview of the Dragonfly Pipeline Software Architecture} \label{sec:pipelinearchitecture}
The Dragonfly Pipeline takes in raw dark, flat and science exposures that have never been looked at by a human, and produces a combined image in both \textit{g} and \textit{r}-band that can be used for scientific analysis. It is designed to automatically calibrate, sky subtract, embed a world coordinate system (WCS) solution into, re-sample all science exposures to a common grid and finally combine images. At appropriate points, bad dark, flat, and science exposures are identified and removed so as not to contaminate the final data product.

All of the numerous steps of the Pipeline are done without the need for human intervention, however, critical to the Pipeline design is the ease of human inspection of the pipeline steps. This is why the Pipeline is designed with modular scripts, with clear inputs and outputs, where each subroutine carries out conceptually distinct processes on images. The modular design and ease for human inspection of Pipeline steps enable several important design goals:
\begin{itemize}
    \item Verification: given an input into a modular script, it is easy to check if the desired output is achieved. 
    \item Customization: the pipeline can be modified to allow specialized reduction of non-canonical data (e.g. cirrus data).
    \item Training/ Use: New collaborators can be trained to conceptualize the pipeline as a series of functionally distinct steps and inspect the input and outputs of each step. 
    \item Debugging: if the Pipeline malfunctions and the final output is problematic, the source of the problem is traceable to a specific step in the process, or at least a combined effect of a subset of poor data together with a pipeline step. 
    \item Development: New processes that improve or further customize pipeline function can be easily inserted, no matter where in the pipeline this is desired.
\end{itemize}

\begin{figure*}[!htbp]
\centering
\includegraphics[width=0.9\textwidth]{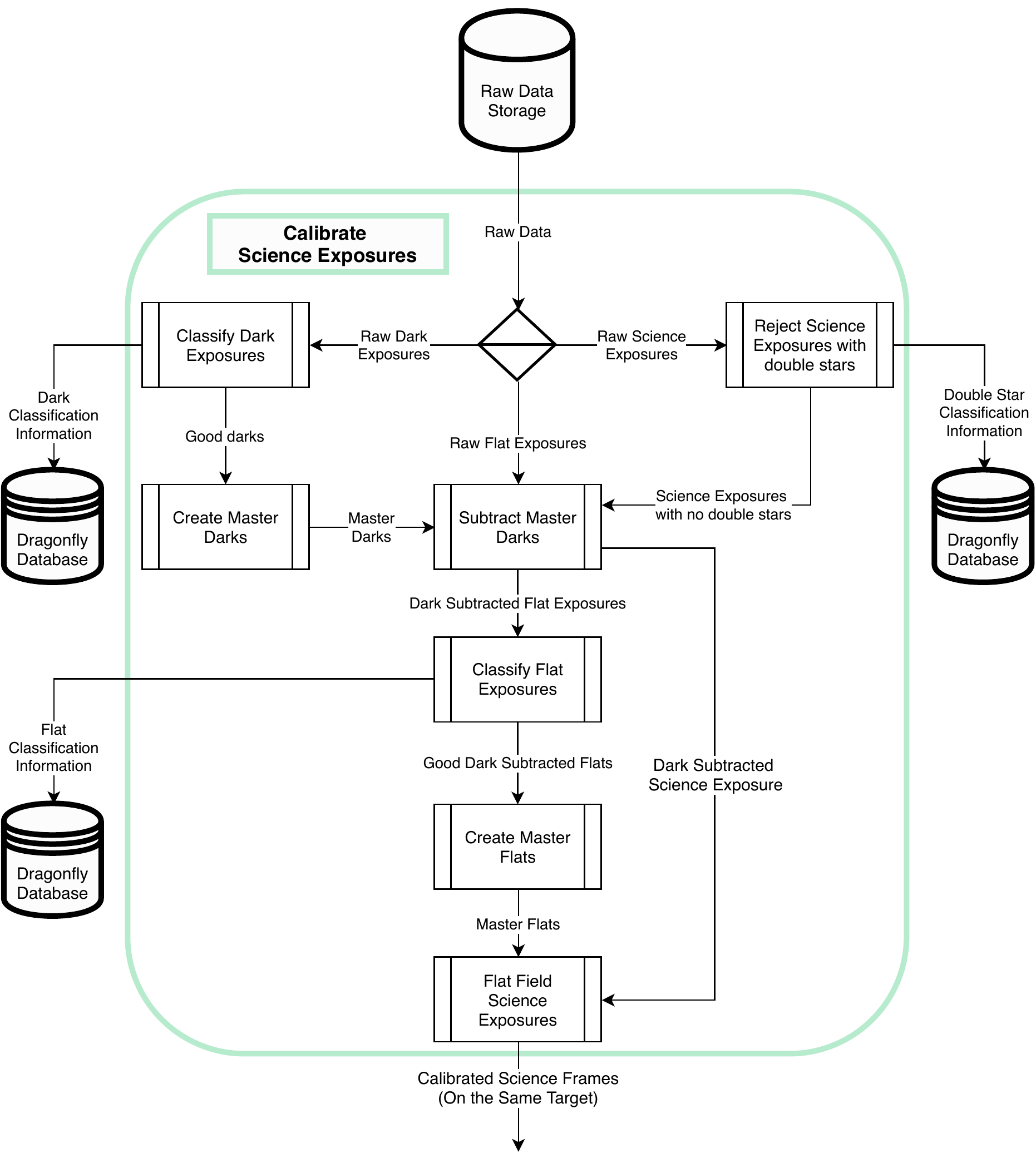}
\caption{Flowchart for the section of the Dragonfly Pipeline that does science exposure calibration and rejection of double star frames. See Section~\ref{sec:doublerejection} for an explanation of the origin of double star frames.}
\label{fig:pipelineflow_cal}
\end{figure*}
\begin{figure*}[!htbp]
\centering
\includegraphics[width=0.9\textwidth]{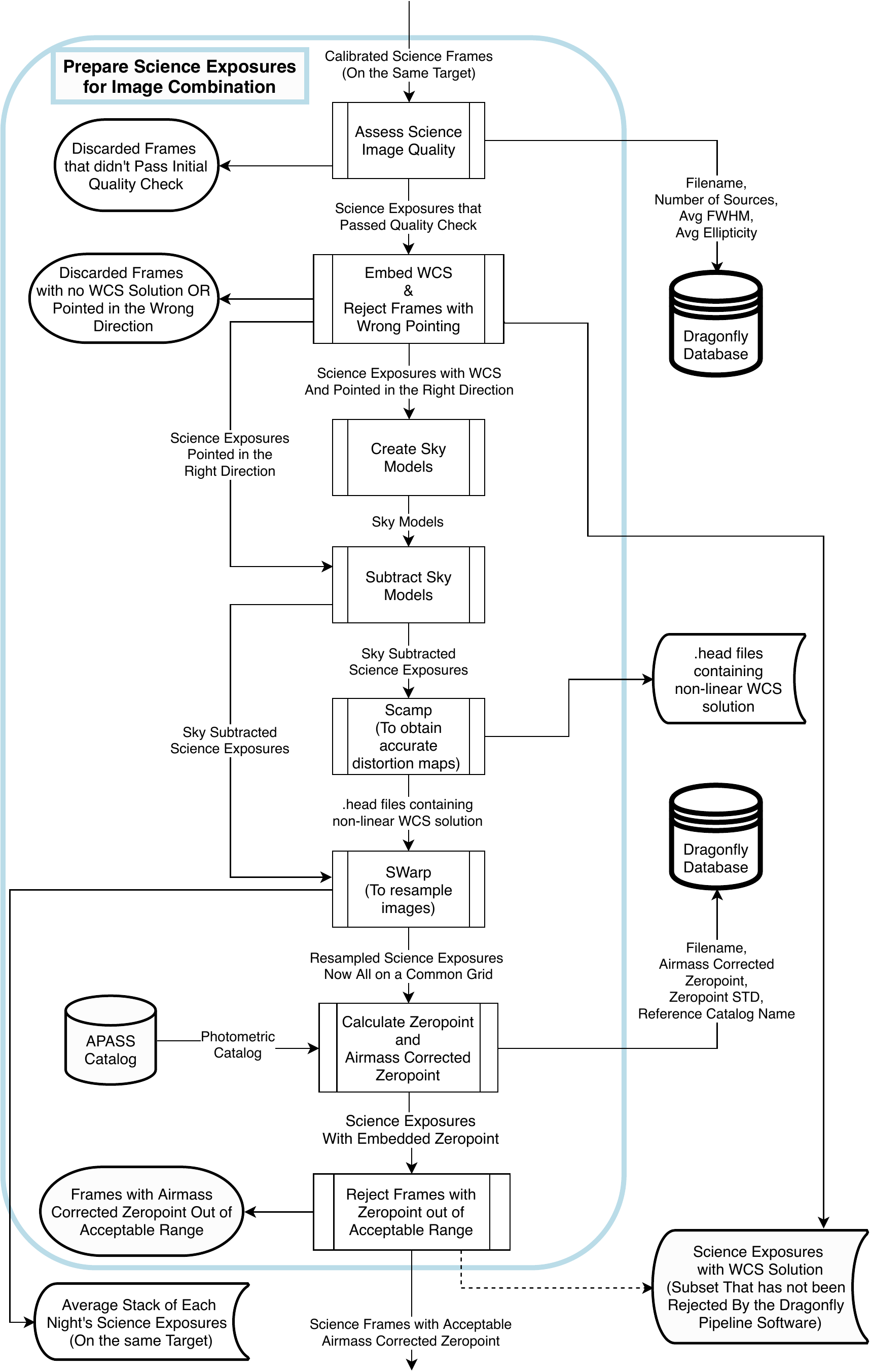}
\caption{Flowchart for the section of the Dragonfly Pipeline that does the initial preparation for image combination.}
\label{fig:pipelineflow_init}
\end{figure*}
\begin{figure*}[!htbp]
\centering
\includegraphics[width=0.9\textwidth]{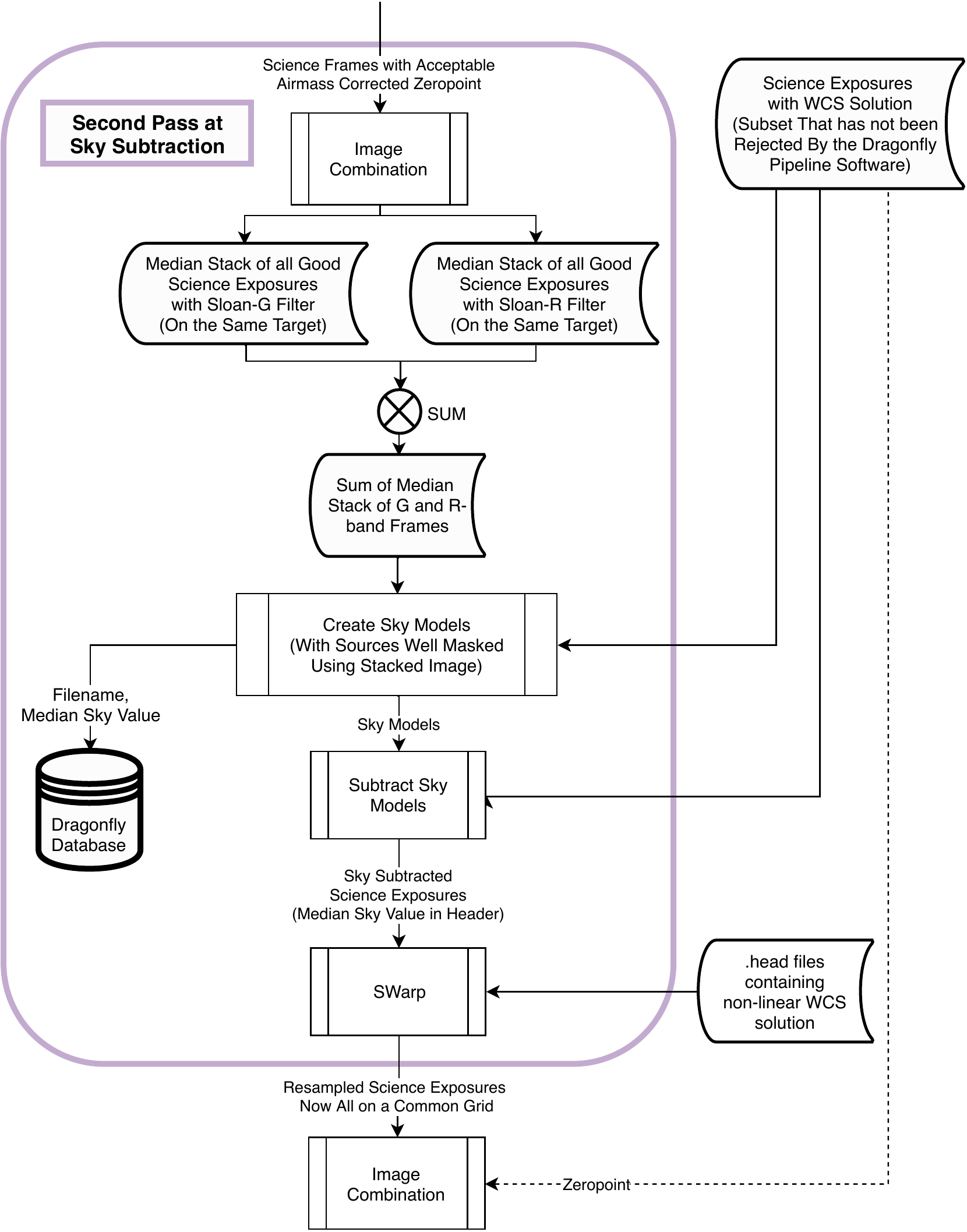}
\caption{Flowchart for the section of the Dragonfly Pipeline that does the second pass at sky subtraction and the final preparation for and carrying out of image combination.}
\label{fig:pipelineflow_final}
\end{figure*}

A diagrammatic representation of the architecture of the Dragonfly Pipeline Software is presented as a set of flowcharts in Figures~\ref{fig:pipelineflow_cal},~\ref{fig:pipelineflow_init}, and~\ref{fig:pipelineflow_final}. Each conceptually-independent modular script is represented with a rectangular box. Inputs and outputs are represented by labeled arrows. Notice that sky subtraction procedures appear twice in the flowcharts, i.e. in both Figure~\ref{fig:pipelineflow_init} and~\ref{fig:pipelineflow_final}. This is because sky subtraction is done in two stages. More details on this are provided below. Each of the three figures illustrate the data flow of each one of the three major sections of the pipeline, which are as follows: 
\begin{enumerate}
    \item Primary rejection of problematic science exposures, and calibration of remaining science exposures
    \item Secondary rejection of problematic science exposures; First stage of sky subtraction and image registration.
    \item Initial image combination, final sky subtraction, image registration and final image combination
\end{enumerate}

Individual modules of the Dragonfly Pipeline are written in Python 2.7 (mostly) and Perl. These are orchestrated by a top-level BASH shell script. The shell script times the operation of each step, and tracks the number of input, output and rejected frames at each step in the data flow. This timing and frame number tracking information is saved in a log file. An overview of each conceptually independent step will now be given in sequential order. To orient the reader, a miniature version of the flowchart containing the step of interest, with the step highlighted, will be shown to the left of the overview description of each step.


\newpage

\section{Overview of Steps in the Dragonfly Pipeline} \label{sec:pipelinesteps}
There are several image assessment gates throughout the pipeline which stop problematic science frames from moving forward in the pipeline flow. While an overview of all these gates are provided in this section, full details are provided separately in Section~\ref{sec:rejection}. This is to ensure the reader is oriented on an overview of how images are processed throughout the pipeline before being made aware of all the details regarding issues with Dragonfly data, as well as how they are dealt with. Similarly full details on image registration, scaling and stacking are presented in Section~\ref{sec:2p6}.

Data reduction is carried out for one target at a time. All raw science exposures taken of the target on all nights are collected, together with all the raw dark and flat exposures from corresponding nights. A full description of how this is done is reserved for later in the chapter as those details are not necessary to understand how the Pipeline works overall. See the dedicated ``Data Backup and Storage Structure" (Section~\ref{sec:databackupandaccess}) and ``Cloud-Orchestration of Software" (Section~\ref{sec:cloud-orch}) sections for these details. 

The steps shown in Figure~\ref{fig:pipelineflow_cal} will now be described. Namely, the rejection of bad dark, flat and a subset of science exposures, and science exposure calibration.

\begin{wrapfigure}{l}{0.3\textwidth}
  \begin{center}
  \vspace{-20pt}
    \includegraphics[width=0.3\textwidth]{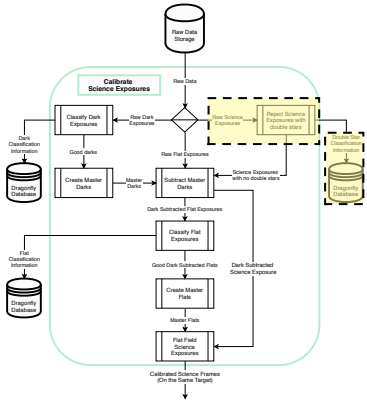}
  \end{center}
  \vspace{-20pt}
  \caption{Reject Science Exposures with Double Stars. This miniaturized flowchart is provided to orient the reader which step in pipeline flowcharts is being discussed in the text to the right of this Figure. The steps being discussed are highlighted using yellow boxes. See Figure~\ref{fig:pipelineflow_cal} for the full-size version of this flowchart.}
\end{wrapfigure}
\textbf{Reject Science Exposures with Double Stars}:
Occasionally a science exposure image taken by Dragonfly looks like there are two or more copies of every source in the image. We call these images exposures with ``double stars". The observing scenario where science exposures display this phenomenon is when the telescope is observing near the meridian. The likely cause of these ``double star" exposures is the telescope subsystem mounting hardware shifting or moving components internal to the lenses shifting, or both. When the telescope slews across the meridian, the physical load on the system switches directions, and may cause shifts in hardware. This automatic classification of whether a science exposure has ``double stars" can be done on images before calibration, so this piece of code is included in the ``calibration and rejection of double star science frames" stage of the pipeline. The algorithm for determining whether a science exposure has double stars is based on identifying the signal in the autocorrelation of an image with the locations of all sources in the original image marked. The exposures that were rejected and so do not continue onward in the pipeline are recorded in the Dragonfly Database, including the reason for rejection. Full details on how images with double stars are identified can be found together with details on other image assessment gates in Section~\ref{sec:rejection}.
\\

\newpage
\begin{wrapfigure}{l}{0.3\textwidth}
  \begin{center}
  \vspace{-20pt}
    \includegraphics[width=0.3\textwidth]{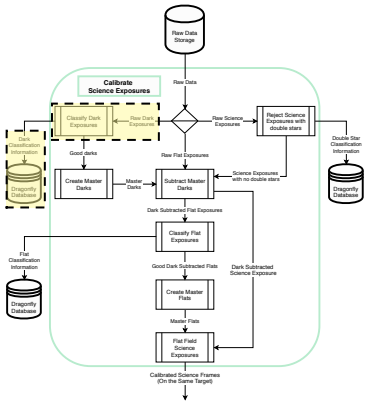}
  \end{center}
  \vspace{-20pt}
  \caption{Classify Dark Frames. See Figure~\ref{fig:pipelineflow_cal} for the full-size version of this flowchart.}
\end{wrapfigure}
\textbf{Classify Dark Frames}:
Dark frames are taken every night there is an observing run, with integration times that match the flat and science exposures taken. This is because dark exposures can vary both with temperature and time. Dark exposures taken on different nights might not match the flat and science exposures taken on any given night. A typical number of dark exposures taken on a night is between one and two thousand, spread across all the camera-lens subsystems. This means we need to be able to automatically classify whether a dark is good or bad. The rejection of bad dark exposures is based on measuring the RMS, median and structure in each image. The exposures that were rejected and so do not continue onward in the pipeline are recorded in the Dragonfly Database, including the reason for rejection. Full details on dark exposure classification can be found together with details on other image assessment gates in Section~\ref{sec:rejection}.\\

\begin{wrapfigure}{l}{0.3\textwidth}
  \begin{center}
   \vspace{-15pt}
    \includegraphics[width=0.3\textwidth]{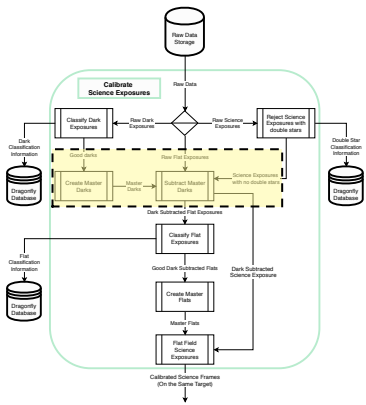}
  \end{center}
  \vspace{-15pt}
  \caption{Create Master Darks and Subtract Master Darks. See Figure~\ref{fig:pipelineflow_cal} for the full-size version of this flowchart.}
\end{wrapfigure}
\textbf{Create Master Darks and Subtract Master Darks}:
A single dark has the typical RMS of 30-40 ADU, this is about 2\% of a typical sky level in a science exposure on a moonless night. To minimize this contribution of noise to science exposures, at least four dark frames of the same integration time (and of course, the same camera-lens subsystem) are sigma clip average combined into a master dark. The sigma clipping ensures that any pixels affected by cosmic rays do not appear in the master dark. It is this master dark image that is subtracted from raw science exposures in the ``Subtract Master Darks" step. 

Flat exposures are also dark subtracted. The master dark for each flat exposure is created from dark exposures of matching integration time and lens-camera subsystem. However, unlike master dark frames for science exposures, the minimum number of raw dark exposures required to dark subtract flats is one. This is because the typical RMS in a flat exposure is around 200, and hence the noise contribution from the dark subtraction process is small, even if only a single dark is used.  \\

\newpage
\begin{wrapfigure}{l}{0.3\textwidth}
  \begin{center}
    \includegraphics[width=0.3\textwidth]{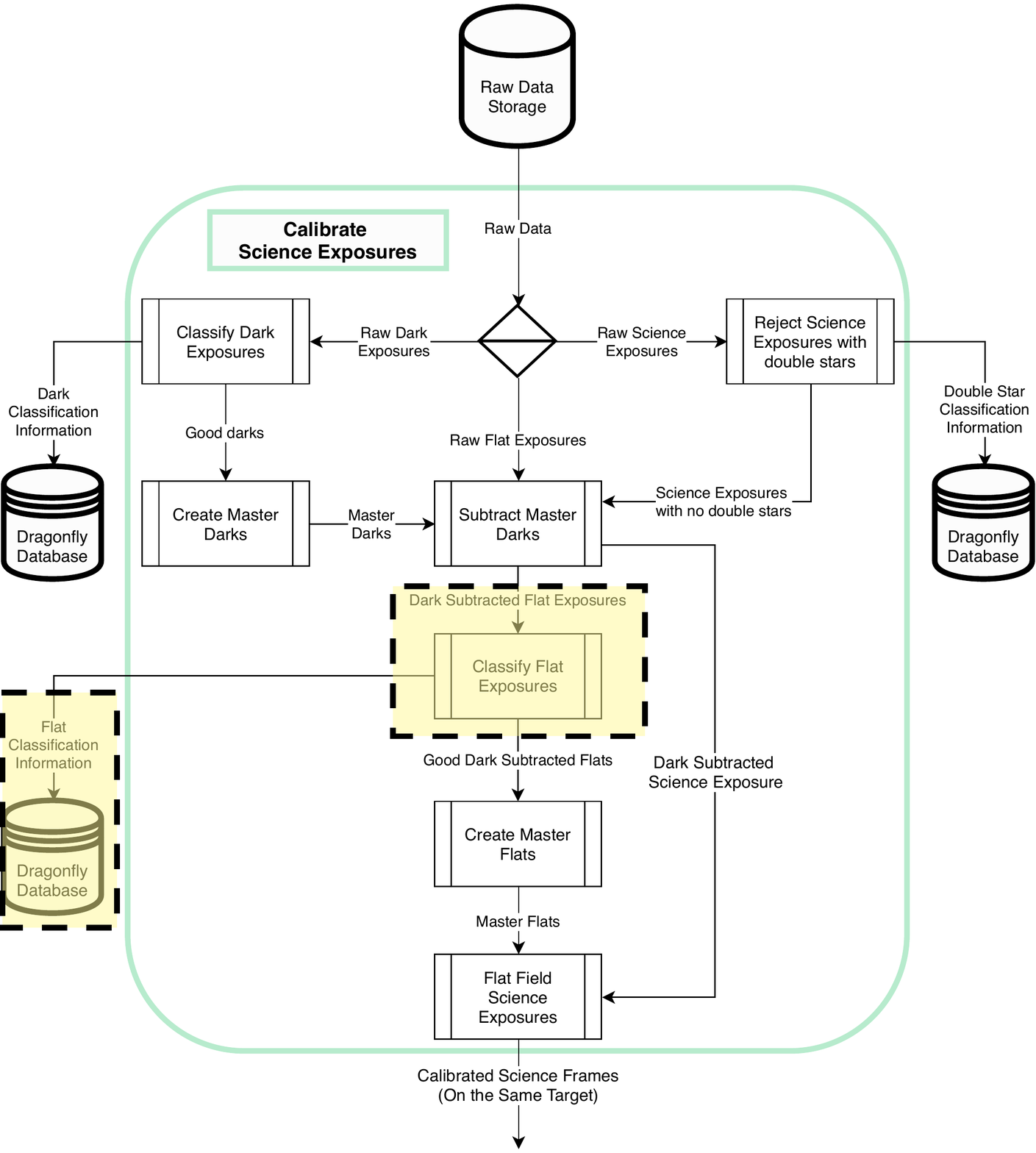}
  \end{center}
  \caption{Classify Flat Exposures. See Figure~\ref{fig:pipelineflow_cal} for the full-size version of this flowchart.}
\end{wrapfigure}
\textbf{Classify Flat Exposures}:
Flat exposures are taken every night there is an observing run. This is because over time lenses can collect dust, and hardware within each lens-camera subsystem on Dragonfly might shift slightly relative to each other, and so the flat field or illumination pattern on the CCD will also change with time. Furthermore, components of each lens-camera subsystem might be switched out occasionally to be inspected or fixed. Around 800 flat exposures are taken on a typical observing night, spread across the camera-lens subsystems. This means we need to be able to automatically classify whether a flat is good or bad. The rejection of bad flat exposures is done after dark subtraction. It is done based on assessing whether the moon is up, each image's median value, if there are too many stars in the image, and whether there are thin clouds or other large scale structures (other than the illumination pattern) in the image. A conservative approach is taken when classifying flats, if more than half of the flat exposures taken at the same time with the 48 lens-camera subsystems are classified as bad, they will all be classified as bad. This is because every night we observe, eight flat exposures per lens-camera subsystem are taken at twilight and again at dawn. We can afford to be conservative and throw out some flats that might be acceptable, but we cannot afford to include flats that might be bad, because incorrect flat fielding can cause science exposures to have uneven sky background, which is hard to model and subtract. 

The exposures that were rejected and so do not continue onward in the pipeline are recorded in the Dragonfly Database, including the reason for rejection. Full details on flat exposure classification can be found together with details on other image assessment gates in Section~\ref{sec:rejection}. \\

\begin{wrapfigure}{l}{0.35\textwidth}
  \begin{center}
    \vspace{-30pt}
    \includegraphics[width=0.35\textwidth]{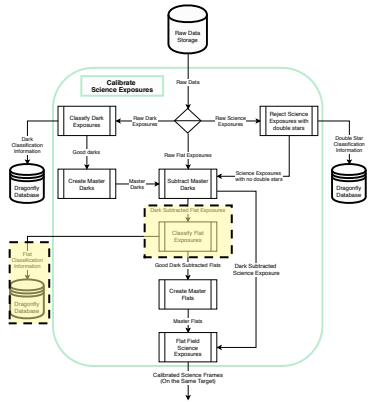}
  \end{center}
  \vspace{-30pt}
  \caption{Create Master Flats and Flat Field Science Exposures. See Figure~\ref{fig:pipelineflow_cal} for the full-size version of this flowchart.}
\end{wrapfigure}
\textbf{Create Master Flats and Flat Field Science Exposures}:
A single dark subtracted flat has a typical RMS of around 200 ADU, which is $\sim$1.5\% of the flat frame signal. To minimize the contribution of noise to science exposures during the flat division process, a master flat composed of at least seven dark subtracted flat exposures is created for each lens-camera subsystem. Flat exposures will typically have some stars in them, which should not be included when creating the master flat. The master flat creation process first masks any sources detected in each single dark subtracted flat exposure. After masking, each dark subtracted exposure is normalized such that the median value in each image is one. Then a median combine of all normalized dark subtracted flats taken with the same lens-camera subsystem is done to create the master flat.  

All dark subtracted science frames are then divided by the matching master flat of the same lens-camera subsystem using the ``Flat Field Science Exposure" script. \\

After this step in the pipeline, we have a set of ``calibrated" science exposures. \\

From here, calibrated science exposures are further prepared for image calibration. The steps shown in Figure~\ref{fig:pipelineflow_init} will now be described. These include the further assessment of science exposure quality, determination of a World Coordinate System (WCS), the first stage of sky subtraction, image registration and zeropoint calculation.\\

\begin{wrapfigure}{l}{0.3\textwidth}
  \begin{center}
  \vspace{-35pt}
    \includegraphics[width=0.3\textwidth]{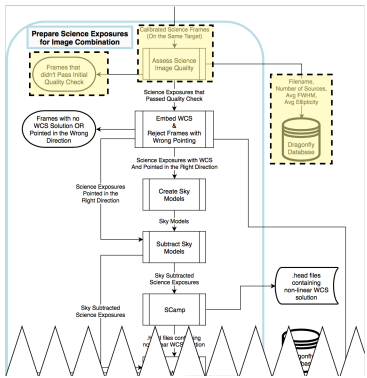}
  \end{center}
  \vspace{-25pt}
  \caption{Assess Science Image Quality. See Figure~\ref{fig:pipelineflow_init} for the full-size version of this flowchart.}
\end{wrapfigure}
\textbf{Assess Science Image Quality}:
Calibrated science frames that have gotten to this part of the pipeline can have a host of potential issues, making them unsuitable to be used at part of the final stacked image. At this stage, images are assessed according to the number of sources detected in the image, their average full-width half-max (FWHM) and their average ellipticity. These values are calculated using SExtractor, a software package that specializes in the creation of catalogs of sources in astronomical images~\citep{SExtractor}. This step removes from the pipeline flow images that are out of focus, taken with the shutter or dome closed or partially closed, taken while thick clouds covered the sky or are severely under-exposed. 

The calculated values of average FWHM and ellipticity, the number of detected sources in the image, and whether the image was rejected from continuing onward in the pipeline are all recorded in the Dragonfly Database. 
\\

\begin{wrapfigure}{l}{0.3\textwidth}
  \begin{center}
  \vspace{-20pt}
    \includegraphics[width=0.3\textwidth]{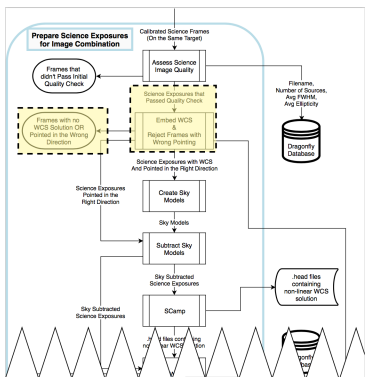}
  \end{center}
  \vspace{-20pt}
  \caption{Embed WCS and Reject Frames with Wrong Pointing. See Figure~\ref{fig:pipelineflow_init} for the full-size version of this flowchart.}
\end{wrapfigure}
\textbf{Embed WCS and Reject Frames with Wrong Pointing}:
Science exposures that have survived various pipeline image assessment gates up to this point are embedded with a WCS solution using Astrometry.net~\citep{astrometrynet}. Astrometry.net requires a set of index files, which contain a set of geometric hash codes that describe the relative positions of sets of (mostly) four stars in the image. Different sets of index files are needed depending on the field of view of the input image. For the Dragonfly field of view of about two by three degrees, the best index files to use are ``index-4208.fits", ``index-4209.fits" and ``index-4210.fits". Any images with no WCS solution or if the WCS solution of the image shows the lens was not pointing at the intended target is then rejected, and this information is recorded in the Dragonfly Database. 

Typically, the frames for which a WCS solution cannot be found are those which were not exposed to a sky with stars. Most of these frames should already be rejected in the ``Assess Science Image Quality" step immediately before this step. However, this offers another net to catch any problematic exposures. The number of frames that are rejected for pointing in the wrong direction have decreased with time as the mount control software has become more robust. 
\\

\begin{wrapfigure}{l}{0.3\textwidth}
  \begin{center}
    \includegraphics[width=0.3\textwidth]{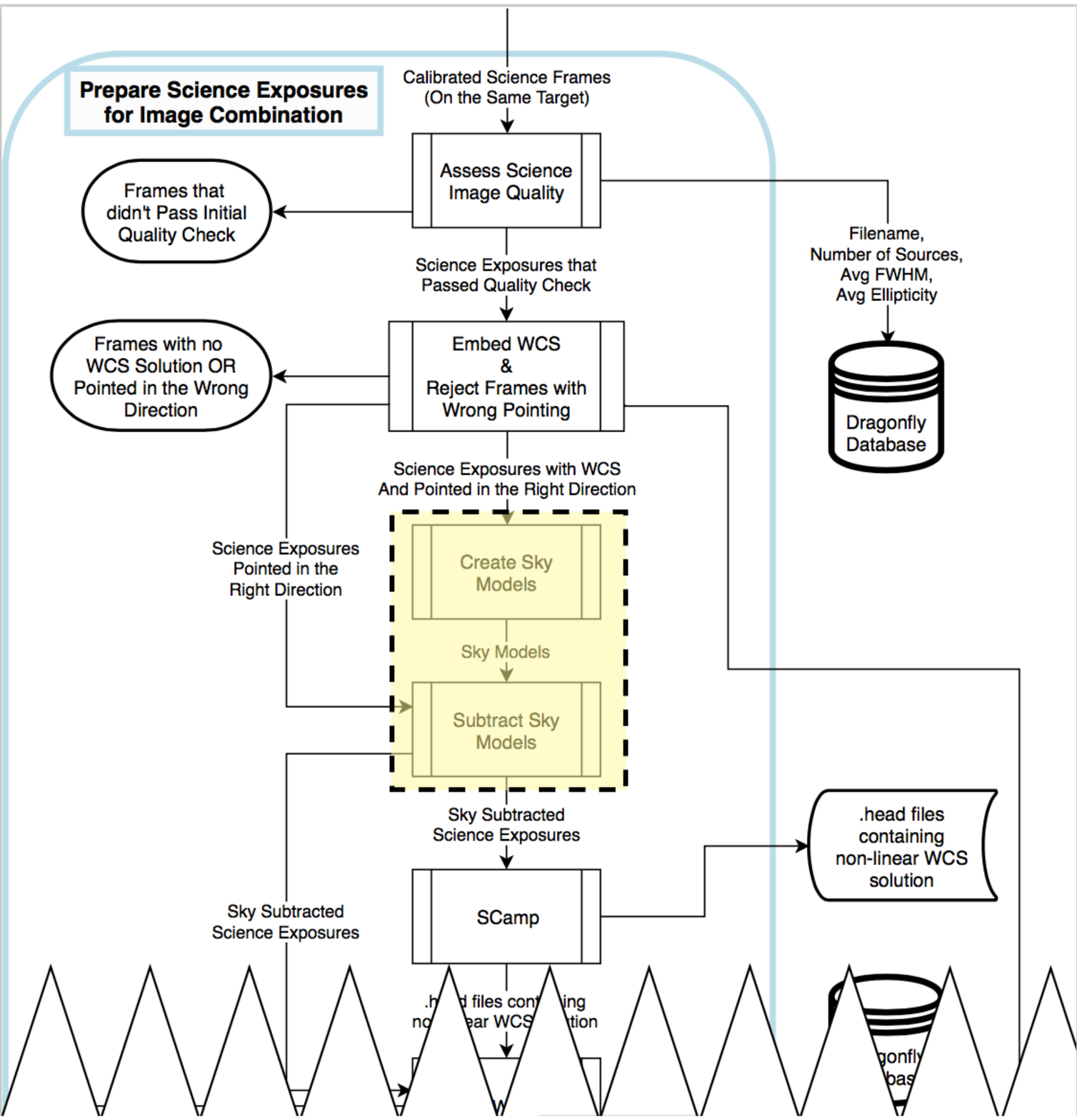}
  \end{center}
  \vspace{-20pt}
  \caption{Create Sky Models and Subtract Sky Models- Stage 1. See Figure~\ref{fig:pipelineflow_init} for the full-size version of this flowchart.}
  \label{fig:skysub_init_boxed}
\end{wrapfigure}
\textbf{Create Sky Models and Subtract Sky Models - Stage I}:
Sky subtraction is critical to enable optimal image combination. This is because image combination should average the astronomical signal separately the sky, and not the signal plus sky to achieve best signal to noise. The sky level cannot be assumed to be uniform across the field of view. Thus, modeling and subtracting the sky is a critical step in the data reduction process of images optimized for low surface brightness studies, where the signal is $\sim$0.05\% the brightness of a dark, moonless night sky. 

The main goal is to not over or under-subtract the sky, as that can have a large effect on the image combination procedure, as well as the measured value of the astronomical signal. To ensure this, the sky model should only be fit to pixels with no astronomical signal. This means all pixels containing sources need to be masked before fitting the sky model. This is especially important in the region around the galaxy of interest. There is a real risk of over-subtracting the local background around bright galaxies. This is because in a single exposure, it is difficult to determine the extent of the mask around individual galaxies, which may have low surface brightness regions that are below detection. These low surface brightness pixels, if left unmasked, will systematically raise the value of the measured sky background in that region. The counter situation of over-masking should also be avoided, if too much of the area around the galaxy is masked, then it is difficult for the model to be able to accurately account for the background in the region around the galaxy.

This is the reason why, as noted above, the ``Create Sky Models" and ``Subtract Sky Models" steps occur twice in the Pipeline, once in the flowchart shown in Figure~\ref{fig:pipelineflow_init}, and again in Figure~\ref{fig:pipelineflow_final}. In this first stage shown in Figure~\ref{fig:pipelineflow_init} (replicated in miniature in Figure~\ref{fig:skysub_init_boxed}), the masks for each image are created based on single exposures. After the first stage, images are processed up till image combination, then a median image is created out of all good science exposures. This intermediate median stacked image is used to create source masks in the second stage of sky subtraction in Figure~\ref{fig:pipelineflow_final}. Improper sky subtraction is a source of systematic error, and so this is a critical step in the Dragonfly Pipeline. Full details on how this is done can be found in the ``Controlling Systematic Errors" section of this chapter, namely Section~\ref{sec:skysub}. \\

\newpage
\begin{wrapfigure}{l}{0.3\textwidth}
  \vspace{-20pt}
  \begin{center}
    \includegraphics[width=0.3\textwidth]{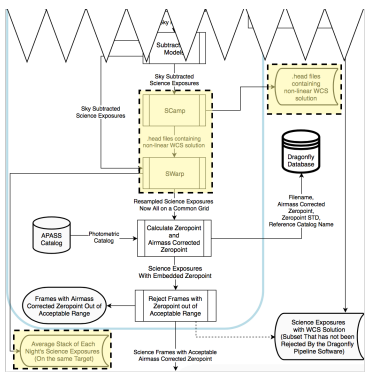}
  \end{center}
  \vspace{-16pt}
  \caption{Scamp and SWarp. See Figure~\ref{fig:pipelineflow_init} for the full-size version of this flowchart.}
  \label{fig:scampswarp1}
\end{wrapfigure}
\textbf{Scamp and SWarp}:
After sky models are subtracted, the science exposures are ready for image registration. Image registration is done using three software packages provided by astromatic.net. The software packages are SExtractor~\citep{SExtractor}, SCAMP~\citep{Scamp} and SWarp~\citep{Swarp}. Inside the ``Scamp" modular script, SExtractor is run to extract sources and their positions into a catalog. SCAMP then uses the catalog of source positions to calculate a non-linear astrometric solution by comparing source these positions to those in an online catalog. This solution is stored in a text file with a .head file extension in standard WCS format. Inside the ``SWarp" modular script, SWarp takes the .head files, each of which contains the astrometric solution calculated by SCAMP for an image, and resamples the input images to a common grid. 

SWarp is not only able to register images to a common grid, it is also able to stack images. At this step of the pipeline, SWarp is also configured to output an average combined frame per night of data taken of the target being reduced. This is useful for inspecting whether a subset of the data is behaving well. For example, if the final combined image is problematic, it is easier to locate which subset of data might be the culprit by inspecting these nightly coadds. 

More details on SCAMP, SWarp, and image registration can be found in Section~\ref{sec:imagereg}. \\

\begin{wrapfigure}{l}{0.3\textwidth}
  \begin{center}
  \vspace{-15pt}
    \includegraphics[width=0.3\textwidth]{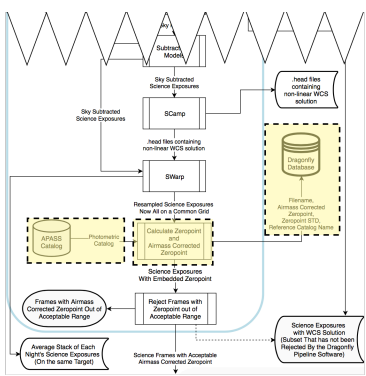}
  \end{center}
  \vspace{-15pt}
  \caption{Calculate Zeropoint and Airmass Corrected Zeropoint. See Figure~\ref{fig:pipelineflow_init} for the full-size version of this flowchart.}
  \label{fig:zpstepoverview}
\end{wrapfigure}
\textbf{Calculate Zeropoint and Airmass Corrected Zeropoint}:
So far, all images are saved with ADU units. Exposures taken with the 48 different lens-camera subsystems will have different ADU levels, even though the exposure time and the part of the sky being observed is the same. This is because each CCD has its own slightly different sensitivity, coupled with the fact that each optical system will have slightly different attenuation of incoming light. The flux of sources in ADU in exposures taken at different times with the same camera-lens system will also differ, for example, when observing through different amounts of airmass. This means, before an average value can be found of the astronomical signal (after sky subtraction), all images need to be normalized to a common flux level. This is done using the zeropoint of each image. 

The zeropoint of each registered science exposure is calculated by comparing the photometry of all point sources in the image to a reference photometric reference catalog. In our case, the photometric catalog used is the AAVSO Photometric All-Sky Survey\footnote{APASS catalogs contain photometry for five filters: Johnson B and V, and Sloan \textit{g}, \textit{r}, and \textit{i}. It covers a magnitude range from about 7th magnitude to about 17th magnitude.} (APASS,~\cite{APASS2016}). Typically, the zeropoint is calculated using about 1000 sources. 

The extinction due to the Earth's atmosphere in each of the \textit{g} and \textit{r}-bands can be used to find an airmass corrected zeropoint for each image. This zeropoint will no longer include variations due to observing through different airmass. The zeropoint, airmass corrected zeropoint, zeropoint standard deviation and reference catalog name are written into the Dragonfly Database.\\

\begin{wrapfigure}{l}{0.3\textwidth}
  \begin{center}
  \vspace{-25pt}
    \includegraphics[width=0.3\textwidth]{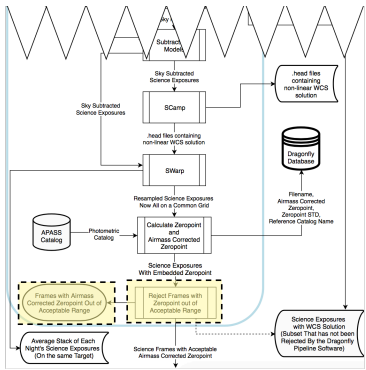}
  \end{center}
  \vspace{-20pt}
  \caption{Reject Frames with Zeropoint out of Acceptable Range. See Figure~\ref{fig:pipelineflow_init} for the full-size version of this flowchart.}
\end{wrapfigure}
\textbf{Reject Frames with Zeropoint out of Acceptable Range}:
The airmass corrected zeropoint should be quite stable for each lens-camera subsystem through time. Any changes in this value could be due to exposure time change, or hardware changes in the subsystem. These are carefully tracked. Changes in atmospheric conditions can also change the airmass corrected zeropoint of images taken with the same lens-camera subsystem. Images taken through a certain type of atmospheric condition produces a lot of power in the wide-angle PSF described in the introduction. These images should be rejected from continuing in the pipeline. The purpose of this step in the pipeline is to exclude these frames from continuing in the pipeline. Anomalous wide-angle PSFs are a source of systematic error, and so this is a critical step in the Dragonfly Pipeline. Full details on how this is done can be found in ``Controlling Systematic Errors" section of this chapter, namely Section~\ref{sec:pipe:wide-angle PSF}.  \\

After this step, we have a set of registered science exposures that are taken in nominal atmospheric conditions. 

The steps shown in our third flowchart presented in Figure~\ref{fig:pipelineflow_final} will now be described. This covers how registered images are combined, and how this initial combined image is used to do a second stage of sky modeling and subtraction, followed by image registration and final image combination. 

\newpage
\begin{wrapfigure}{l}{0.35\textwidth}
\vspace{-25pt}
  \begin{center}
    \includegraphics[width=0.35\textwidth]{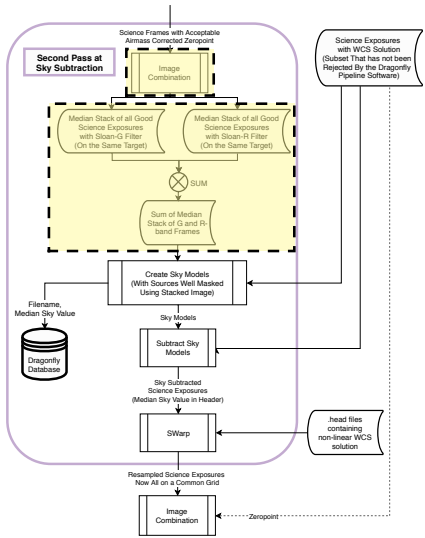}
  \end{center}
  \vspace{-25pt}
  \caption{Image Combination. See Figure~\ref{fig:pipelineflow_final} for the full-size version of this flowchart.}
  \label{fig:imagecomb1_boxed}
\end{wrapfigure}
\textbf{Image Combination Part 1: The Giant Median Coadd}:
\noindent
This initial image combination step is straightforward. A median image is calculated for each of the \textit{g} and \textit{r} filter bands, then summed as indicated in the flowchart in Figure~\ref{fig:pipelineflow_final} (reproduced to the left in miniature in Figure~\ref{fig:imagecomb1_boxed}). A median image is created as this ensures satellite trails, cosmic rays and other image artifacts will not be present in the combined image. The reason that \textit{g} and \textit{r}-band images need to be separately median combined is because sources have different flux in the two bands. Taking a median of values that are not drawn from the same distribution is pointless. 

We refer to this sum of the two median stacks of  \textit{g} and \textit{r}-band frames as the ``giant median coadd". This term will be referred to below in the sky modeling step. 

\vspace{50pt}

\begin{wrapfigure}{l}{0.35\textwidth}
  \begin{center}
  \vspace{-25pt}
    \includegraphics[width=0.35\textwidth]{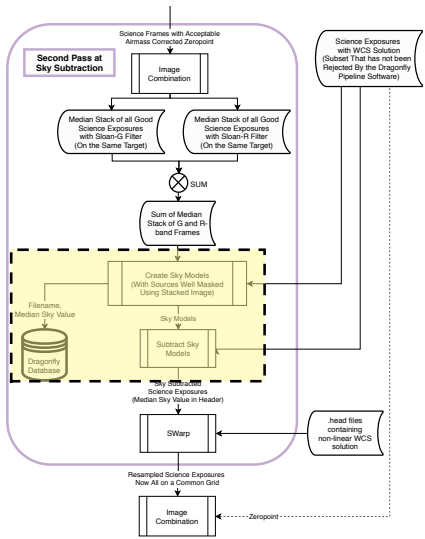}
  \end{center}
  \vspace{-25pt}
  \caption{Create Sky Models and Subtract Sky Models Stage 2. See Figure~\ref{fig:pipelineflow_final} for the full-size version of this flowchart.}
\end{wrapfigure}
\textbf{Create Sky Models and Subtract Sky Models- Stage II}:
As in the first stage of sky modeling, the sky model should only be fit to pixels with no astronomical signal. This means all pixels containing sources need to be masked before fitting the sky model. The difference between this second stage and the first stage of sky modeling is that here, the source mask is developed from the giant median coadd (produced in the step immediately above this step). The faint outskirts of galaxies is much more pronounced in this giant median coadd image. 

The inputs to this stage of sky subtraction are the science exposures immediately after being embedded with a WCS solution. This means they have not been registered to the common grid of the giant median coadd. To create the mask for each science exposure, the giant median coadd is re-sampled to grid of that exposure. Once new sky models are created, they are subtracted from the science exposures. The newly sky-subtracted science images are ready for registration and the final image combination step. Details on exactly how the mask is created out of the giant coadd and the whole two stage sky subtraction procedure can be found in the ``Controlling Systematic Errors" section of this chapter, namely Section~\ref{sec:skysub}. \\

\newpage
\begin{wrapfigure}{l}{0.3\textwidth}
  \begin{center}
    \vspace{-25pt}
    \includegraphics[width=0.3\textwidth]{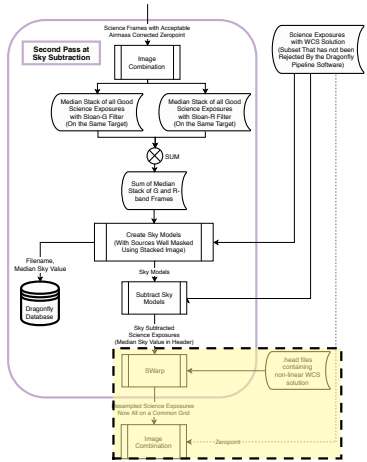}
  \end{center}
  \vspace{-25pt}
  \caption{SWarp and Image Combination Final. See Figure~\ref{fig:pipelineflow_final} for the full-size version of this flowchart.}
\end{wrapfigure}
\textbf{SWarp and Image Combination Final}:
The images are registered to a common grid in the same way they are immediately after the first stage of sky subtraction (see Figure~\ref{fig:scampswarp1} and associated text). Notice that here, the step labelled ``SCAMP" is not called again. This is because the astrometric distortions do not change even if the sky model subtracted from the image is slightly different to the model calculated in the first stage of sky subtraction. This means that the .head files calculated in the ``SCAMP" step in Figure~\ref{fig:pipelineflow_init}, and reproduced in miniature in Figure~\ref{fig:scampswarp1} is used here again, and not re-calculated. SWarp takes in the .head files produced earlier on in the pipeline and registers the newly sky subtracted images to a common grid. 

Once the science exposures are registered, the non-airmass-corrected zeropoints used to scale frames for image combination (calculated in the ``Calculate Zeropoint and Airmass Corrected Zeropoint" step in Figure~\ref{fig:pipelineflow_init}, and reproduced in miniature in Figure~\ref{fig:zpstepoverview}) are also copied over from the previous calculation. Now these registered science exposures are ready for final image combination. An average combined image produces the highest signal to noise. Sigma clipping is used to deal with satellite trails and cosmic rays. The images are combined using a weighted average, where the weight is the signal to noise of each image. Details on exactly how this is all done can be found in Section~\ref{sec:imagecombination}.

Having sketched out the operation of the pipeline, we now turn to the rationale for some of the choices made in the description given above. 

\section{Controlling systematic errors} \label{sec:systematics}
\subsection{Sky background subtraction} \label{sec:skysub}
The purpose of Dragonfly is to observe and characterize low surface brightness regions, the aim is to be able to routinely detect signal below 30 mag arcsec$^{-2}$. The absolute flux or photometry of low surface brightness regions can be greatly affected by the accuracy of the sky measurement. This is because the signals we are trying to measure are about 10,000 times dimmer than the surface brightness of the night sky on a moonless night (in the \textit{g}-band, this is about 22 mag arcsec$^{-2}$ at New Mexico Skies). In other words, if we get the sky level slightly wrong, the brightness of the astronomical source will come out very wrong. This is a particularly significant problem in the low surface brightness outskirt regions of galaxies because bad sky subtraction can mimic a host of physical effects, such as disk truncation or the existence of a stellar halo.

As already described, for each individual Dragonfly science exposure, a sky background is spatially modeled and subtracted. High precision sky subtraction and scaling of images to a common zeropoint are needed before images can be median combined (or sigma clipped before averaging). A median combination (or sigma clip average) of sky-subtracted images allows removal of anomalous features such as satellite trails, cosmic rays and scattered light particular to any lens-camera subsystem. It is critical not to over-subtract or under-subtract the sky, as that can have a large effect on the measured value of the astronomical signal (as described in the introduction of this thesis). To ensure that the sky background model is only being fit to pixels with no signal from sources aside from the sky, all pixels containing sources need to be masked. 
This is especially important in the local region around the galaxy of interest. There is a real risk of over-subtracting the local background around the respective galaxy. This is because in a single exposure, it is difficult to determine the extent of the mask around individual galaxies, which may have low surface brightness regions that are below detection. These low surface brightness pixels, if left unmasked, will systematically raise the value of the measured sky background in that region. The counter situation of over-masking should also be avoided, if too much of the area around the galaxy is masked, then it is difficult for the model to be able to accurately account for the background in the region around the galaxy. As has already been noted, to ensure optimal masking and that the most accurate sky model is subtracted, this process is done in two stages in the Dragonfly Pipeline Software. We now provide details for this process. 

The first stage of sky removal occurs directly after the calculation of a WCS solution (see Figure~\ref{fig:pipelineflow_init}). SExtractor\footnote{SExtractor is part of the astromatic.net suit of software. It is designed to create catalogs of sources and their photometry parameters for an input image. Part of this photometry procedure is background subtraction.}~\citep{SExtractor} is used to create a background map of each science exposure with a background mesh size of 128 by 128 pixels. SExtractor finds the sky level within each background mesh cell by iteratively sigma clipping until the spread of remaining pixel values is within three sigma of the median of these values. This does a good job at rejecting bright point source pixels within the cell. Then, a third order polynomial is fitted to this background map and subtracted from each individual exposure (the ``create sky model" and ``subtract sky model" procedures in Figure~\ref{fig:pipelineflow_init}). The mesh size of 128 by 128 pixels was chosen to minimize the effect that sources have on the background estimation within each mesh cell\footnote{If the mesh size is too small, then there may not be enough sky pixels in the mesh cell for the iterative sigma clipping method to identify the sky value in the cell.}, while still retaining information about the background variation on a small enough scale\footnote{The mesh cell size should be smaller than the scale on which the fitted polynomial varies.}. After the first stage of sky modeling and subtraction, an initial combined image is produced. This combined image is the sum of the median of all images taken with a \textit{g}-band filter and the median of all images taken with a \textit{r}-band filter, as shown in the top half of the flowchart in Figure~\ref{fig:pipelineflow_final}. Despite SExtractor's source rejection algorithm during its calculation of the background in this first stage of sky subtraction, pixels belonging to low surface brightness sources extended over large angular scales will not be rejected, and instead, will likely be used to form the background map that is fit and subtracted. This is why a second stage of sky subtraction is needed.

In the second stage of sky subtraction (illustrated in the bottom half of the flowchart in Figure~\ref{fig:pipelineflow_final}), a sky model is again fit to each science exposure. The sky background modeling is also carried out using a similar process to the procedure in the first stage. SExtractor~\citep{SExtractor} is used to create a background map of each science exposure with a background mesh size of 128 by 128 pixels. The difference in this second stage is that a weight map is also input into SExtractor during background map making. The weight map masks all sources all the way out to their low surface brightness outer edges using the information from the initial combined image (the ``giant median coadd"). The weight map is 1 for sky pixels, and 0 for pixels where a source is detected. In the giant median coadd, a pixel is considered to not include only sky signal, but also signal from a source if it is brighter than the median pixel value of the giant median coadd from stage one or within 12 pixels of such a pixel. This careful masking procedure ensures we do not over subtract the sky by fitting a sky model to ultra-faint galaxy light that is undetected in individual exposures. The median sky level is calculated in this second stage of sky subtraction, and this value is recorded in the Dragonfly Database. 

\subsection{The wide-angle point-spread-function} \label{sec:pipe:wide-angle PSF}

On its own, careful sky subtraction is still not enough to fully exploit the potential of Dragonfly images. A critical ingredient in the Dragonfly Pipeline is to carefully account for what may be the ultimate limiting factor in low surface brightness observations, namely the wide-angle (degree-scale) PSF. As described in the introduction of this thesis, this largest-scale component of the PSF is also named the ‘aureole’ in literature~\citep{King1971}. If a large fraction of light is distributed at wide angles away from the central bright region of the point-spread function, it can significantly affect the measured surface brightness profile of galaxies. 

\subsubsection{Effect of a significant wide-angle point-spread function}

\begin{figure*}[!htbp]
\centering
\includegraphics[width=1.0\textwidth]{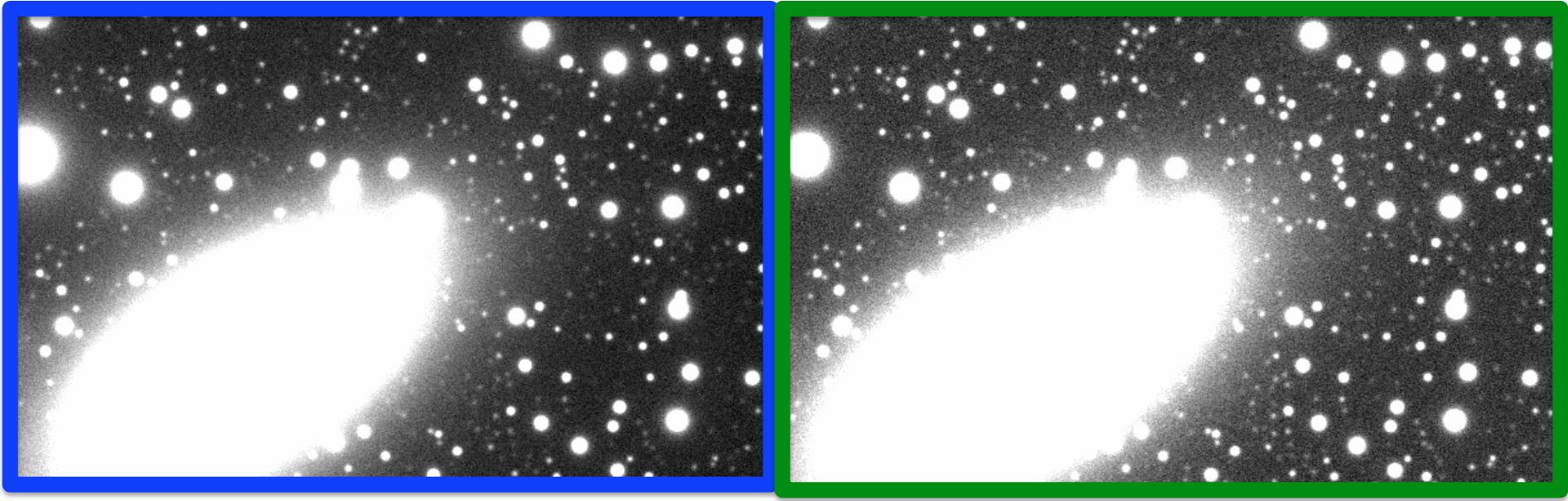}
\caption{Simulated image containing a galaxy and a field of stars. The parameters for the simulated galaxy are based on those of NGC 2841, and the position and magnitudes of stars simulated are those around NGC 2841. The image boxed in blue has a brighter stellar aureole, lowering the zeropoint of this image by 0.1 mag.}
\label{fig:SimGals}
\end{figure*}

\begin{figure*}[!htbp]
\centering
\includegraphics[width=1.0\textwidth]{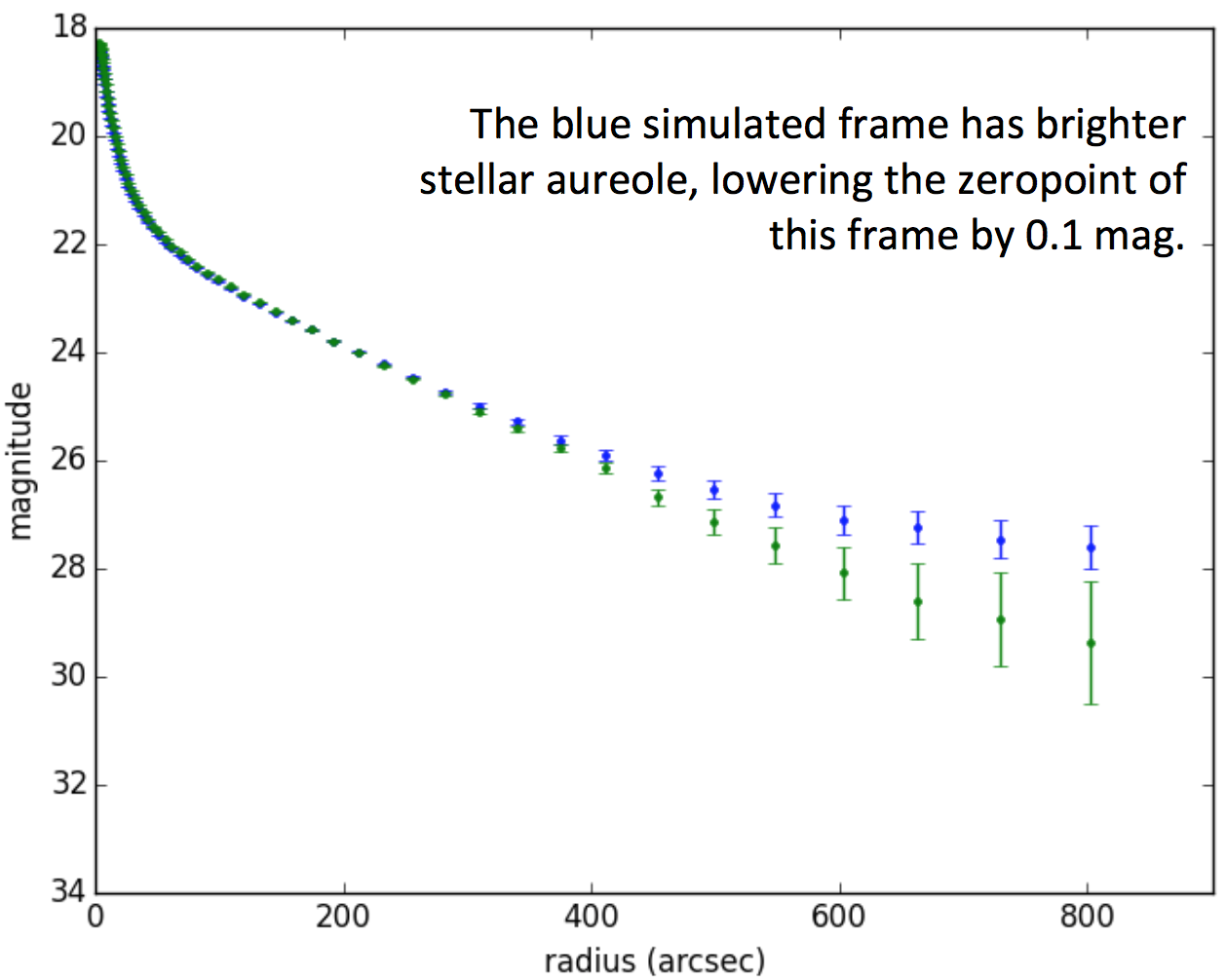}
\caption{Surface brightness profiles of two simulated galaxy images.}
\label{fig:SimGalProfiles}
\end{figure*}

To illustrate the importance of the aureole to modeling the outskirts of galaxies, two images were simulated using SkyMaker\footnote{SkyMaker is part of the astromatic.net suit of software and simulates realistic astronomical images.}~\citep{SkyMaker}. Each image simulates a stack of 400 Dragonfly science exposures, where the exposure time of each frame is 600 seconds. The simulated images each contain the same model galaxy but the point spread functions that are convolved with the model contain different levels of stellar aureole brightness for the two simulated images. Cutouts of the simulated images are shown in Figure~\ref{fig:SimGals}. The image boxed in blue was simulated with a stellar aureole component, while the image boxed in green was simulated with no stellar aureole component. Surface brightness profiles for the two model galaxies in each image are shown in Figure~\ref{fig:SimGalProfiles}. The blue surface brightness profile corresponds to the model galaxy in the simulated image boxed in blue in Figure~\ref{fig:SimGals}. The surface brightness profile of the galaxy in the simulation with a stellar aureole component appears to contain an abundance of stellar light at large radius, even though in reality it has none. The spurious profile light is simply contamination from the wings of the wide-angle PSF. 

\subsubsection{Rejecting science frames affected by a significant wide-angle point-spread function} \label{sec:aureole}

In conventional telescopes, the aureole is dominated by scattered light from internal optical components (e.g. The Burrell Schmidt Telescope~\cite{Slater2009}). In Dragonfly, the stellar aureole varies on a timescale of minutes, and so its origin is most likely atmospheric such as the presence of high atmospheric ice crystals~\citep{AtmIceCrystals2013}. Based on simulations such as that just shown, it is evident that science frames whose PSF has a significant aureole component should not go into the final stack of images. An efficient method for determining which frames should go into the final stack utilizing the photometric zeropoints of images is used in the Dragonfly Pipeline Software to detect the existence of atmospheric conditions which result in prominent stellar aureoles. 

It turns out that the image boxed in blue in Figure~\ref{fig:SimGals} has a zeropoint that is 0.1 mag lower than the image boxed in green. The zeropoint of each image was calculated using a Dragonfly Pipeline subroutine that compares stellar magnitudes in the image to those in a catalog. The fact that the strength of the wings in the wide-angle PSF leaves a record in the zeropoint determined by SExtractor provides us with a means for automatically identifying frames that are problematic~\footnote{The reason the image boxed in blue has a lower zeropoint than the one boxed in green in Figure~\ref{fig:SimGals} is because the sources in the image boxed in blue has some of their light thrown to large angles. The SExtractor routine which calculates the total flux of each source does not do full wide-angle PSF modeling\footnote{this is not done routinely in astronomy, and as seen in the introduction of this thesis, this wide-angle PSF modeling is a research topic in itself!}. SExtractor varies the aperture over which it calculates the flux of sources based on the brightness of the source, but it still does not account for the light thrown to degree scales away from the central source of light. This means if there is a significant aureole component of the PSF, all sources in this image will appear dimmer to SExtractor and the calculated zeropoint will be correspondingly smaller.}. The subroutine source extracts (using SExtractor) all sources in the input frame. The number of sources detected ranges typically from just under 1000 up to a few thousand. these sources are then matched to a photometric catalog using their RA and DEC positions. In most cases, we use the AAVSO Photometric All-Sky Survey (APASS) catalog~\citep{APASS2016}. The zeropoint of each matched source is calculated and the median zeropoint is then chosen to be the zeropoint for the whole image and saved into the Dragonfly Database. Also saved to the Database is the standard deviation of zeropoints. The photometric zeropoints of individual exposures are monitored and those with deviations from the nominal zeropoint for a given camera at a given airmass are identified and rejected. The nominal zeropoint for any camera subsystem is determined by aggregating the data taken from that camera over at least a month. The long time period gives confidence that there has been a range of atmospheric conditions, including a range of images with nominal observing conditions, and hence the highest zeropoint values. 


 
\begin{figure*}[!htbp]
\centering
\includegraphics[width=0.9\textwidth]{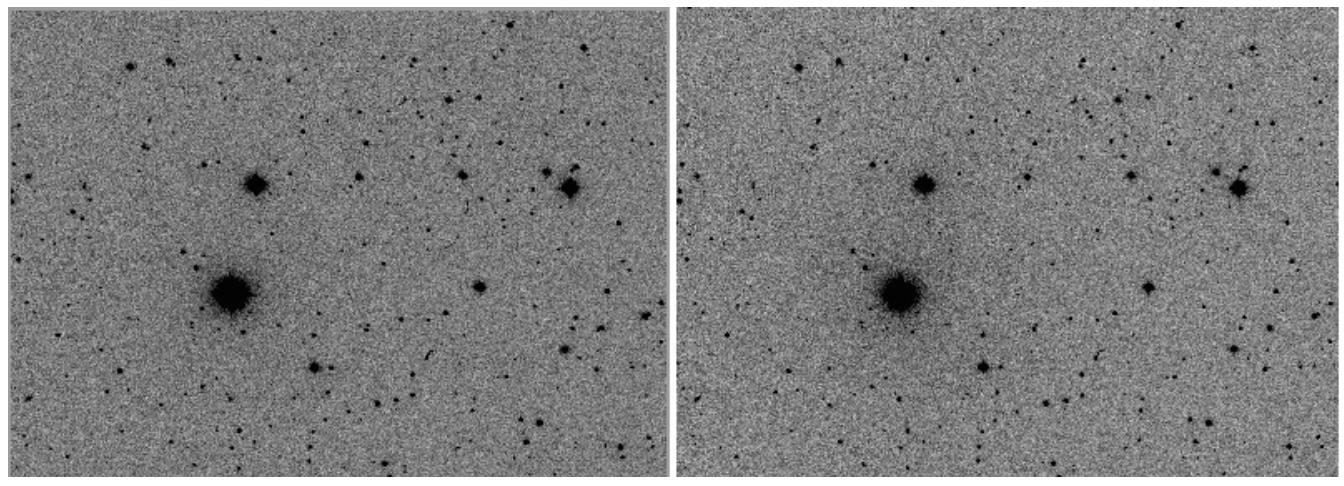}
\caption{Cutouts from two 600 second science exposures. The right hand side cutout is from a exposure that has a zeropoint 0.1 mag arcsec$^{-2}$ higher than the left hand side image.}
\label{fig:RealGalZPdiff}
\end{figure*}

Images with a zeropoint difference of 0.1 mag do not visually look very different, and would be otherwise difficult to separate. Figure~\ref{fig:RealGalZPdiff} shows a common cutout region from two 600 second exposure science images. The two images have an airmass corrected zeropoint difference of 0.1 mag. Just by looking at the images, it is almost impossible to tell that one of the images is hampered by a wide-angle PSF, however the image on the right is excluded from the final stacked image based on its zeropoint. The maximum deviation below the nominal zeropoint allowed in order to be accepted as a satisfactory science exposure is $\sim$0.2 mag. A less stringent upper bound is used to exclude certain frames that have very large zeropoints. Typically 25\% of science exposures are rejected in this step. 

\section{Automatic Rejection of Problematic Frames} \label{sec:rejection}
The Dragonfly Pipeline Software needs to be able to run unsupervised on large sets of data. This means that it needs the capability of rejecting dark, flat and science exposure frames that are in various ways problematic. This section describes how this is done for different types of issues. 

\subsection{Dark frames} \label{sec:pipe_darkrejection}
In theory, dark frames include contributions from several components. A bias structure, dark current and read noise. The bias structure can change with CCD temperature, and with time. The dark current can change with CCD temperature, exposure time, and also with time. Because neither the structure in the dark exposures nor the absolute value of the dark exposure is constant over time, we choose to take dark exposures every night we take science exposures, with integration times that match the flat and science exposures taken. This also means we need to be able to automatically classify whether a dark is good or bad. 

A dark is subjected to several tests in order to decide if it is acceptable or not. A dark is rejected if:
\begin{enumerate}
    \item It has an RMS that is larger than 60 or smaller than 15. 
    \item It has a median value that is larger than 2000 or smaller than 500.
    \item The range of values in the best fit model to the dark frame is too large.
    \item The best fit model to the dark is too different to the dark.
\end{enumerate}

One issue with some dark exposures is an occasional hardware malfunction that causes the shutter to open during a dark exposure. In these cases, the RMS and median value of the dark will be outside of the acceptable ranges in tests 1-2 above. The absolute value of the good dark exposures, with exposure times between 1 second to 600 seconds is relatively stable at around 1000 ADU. 

\begin{figure*}[!htbp]
\centering
\includegraphics[width=1.0\textwidth]{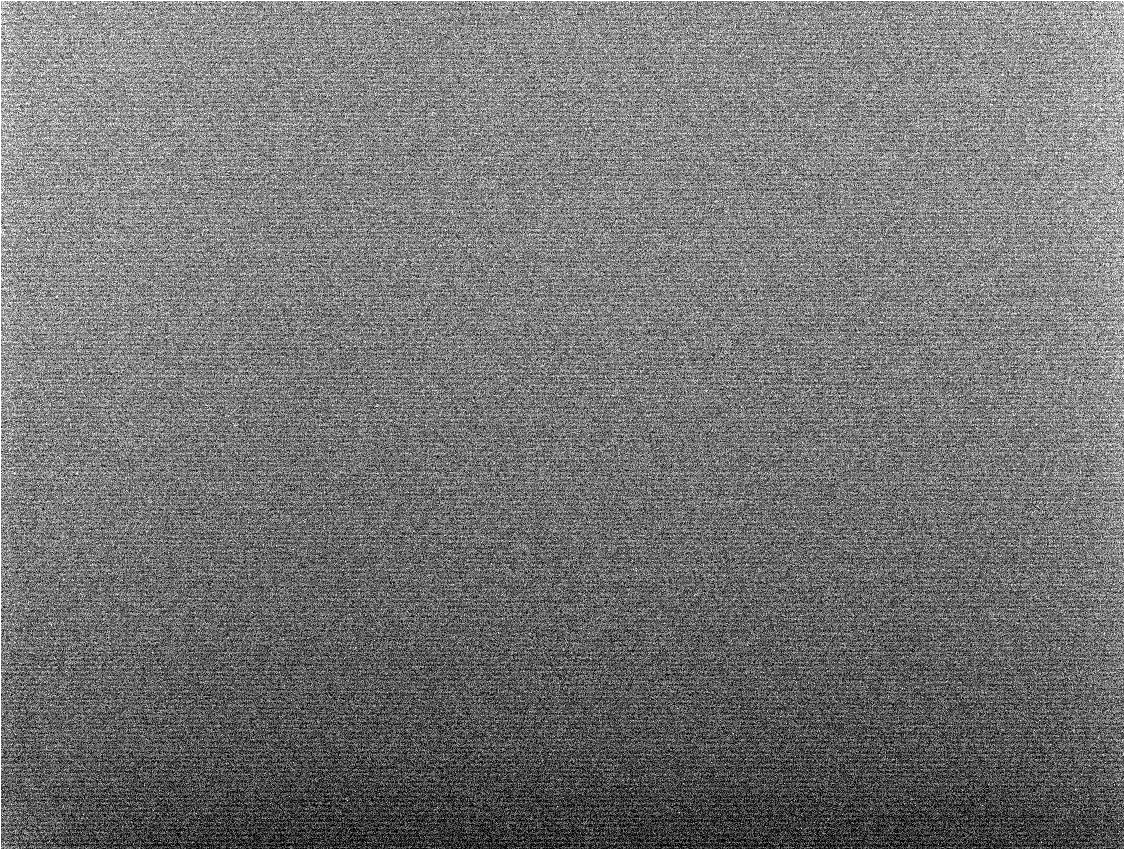}
\caption{An example dark exposure that has ridges in the image.} 
\label{fig:ridgedark}
\end{figure*}

\begin{figure*}[!htbp]
\centering
\includegraphics[width=0.95\textwidth]{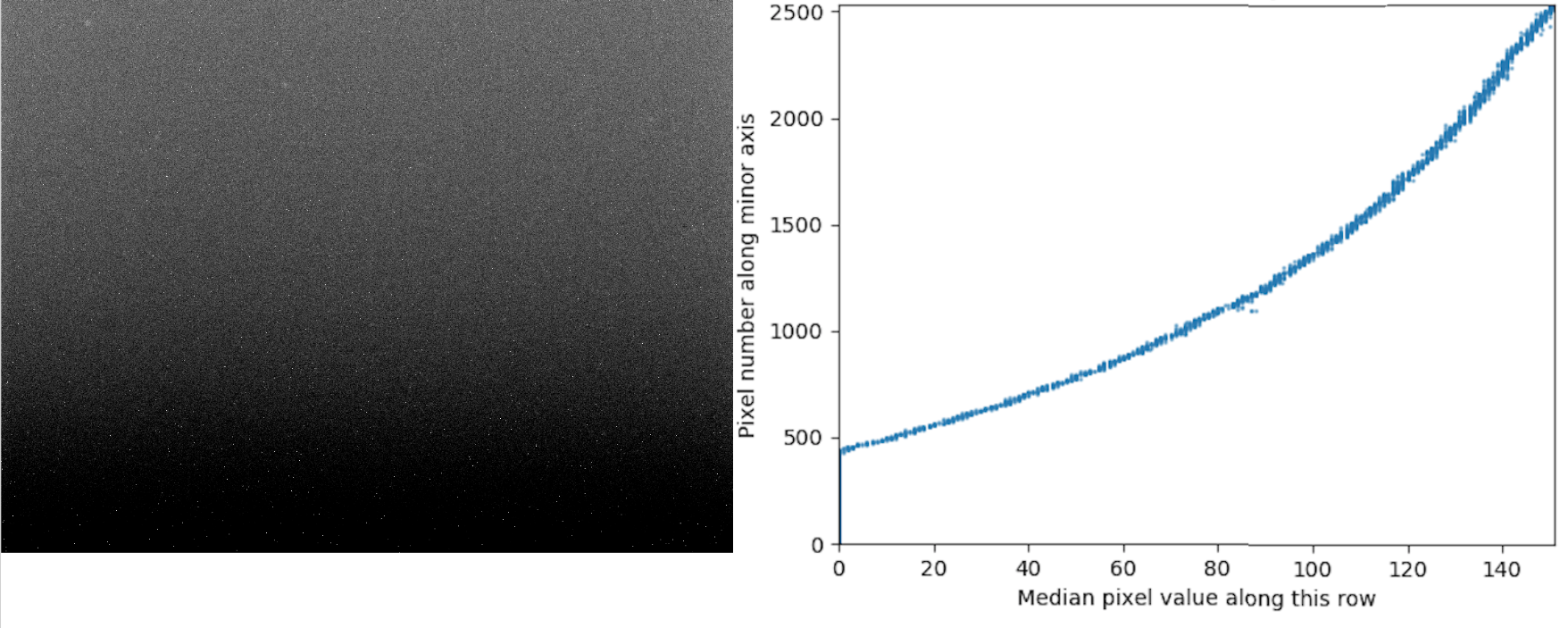}
\caption{An example dark where the large-scale gradient in the image is extreme. The image on the right hand side is a plot of the median value of each row in the dark. The pixel number is plotted on the y-axis to correspond to the vertical axis of the dark image on the left. The gradient is steep, and for a large fraction of rows at the bottom of the dark, the median value is zero.}
\label{fig:toogradientdark}
\end{figure*}

Another issue with some dark exposures is structure in the image that is very different from what it usually is. This can include extreme large-scale gradients or small-scale structure, such as ridges, in the image. The small-scale structure is associated with electrical interference during the exposure. Upon discovery of ridges in dark exposures, the wiring was adjusted and the ridges went away. The cause of large-scale gradients is currently not understood, however, it is often associated with times when the temperature regulation of the camera is not stable. Figure~\ref{fig:ridgedark} and~\ref{fig:toogradientdark} show examples of dark exposures that have ridges in the image, and large-scale gradients that are too steep, respectively. 

In order to reject dark exposures where the large-scale gradient is too steep, a model is fit to each dark. The model is created by fitting a 1D 5th order polynomial function to each vertical column of pixels, and then median filtering the result in the horizontal direction. If the ratio of the maximum to minimum value in the model is too large (larger than 1.1), then the image is rejected. In order to reject dark exposures with ridges or other small scale structure, a maximum limit (1.015) is set on the ratio of maximum to minimum value in the image created by dividing the dark exposure and model. 

\begin{figure*}[!htbp]
\centering
\includegraphics[width=0.95\textwidth]{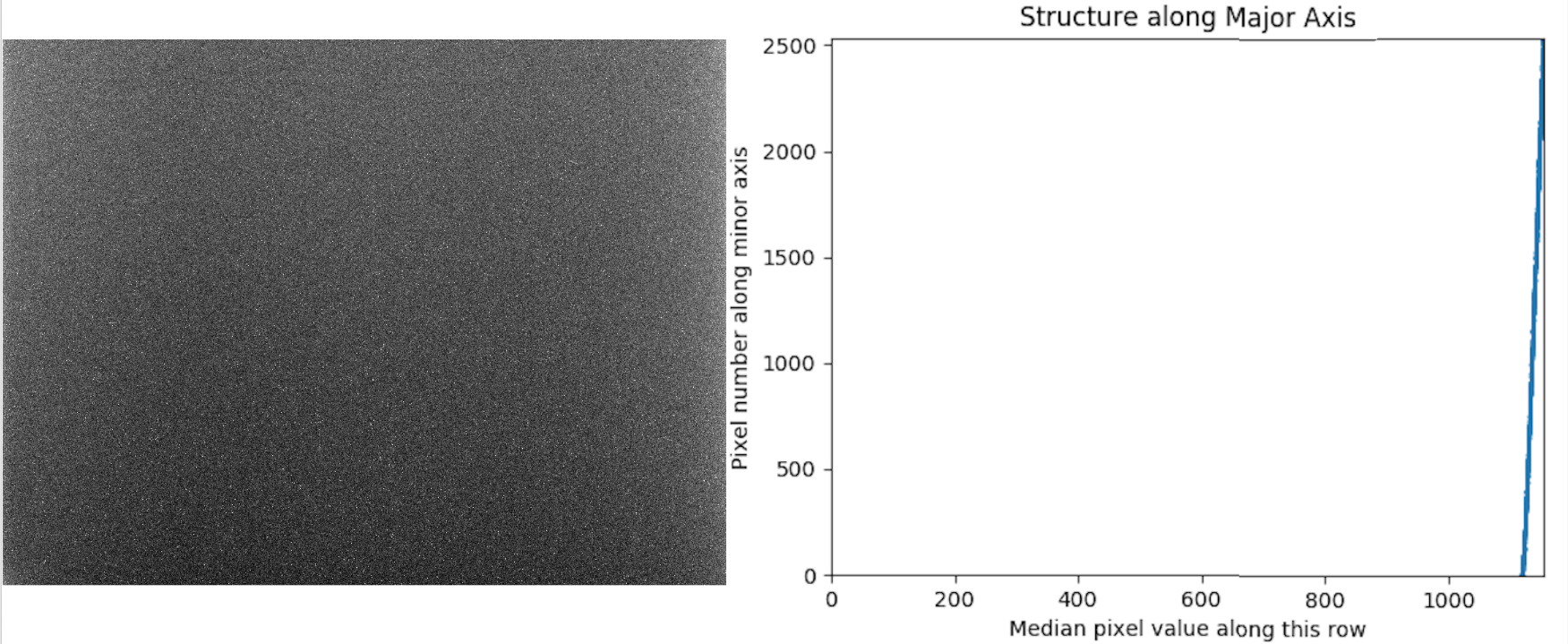}
\caption{An example of a good 600 second master dark. The right hand side is a plot of the median value of each row in the dark. The x and y range in this plot matches that in Figure~\ref{fig:toogradientdark} for easy comparison.} 
\label{fig:goodmasterdark}
\end{figure*}

An example of a good master dark frame is shown in Figure~\ref{fig:goodmasterdark} on the left hand side. The right hand side plots the median value of each row in the master dark. Compared to the bad dark shown in Figure~\ref{fig:toogradientdark}, the gradient in the vertical direction is much smaller.

\subsection{Sky Flats} \label{sec:pipe_flatrejection}
Each subsystem of the Dragonfly Telephoto Array has its own lens, filter, and camera. If the hardware configuration does not change, i.e. specific lenses, filters and cameras are always part of the same subsystem, if no dust ever collects on the lenses or filters, and if no piece of hardware ever shift slightly relative to any other, then the flat field or illumination pattern on the CCD will never change. In such a situation, we can make the perfect master flat for each subsystem and use it for all time. However, this ideal situation is not the case in the real world. Hardware components often need upgrades (since there are so many components, something changes at least once every few months), dust collect onto lenses and cameras can shift slightly. Therefore, we choose to take flats every night and use flats taken close in time to science exposures in order to flat field the science exposures. This means we need to be able to automatically classify whether a flat is good or bad.

A sky flat is subjected to several tests in order to determine its quality. We reject a flat or a series of flats if: 
\begin{enumerate}
    \item If the moon is above an altitude of minus three degrees.
    \item If the median of the flat is less than 5000 ADU.
    \item If the median of the flat is greater than 40,000 ADU.
    \item If there are lots of streaks in the flat.
    \item If a series of flats taken adjacent to each other in time changes significantly.
    \item If the number of flats that pass the above set of tests is less than 4 for the night
    \item If the number of flats exposed at the same time that passed the above tests is less than 50\%. 
\end{enumerate}

The sky background when a sky flat is taken needs to be uniform. This means there should be no clouds, gradients in the sky background or significant light pollution. If the moon is up, the sky background has a detectable gradient across the large field of view of Dragonfly, so flats taken when the moon is above minus three degrees altitude are rejected. If the sky is too dark when flats were obtained, then there will be more stars and unusable pixels, and the signal-to-noise of the flat illumination pattern will be too low. If the flat is too bright, then it is likely taken in the presence of clouds, light pollution or if the Sun hasn't set low enough below the horizon yet, and in any case strong illumination risks characterization of the flat at the non-linear portion of the CCD response curve.  

\begin{figure*}[!htbp]
\centering
\includegraphics[width=0.9\textwidth]{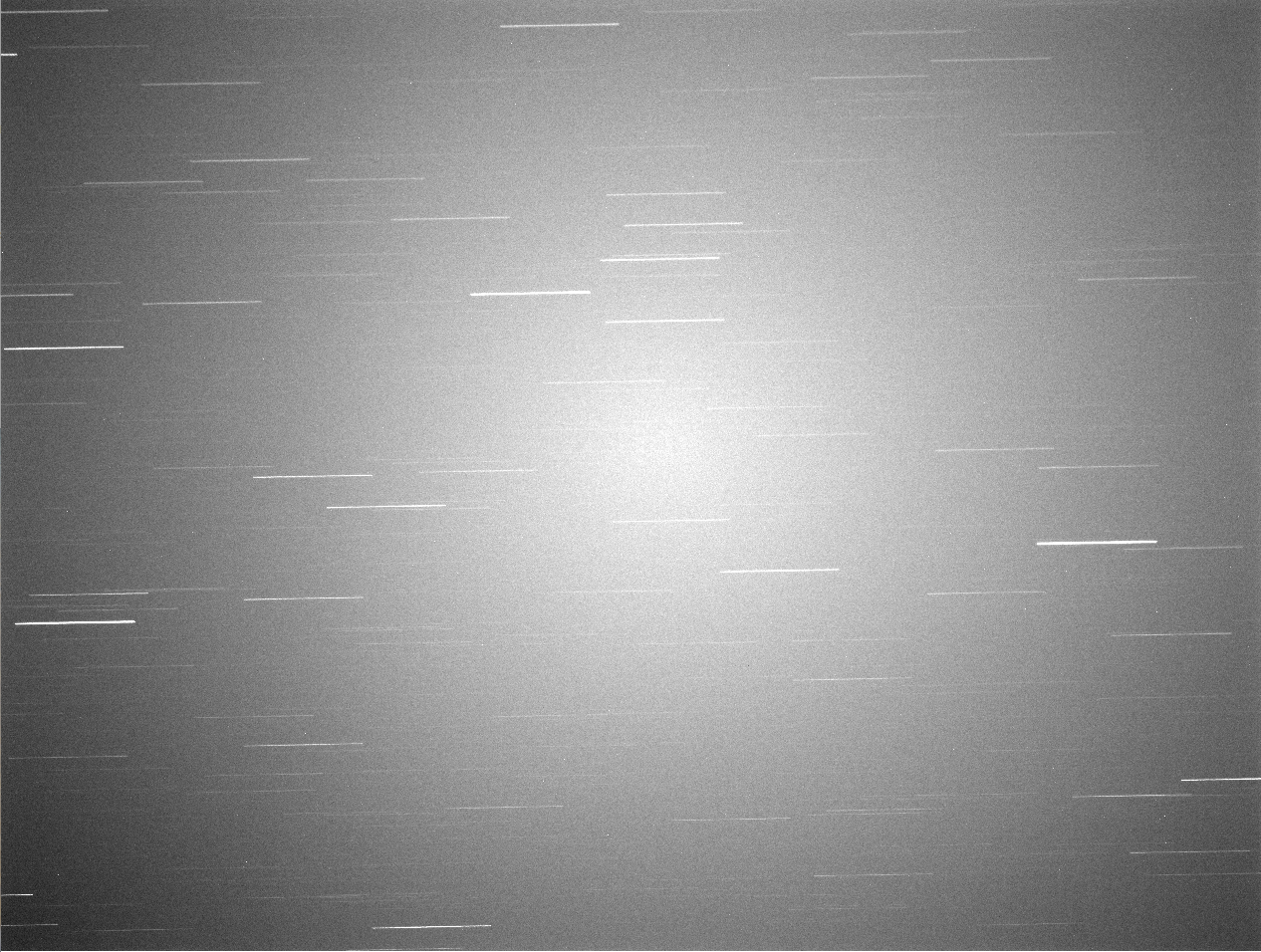}
\caption{An example flat with streaks due to stars being visible when the sky flat was taken. This flat would be rejected because it has too many streaks. The exposure time was 60 seconds, which corresponds to streaks with lengths of about 300 pixels.}
\label{fig:streakyflat}
\end{figure*}

When flats are taken, we track the movement of the sky. However, in certain cases mount tracking problems result in sky flats with streaks. An example image is in Figure~\ref{fig:streakyflat}. If there are not many, these streaks are not problematic because the affected pixels can be easily masked in the process of combining flats to create a master flat. However, if there are too many streaks, a large area of the flat cannot be used, so such flats are rejected. 

Flats taken when there are clouds in the sky can have a median pixel value greater than 40,000 ADU. However, this is not always the case, especially if the clouds are wispy and thin. In order to reject these flats, we rely on the fact that if there are thin clouds in the sky, then they should move, and the low level large scale structure in the flat should also change in time. In order to detect and reject these flats, a master flat is made of the series of flats taken in a single twilight or dawn flat taking session and each exposure in that series is compared to the master to detect changes between images. Master flats are made out of flats that so far have not been rejected due to tests 1-4 above. If there are less than four flats in a series of twilight or dawn flats that have passed tests 1-4, then a master flat is not created, and the whole series of flats is rejected. 

After all the above tests, there is still a piece of information that has not been utilized. Dragonfly has 48 individual subsystems, so 48 sky flats are taken at the same time. If there is a problem with many flats taken at the same time, chances are there is a problem with the other flats taken at the same time too. So the final check done on the flats is, if more than 50\% of flats taken at the same time are classified as bad using all the above tests, then all flats taken at that time are marked as ``bad by association". 

\begin{figure*}[!htbp]
\centering
\includegraphics[width=0.9\textwidth]{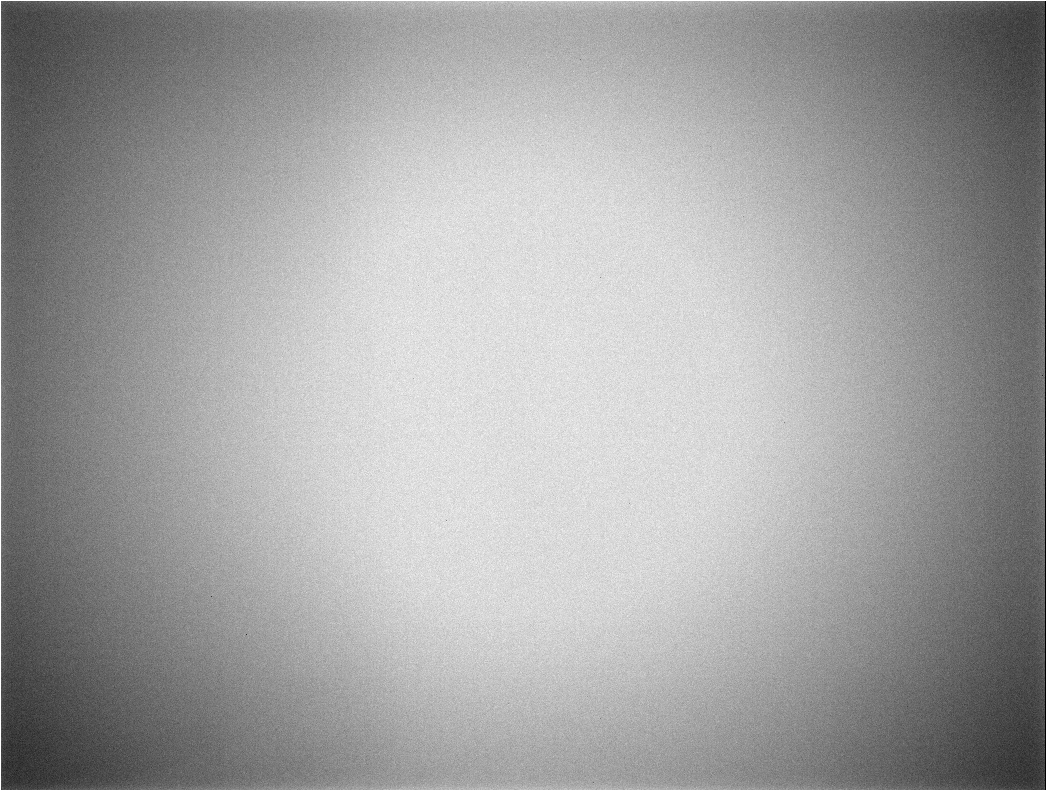}
\caption{An example of a good master flat. This image is a combination of eight flat exposures. Flats that go into a master flat are always from a single camera taken on the same night..}
\label{fig:goodmasterflat}
\end{figure*}

A conservative approach is taken when classifying flats: if there is uncertainty as to whether a flat might be good or bad, then we classify it as bad. This is because every night we observe, 16 flats are taken. We can afford to be conservative and throw out some flats that might be acceptable, but we cannot afford to include flats that might be bad. Incorrect flat fielding can cause science exposures to have uneven sky background, making it harder to model and subtract. Properly doing sky background subtraction is one of the most important steps in the Dragonfly Pipeline Software to ensure the deepest images are output. An example of a good master flat frame is shown in Figure~\ref{fig:goodmasterflat}. 

\subsection{Science Exposures with Double Stars} \label{sec:doublerejection}
\begin{figure*}[!htbp]
\centering
\includegraphics[width=0.9\textwidth]{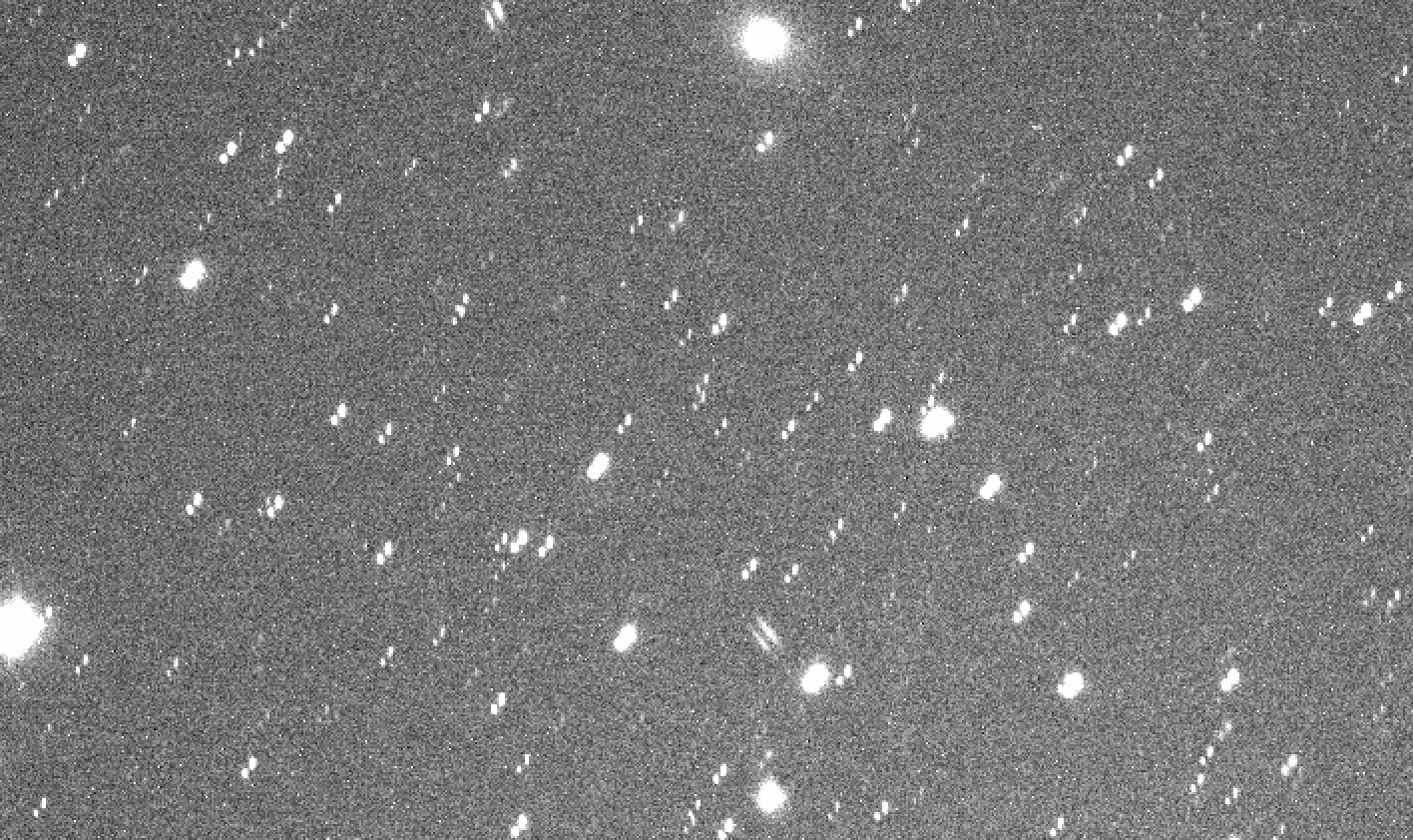}
\caption{An example of an frame where there are two copies of each source in the image.}
\label{fig:doubletar}
\end{figure*}
As described in the pipeline overview (Section~\ref{sec:pipelinesteps}), occasionally a science exposure image taken by Dragonfly looks like there are two or more copies of every source in the image. We call these images exposures with ``double stars". An example of this type of image is shown in Figure~\ref{fig:doubletar}. The observing scenario where science exposures display this phenomenon is when the telescope is observing near meridian. The likely cause is a shift in either the telescope subsystem mounting hardware or the moving components internal to the lenses shifting, or both. When the telescope slews across the meridian, the physical load on the system switches directions, and may cause shifts in hardware. The relevant internal component of the Canon lenses that can move is the component that enables the lens image stabilization function. Upon learning about this lens component, all Dragonfly lenses were sent back to the manufacturer and this lens component was glued into place so that it can no longer move. This reduced the occurrence of images with double stars drastically, but did not eliminate them completely. We are yet to identify the remaining culprit(s) that cause double star images. This means we need to be able to automatically classify whether a science exposure has double stars or not, both for historical data and for the few exposures that continue to display this issue. This automatic classification can be done on science exposures that have not been calibrated, so this piece of code is included in the ``calibration and rejection of double star science frames" stage of the pipeline, the flow chart for this stage of the pipeline is shown in Figure~\ref{fig:pipelineflow_cal}.

\begin{figure*}[!htbp]
\centering
\includegraphics[width=0.741\textwidth]{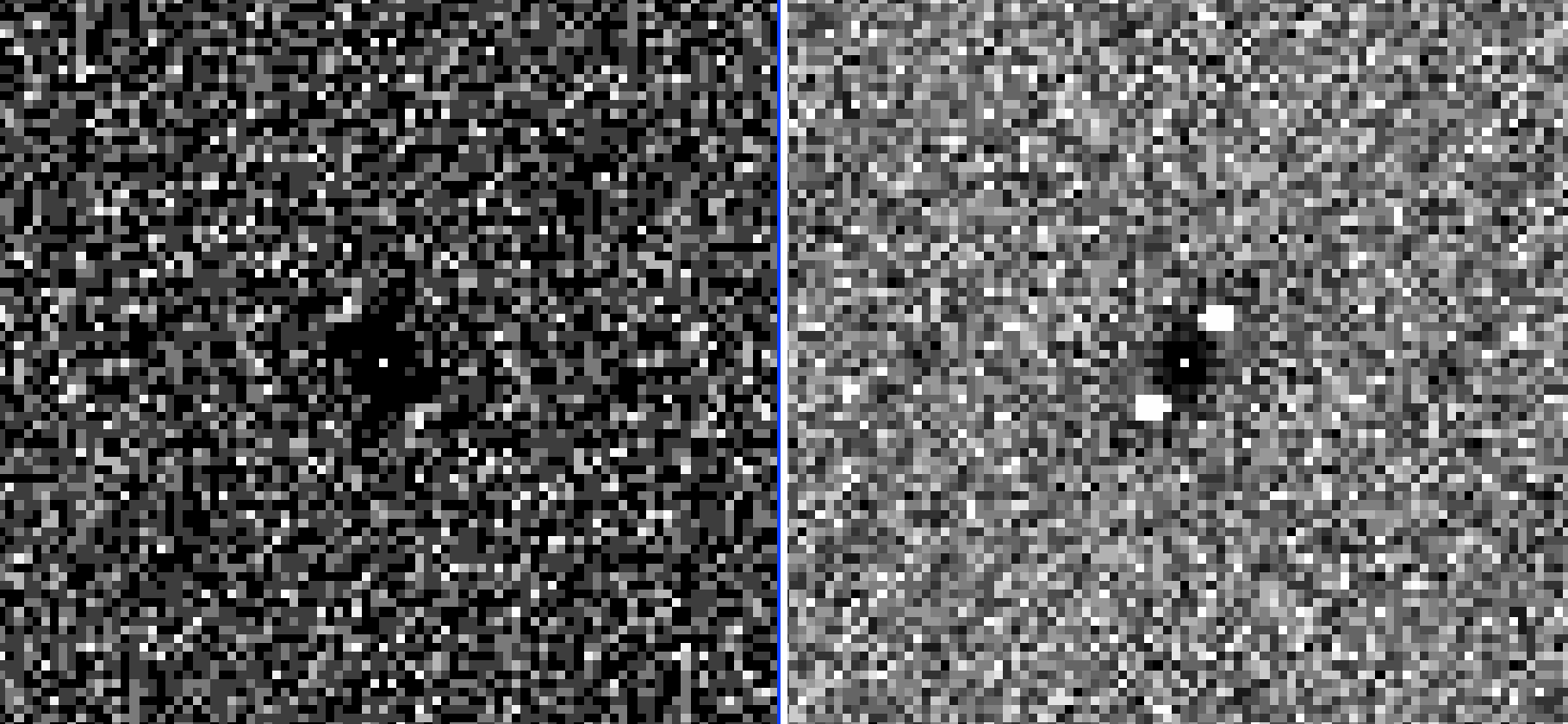}
\caption{Left: example of the central $\sim$100 by $\sim$100 pixels of the auto correlation image produced for an image with no double stars. Right: example of the central $\sim$100 by $\sim$100 pixels of the auto correlation image produced for an image with double stars.}
\label{fig:doubleautocorr}
\end{figure*}

The algorithm for determining whether a science exposure has double stars is based on identifying the signal in the auto correlation of an image with the locations of all sources in the original image marked. SExtractor~\citep{SExtractor} is used to determine the location of sources in the image. Then a new image with zeros and ones is created. This new image is zero in most places and one at pixels that correspond to the central pixel position of each source detected by SExtractor. The auto correlation of this image can be used to determine whether there are double stars in the original image. Figure~\ref{fig:doubleautocorr} shows the difference between the central $\sim$100 by $\sim$100 pixels of the auto correlation images of a science exposure with and without double stars on the right and left respectively. The central pixel of both auto correlation images are the brightest pixel in each image respectively. The two off-centre bright circles in the right hand image is the signal that correspond to the fact there are two copies of each bright source in the original image. 

\begin{figure*}[!htbp]
\centering
\includegraphics[width=0.7215\textwidth]{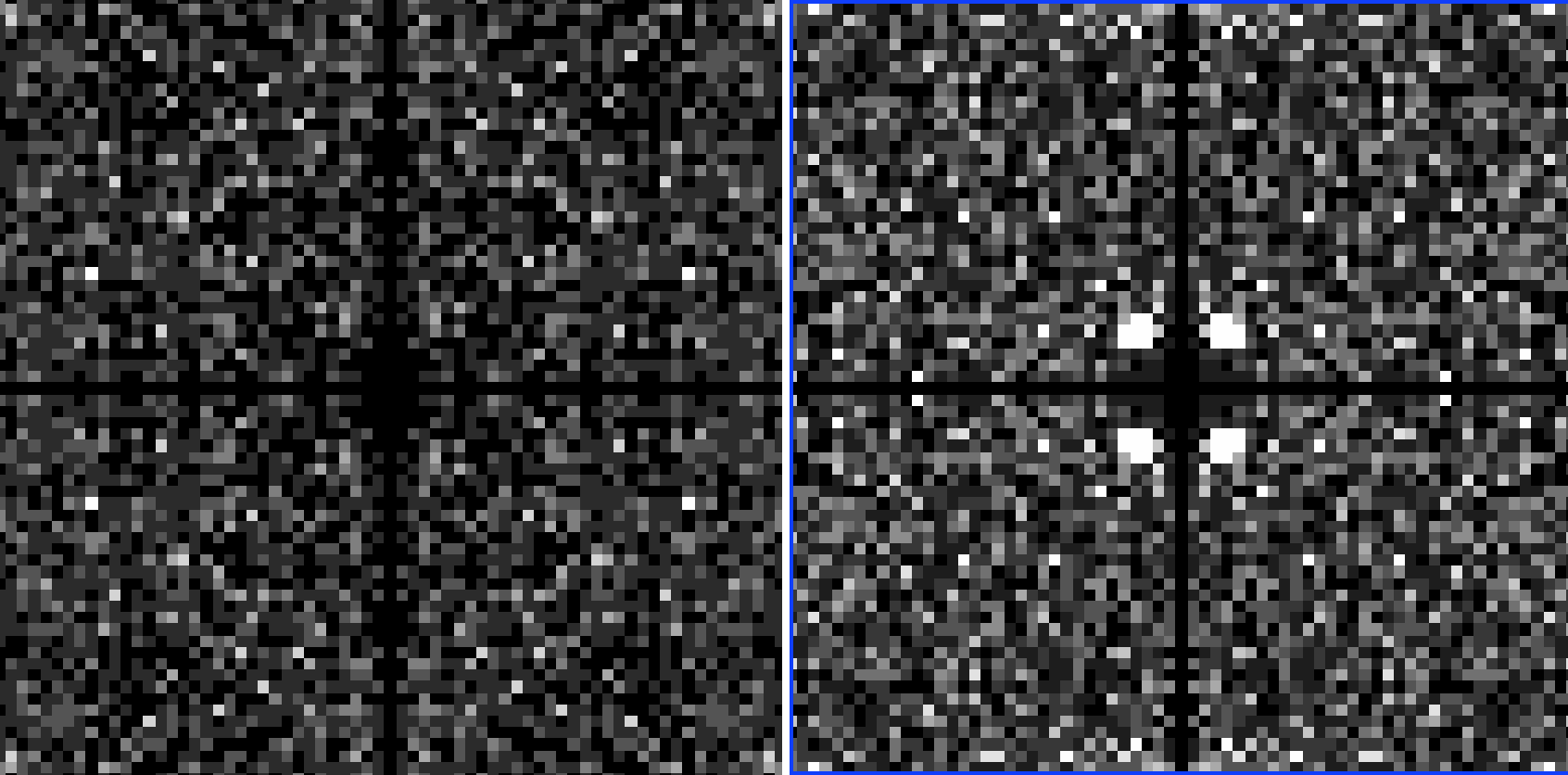}
\caption{Left: Example of the central $\sim$100 by $\sim$100 pixels of the absolute value of the difference between the auto correlation image and flipped auto correlation image produced for an image with no double stars. This is the double star detection image. Right: The double star detection image for an image with double stars.}
\label{fig:doubleautocorrflipdiff}
\end{figure*}

In order to isolate the two bright off-centre bright circles, each auto correlation image is flipped, differenced from the original auto correlation image and then an absolute value of that difference image is the final double star detection image. Figure~\ref{fig:doubleautocorrflipdiff} shows the central $\sim$100 by $\sim$100 pixels of the final detection image for a science exposure without (left) and with (right) double stars. If there are any bright sources in this this image, it indicates that this exposure has double stars. 

\subsection{Other issues with science exposures}
The Dragonfly Pipeline Software has to run unsupervised. In the end, it has the goal of combining only exposures that are of high quality into the stacked images that are used for science. While we cannot ensure that every single exposure that goes into the final stacked image has zero issues, we aim to minimize the number of problematic frames that reach the final combination stage of the pipeline. The issue of rejecting bad dark and flat exposures, as well as science frames with double stars or are affected by atmospheric conditions that result in a significant stellar aureole component in the PSF have been addressed above. Other known issues with science exposures that we have encountered and filter out include images:
\begin{itemize}
    \item that look like a dark
    \item where the exposure was too short
    \item where every pixel value is zero
    \item that contain lots of hot pixels and only a small number of astronomical sources in the image
    \item with unidentified bright light pollution shone into the lens
    \item that have an unidentified obstruction of the camera
    \item with stars that are out of focus
    \item taken when the telescope was pointed in the wrong direction
    \item taken when there were thin or thick clouds in the sky
\end{itemize}

\begin{figure*}[!htbp]
\centering
\includegraphics[width=0.9\textwidth]{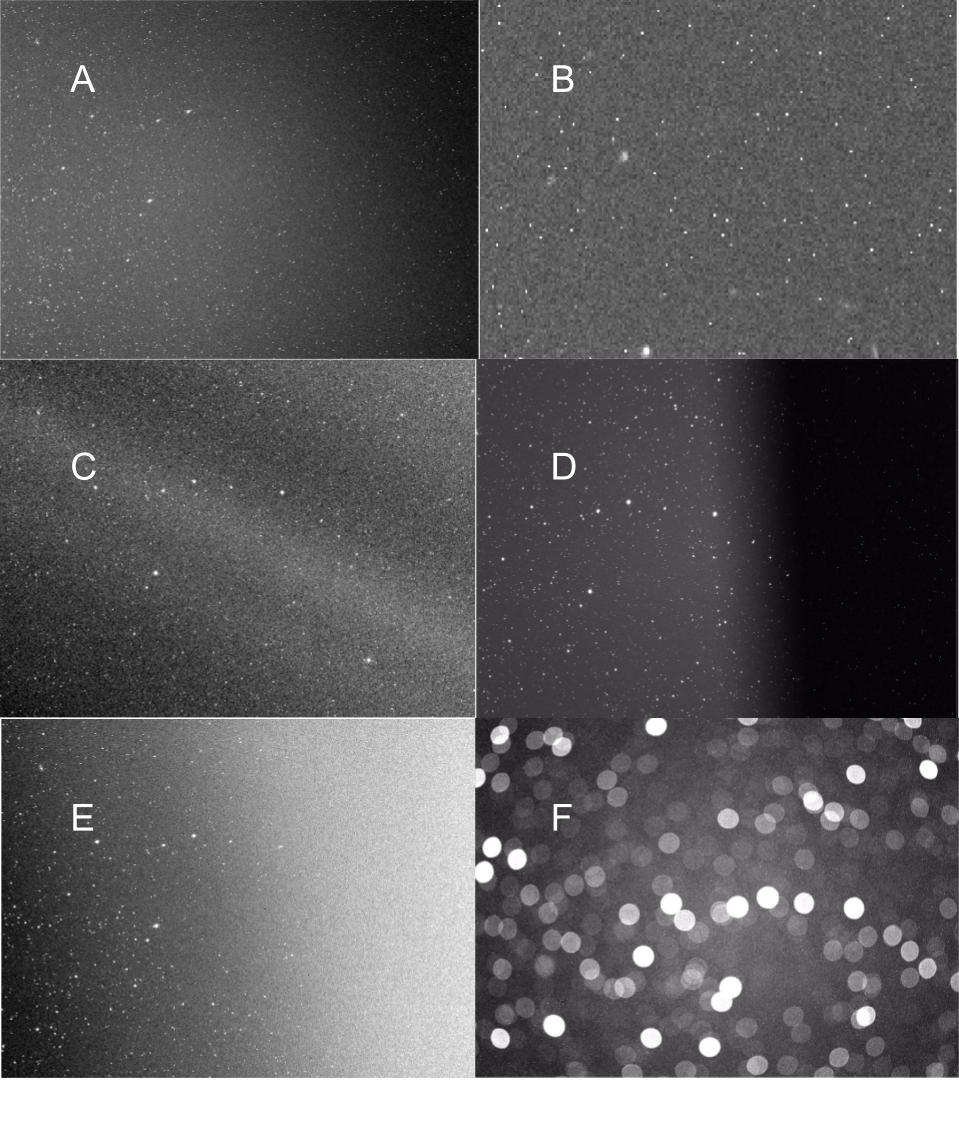}
\caption{Examples of science exposures with various issues. A: Frame contains mostly hot pixels. B: Zoom in on a section of frame shown in A to demonstrate that it has mostly hot pixels. C: Frame was exposed when there were thin clouds in that part of the sky. D: Half of the field of view of this frame is obscured. E: Frame is affected by unidentified light pollution. F: The camera was out of focus.}
\label{fig:otherbadexamples}
\end{figure*}

This rich set of potential problems is a product of hardware failures and the fact that Dragonfly observes nearly every moonless night, even on nights of dubious quality. Figure~\ref{fig:otherbadexamples} show a selection of bad frames. There are no specific algorithms to catch individual issues listed above. However, several steps in the Dragonfly Pipeline Software check for overall data quality and these identify and reject such frames. The checks include: 
\begin{itemize}
    \item The average full-width half-max of sources in the image needs to be within 1.2 - 5.5 pixels (part of the ``Assess Science Image Quality" step in Figure~\ref{fig:pipelineflow_init}.)
    \item There should be at least 1000 sources in the image (part of the ``Assess Science Image Quality" step in Figure~\ref{fig:pipelineflow_init}.)
    \item The average ellipticity of detected sources should be below 0.3 (part of the ``Assess Science Image Quality" step in Figure~\ref{fig:pipelineflow_init}.)
    \item The WCS solution should indicate that the lens was pointed within 45 arc minutes of the source during exposure. 
    \item The airmass corrected photometric zeropoint of the image should be no more than 0.2 mag below the nominal zeropoint for that subsystem. 
\end{itemize}

\section{Details on Image Registration, Scaling, and Combination} \label{sec:2p6}
\subsection{Image Registration} \label{sec:imagereg}
Image registration in the Dragonfly Pipeline Software is done by using SExtractor~\citep{SExtractor}, SCAMP~\citep{Scamp} and SWarp\footnote{SExtractor, SCAMP and SWarp are all part of the astromatic.net suit of software.}~\citep{Swarp}. Images are registered, then scaled to a common flux level and finally a dedicated python script is used to do the actual combination. 

For each image, SExtractor identifies sources and records their positions in a catalog. SCAMP uses the catalog of source positions, compares them to a combination of online catalogs to calculate an astrometric solution, including accounting for non-linear distortions within the image. The online catalogs used include the Guide Star Catalog~\citep{GSC1990}, The U.S. Navy Observatory (USNO) catalog~\citep{USNO2003}, USNO CCD Astrograph Catalog~\citep{UCAC42013} and the Two Micron All Sky Survey~\citep{2MASS2006}. The astrometric solution is written into a text file named with a .head suffix, storing this information using WCS standards. SWarp takes this astrometric solution for each image and re-samples all input images to a common grid.  

\begin{figure*}[!htbp]
\centering
\includegraphics[width=0.7\textwidth]{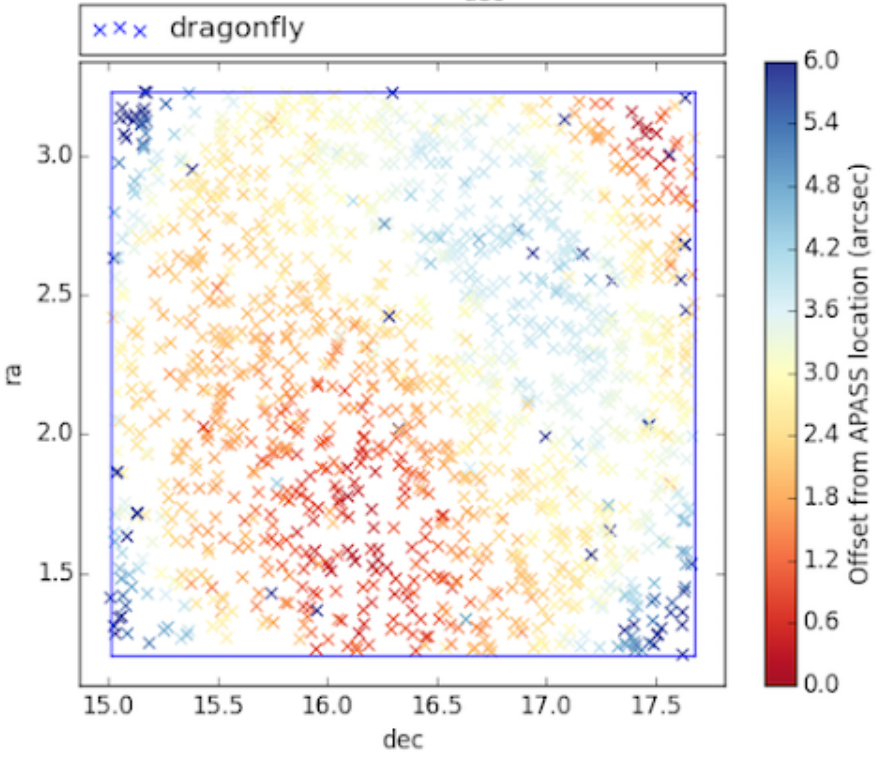}
\caption{Dragonfly astrometric distortions before corrections using a combination of SExtractor, SCAMP and SWarp are typically up to $\sim6$". The color bar on the right hand side indicates the distance between the RA and DEC position of sources in Dragonfly versus the closest matching source in the APASS catalog.}
\label{fig:astrometrybeforeSWarp}
\end{figure*}

\begin{figure*}[!htbp]
\centering
\includegraphics[width=0.7\textwidth]{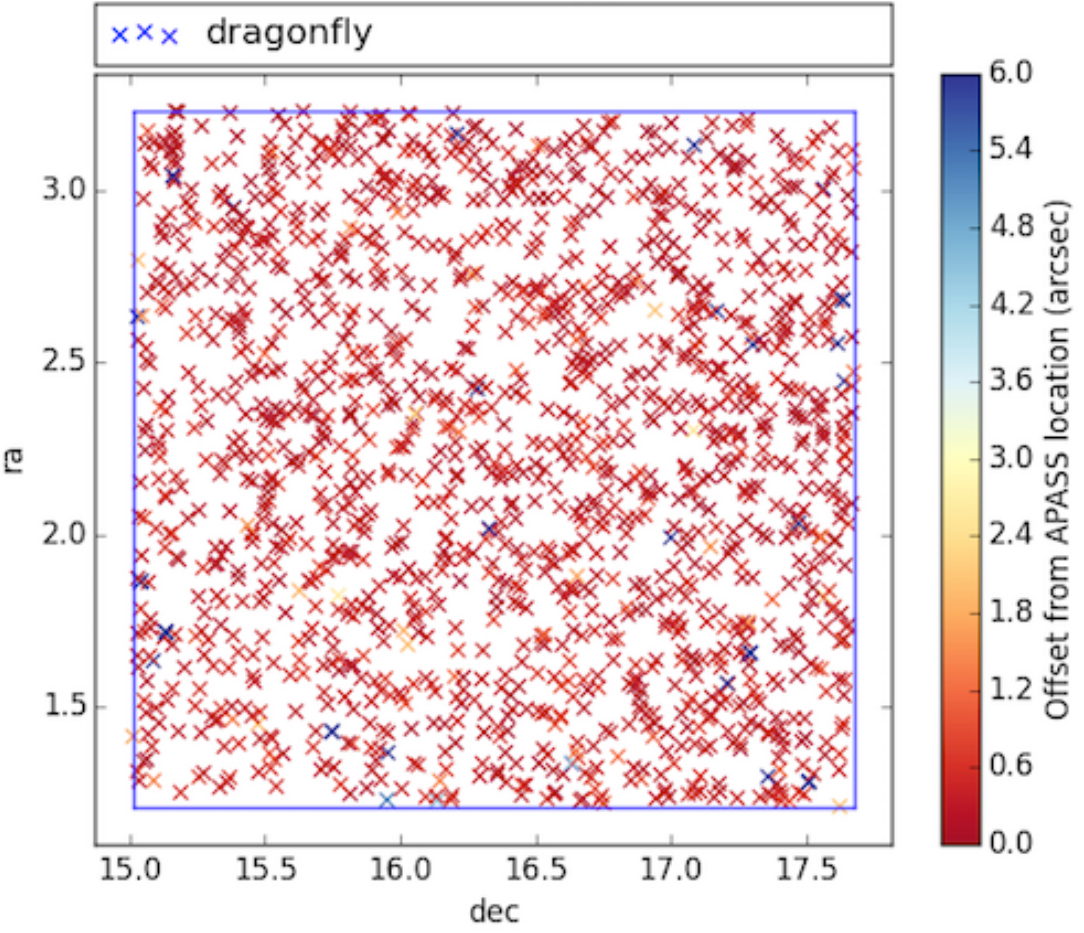}
\caption{Dragonfly astrometric distortions after corrections using a combination of SExtractor, SCAMP and SWarp are much less than 1". The color bar on the right hand side indicates the distance between the RA and DEC position of sources in Dragonfly versus the closest matching source in the APASS catalog.}
\label{fig:astrometryafterSWarp}
\end{figure*}

Astrometric distortions in Dragonfly are corrected reliably with SExtractor, SCAMP and SWarp. Figure~\ref{fig:astrometrybeforeSWarp} shows the distance between Dragonfly sources and the closest APASS catalog source position in RA and DEC for a typical Dragonfly image before SWarp. Distortions are at the level of several arcseconds. After re-sampling with SWarp, distortions are much less than 1" across the entire field of view. This can be seen in Figure~\ref{fig:astrometryafterSWarp}. 
 
\subsection{Zeropoint Determination} \label{sec:zpdet}
Before image combination, each science exposure must be scaled to a common flux level. This can be done using the zeropoint of each image. The zeropoint (ZP) of each image is defined as follows:
\begin{equation}
\text{ZP} = \text{m}_{\text{cat}} + 2.5\cdot log_{10}(\text{m}_{\text{image}}) 
\end{equation}
\noindent
where $\text{m}_{\text{cat}}$ is the magnitude of a source in a photometric reference catalog (such as APASS), and $\text{m}_{\text{image}}$ is the magnitude of the same source in the image before any scaling. This image source magnitude is calculated assuming the zeropoint of the image is zero, in other words using this equation:
\begin{equation}
\text{m}_{\text{image}} = -2.5 \cdot log_{10}(ADU) 
\end{equation}
\noindent
where ADU represents the flux integrated over the source in question in units native to the FITS file. Once the zeropoints of science exposures are calculated, the flux-scale of each image is defined as follows:
\begin{equation} \label{eqn:flux-scale}
\text{flux-scale} = 10^{(\text{ZP}-\text{ZP}_{\text{average}})/-2.5}
\end{equation} 
where ZP is the zeropoint of the image and $\text{ZP}_{\text{average}}$ is the average zeropoint of all frames of that filter band to be combined. Scaling images by this flux-scale ensures that even if different lens subsystems may have slightly different ADU values while looking at the same position on the sky, their pixel values are all redistributed to a common mean before median combining. When a median stacked image is produced, satellite trails, cosmic rays and other problematic pixel values do not affect the final stacked image.

\begin{figure*}[!htbp]
\centering
\includegraphics[width=0.7\textwidth]{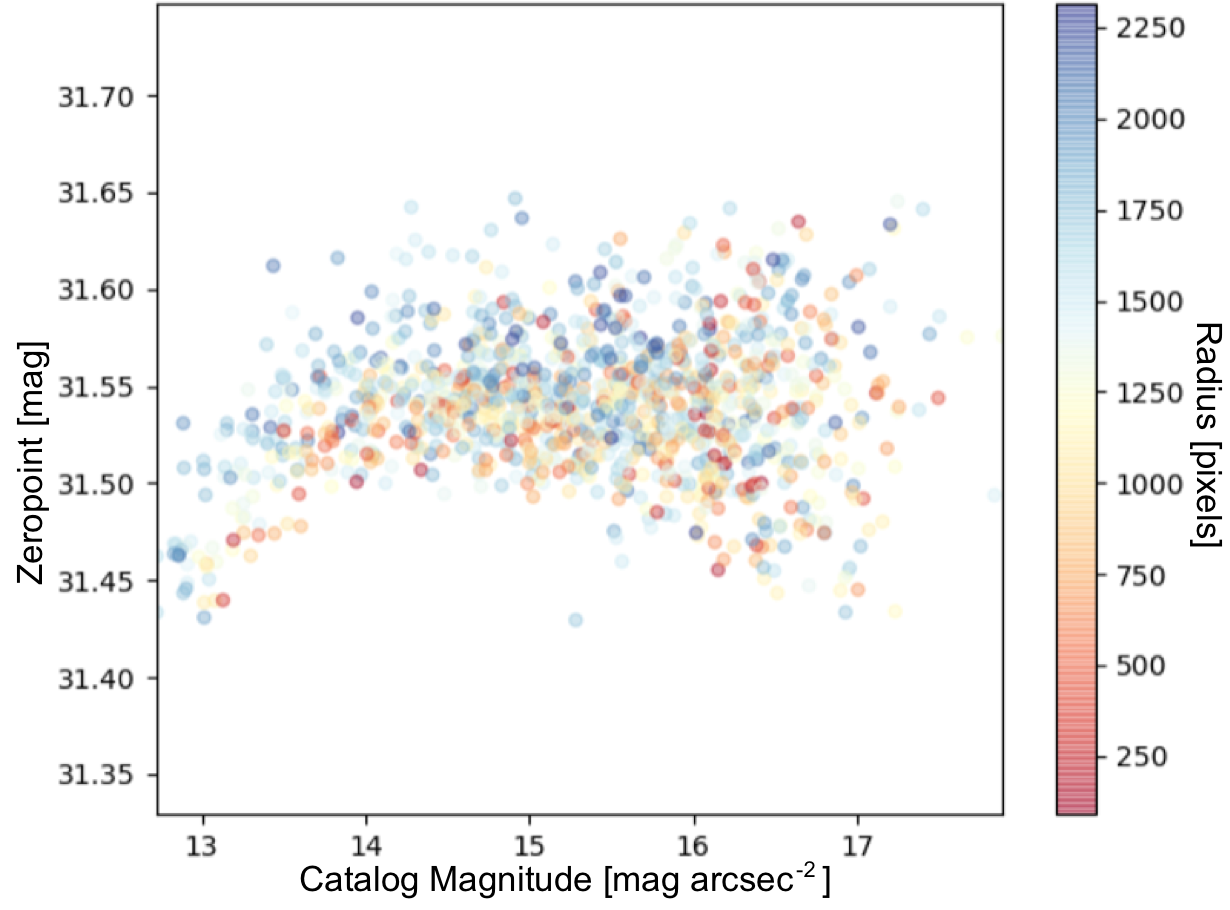}
\caption{The zeropoint for all detected sources in a Dragonfly \textit{g}-band image plotted against the APASS catalog \textit{g}-band magnitude. Based on this plot, it can be seen that sources saturate at $\sim$14 mag, and the signal-to-noise of sources becomes large above $\sim$16 mag. This is typical for all Dragonfly images.}
\label{fig:zp_magcuts}
\end{figure*}

In order to determine the zeropoint of a Dragonfly image, the photometry of thousands of stars in each exposure are compared to those in the APASS photometric catalog. Figure~\ref{fig:zp_magcuts} is a plot of the catalog magnitude of the star versus the zeropoint calculated for each star in a Dragonfly image. For sources brighter than $\sim$14 mag, the zeropoints start to drop. This is because those stars are saturated in Dragonfly images. Towards the faint end of this plot, above $\sim$16 mag, the photometry has a relatively large scatter, therefore the zeropoint of the image is calculated based on unsaturated stars brighter than 16 mag. The data points in Figure~\ref{fig:zp_magcuts} are color coded by the distance in pixels away from the centre of the image to give an indication if there is some sort of radial dependence of the zeropoint. If there is, then this is a signature that flat fielding was not done well, and those frames are flagged in the database and eliminated from the final stack. 
 
Flux scales based on image zeropoints are calculated using Equation~\ref{eqn:flux-scale} for each image, and this information is written into the FITS header of each image. 

\subsection{Image Combination} \label{sec:imagecombination}
Calibrated, sky subtracted, quality science exposures that have been re-sampled to a common grid and embedded with a flux-scale fits header are now ready for image combination. The goal of image combination is to produce a weighted average combined image, where the weights used are the signal-to-noise values of individual exposures. In this process, satellite trails and cosmic rays are removed using sigma clipping.

In order to produce the highest signal-to-noise combined images (by filter) flux-scaled exposures are weighted by 1/(flux-scale) multiplied by 1/(sky brightness). Each individual exposure's median sky value is also weighted-average combined using the same weights as their corresponding images. This uniform sky value is added back into the final combined image, to preserve information on how bright the sky was during observations. 

\section{Data Backup and Access} \label{sec:databackupandaccess}
\subsection{Data backup and storage structure} \label{sec:databackup}
During observations, data collected by each camera-lens subsystem is stored on that subsystem's Intel Compute Stick. Each Intel Compute Stick is accessible via the network at the New Mexico Skies Observatory using a central control PC. There is only enough storage on each Intel Compute Stick to store about one week's worth of data if observing conditions are ideal each night of the week, or for a longer period of time if telescope domes were closed during some nights. Each morning, after observations have finished, the data is automatically copied onto a Redundant Array of Independent Disks (RAID) storage, located at the University of Toronto. This is done via a crontab job, which is a method to schedule processes to run automatically at certain repeating intervals of time (e.g. every day at 8am in our case). Two copies of the raw data stored on the University of Toronto RAID are made. One copy is kept at the Canadian Institute of Theoretical Astrophysics (CITA). A second copy is uploaded into the cloud data storage facilities provided by CANFAR. 

The CANFAR cloud storage system is called VOSpace. VOSpace provides a standardized interface that supports uploading and downloading of data from any computer connected to the Internet. The software is easy to install as a Python module named ``vos", which includes a command line interface. Files stored on the VOSpace system are mirrored in four physical locations, ensuring adequate backup. Information on registering for an account to access VOSpace, how to install it and user documentation can be found at \url{http://www.canfar.net/en/docs/storage/}. 

Each exposure's FITS file includes the serial number of the camera that took the image, the exposure type (i.e. whether it is a dark, flat or science exposure) and a sequential exposure number that is reset at the start of each night for every camera. Important header keywords for each raw exposure fits file are stored in the fits headers, such as the target name, exposure time, camera serial number, temperature of the CCD at the time the exposure was taken, image type (dark, flat or science exposure), name of the filter used, date and time the exposure was taken, and the central RA, DEC, altitude, and azimuth of the image.  

\subsection{Accessing auxiliary information about data} \label{sec:database}
There are two main categories of auxiliary information about each raw image taken with Dragonfly. The first category is real-time observing information that is written into the header of the image FITS file during the observation, by the Intel Compute Stick of each lens-camera subsystem. The second category is qualitative or quantitative information regarding each exposure determined by the Dragonfly Pipeline. The observing information can be accessed in two ways: via the Dragonfly Database or by pulling the fits header information directly from the CANFAR cloud. The Dragonfly Pipeline derived information can only be accessed via the Dragonfly Database.  

The Dragonfly Database\footnote{The Dragonfly Database was set up by my advisor Prof. Roberto Abraham. My role in setting up the Dragonfly Database was to consult on its format and help integrate it into the Dragonfly Pipeline workflow.} is a MySQL database, and is hosted on a virtual machine on the CANFAR cloud. It can be queried (via an SQL query) from any computer with the right authentication credentials. There are currently two SQL tables in the database, named FITSHEADERS and METADATA. These store observing and pipeline-added information, respectively. The FITSHEADERS table has the columns listed in Table~\ref{table:FITSHEADERSschema}. The METADATA table has the columns listed in Table~\ref{table:METADATAschema}.  

Each raw image uploaded into the CANFAR cloud has its fits header information uploaded into the FITSHEADERS table, which can be queried using a set of Dragonfly Database scripts. Fits header information of a specific raw exposure can also be accessed directly using the VOSpace's API for web access. For example, the following command can be used to access an image header:
\begin{lstlisting}[numbers=none]
curl -L -n http://www.canfar.phys.uvic.ca/data/auth/vospace/dragonfly/Dragonfly/Data/Dragonfly001/2017-10-26/T13080513_1_flat.fits?fhead=True -o h.txt
\end{lstlisting}
This command writes information to a local text file named h.txt and this file will contain the fits header of an image file named T13080513\_1\_flat.fits, which was taken on the night of 2017-10-26 by Dragonfly lens-camera subsystem one. In order for this query to work, the right authentication credentials need to be entered, or cached on the machine executing the command.

\begin{table}[h!]
\begin{center}
\caption{The FITSHEADERS table in the MySQL Dragonfly Database.} \label{table:FITSHEADERSschema} 
This is the MySQL table schema for the FITSHEADERS table in the Dragonfly Database.\\ 
.\newline
\begin{tabular}{|l|l|l|l|}
\hline
Field & Type & \multicolumn{1}{m{1cm}|}{Default Value} & Description \\ 
\hline
\hline
FILENAME	& varchar(256) & Null	& Name of fits file	(Primary Key) \\
DIRNAME	& varchar(256) & Null & The name of directory the file is stored under	\\
BZERO &	float & Null	& Zero point for pixel values \\
BSCALE &	float & Null	& Scaling factor for pixel values \\
EXPTIME &	float & Null & Exposure time (seconds) \\	
TEMPERAT	& float & Null	& Temperature of the CCD (degrees Celcius) \\
IMAGETYP &	varchar(8) & Null & Dark, flat or science exposure \\
FILTNUM	& int(11) & Null	& Filter number \\
DATEORIG &	varchar(20) & Null	& The date of the observation (format: YYYY-MM-DD)\\
DATE &	datetime & Null & The date time of the observation \\
SERIALNO &	varchar(20) & Null	& The serial number of the camera \\
TARGET &	varchar(64) & Null	& The name of the target of observation \\
RA &	varchar(64) & Null	& Right ascension \\
DECL &	varchar(64) & Null	& Declination \\
EPOCH &	varchar(8) & Null	& Epoch, e.g. J2000 \\
OBJCTRA	& varchar(64) & Null & Right ascension (same as above) \\	
OBJCTDEC &	varchar(64) & Null	& Declination (same as above) \\
ALTITUDE &	float & Null & Altitude (degrees) \\
AZIMUTH &	float & Null & Azimuth (degrees) \\
MEAN &	float & Null	& The mean value of the image \\
MODE &	float & Null & The most common pixel value of the image \\	
FILTNAM	& varchar(12) & Null & Filter name \\	
\hline
CTYPE1	& varchar(12) & Null & World Coordinate System \\	
CRPIX1 &	float & Null &	World Coordinate System \\
CRVAL1 &	float & Null & World Coordinate System \\
CTYPE2 &	varchar(12) & Null	& World Coordinate System \\
CRPIX2 &	float & Null	& World Coordinate System \\
CRVAL2 &	float & Null	& World Coordinate System \\
CD1\_1 &	float & Null & World Coordinate System \\
CD1\_2 &	float & Null & World Coordinate System \\
CD2\_1 &	float & Null & World Coordinate System \\
CD2\_2 &	float & Null & World Coordinate System \\
RADECSYS &	varchar(12) & Null	& World Coordinate System \\
EQUINOX &	float & Null	& World Coordinate System \\
\hline
FWHM &	float & Null	& The average full width half max of sources \\ 
SSIGMA &	float & Null	& The standard deviation of FWHM \\
NOBJ &	float & Null	& The number of detected sources \\
ELLIP &	float & Null	& The average ellipticity of sources \\
BOVERA &	float & Null	& 
    \multicolumn{1}{m{8cm}|}
    {The average value of the minor divided by major axis of sources} \\
\hline
\end{tabular}
\end{center}
\end{table}

\begin{table}[h!]
\begin{center}
\caption{The METADATA table in the MySQL Dragonfly Database.} \label{table:METADATAschema} 
This is the MySQL table schema for the METADATA table in the Dragonfly Database.\\ 
.\newline
\begin{tabular}{|l|l|l|l|}
\hline
Field & Type & Default Value & Description \\ 
\hline
\hline
FILENAME &	varchar(256) &	NULL & Name of the fits file (Primary Key) \\
TAGNUMBER &	int(64) &	0 & Tags that indicate quality of the image \\
ZPSPACE	& float	& 99999 & Airmass corrected zeropoint \\
ZPCAT &	varchar(20) & None & Reference catalog used to obtain zeropoint \\	
ZPSTD &	float &	99999 & Standard deviation of zeropoint \\
MEDSKY &	float &	99999 & Median sky background (ADU) \\
DIRDATE &	varchar(20) &	None & The date folder under which the file is stored \\
FWHM &	float & None	& The average full width half max of sources \\ 
NOBJ &	float & None	& The number of detected sources \\
ELLIP &	float & None	& The average ellipticity of sources \\
\hline
\end{tabular}
\end{center}
\end{table}

\subsection{Accessing Data} \label{sec:dataaccess}
Raw Dragonfly data can be easily queried and accessed via customized scripts that query the Dragonfly Database and downloaded the outputs of the queries. For example, to download all science exposures (we name them ``light" exposures within the team) of target XYZ on 2017-01-01 that have not been classified as images with double stars, the following terminal command can be used on any computer with the Dragonfly Database software installed and CANFAR authentication credentials cached:

\begin{lstlisting}[numbers=none]
df_download_nodoublestars 2017-01-1 --target XYZ --type "light" --verbose
\end{lstlisting}

\section{Cloud-orchestration of software} \label{sec:cloud-orch}
The process of automatically creating, running, and destroying virtual machines is central to cloud computing. Doing these steps for thousands of virtual machines would be tedious, so automated ``cloud orchestration" software has been developed which handles these steps relatively easily. Cloud-orchestration also avoids the need for all collaborators to either have to wait for data to be reduced after a machine becomes available or to purchase machines for data processing that sit idle for large stretches of time. This section describes the CANFAR cloud computing capabilities.  

CANFAR cloud computing services allow the creation of persistent virtual machines (VMs) and batch processing. During batch processing, temporary VMs are created and deleted once the queued process is finished. The creation, deletion and management of virtual machines is done through OpenStack software. Persistent VMs can be managed using an OpenStack web interface dashboard, or an OpenStack command line interface. This OpenStack command line interface can be easily installed as a Python module. VMs can be accessed via SSH using SSH key pairs. Each research project or group is given a single public IP with which they can connect to a VM. Through this ``relay" VM, all other VMs can be accessed via SSH. Information for registering for an account to access CANFAR cloud computing open stack services, and user documentation can be found at \url{http://www.canfar.net/en/docs/quick_start/}.

CANFAR allocates a maximum number of persistent VM instances, virtual processors that the VMs can be equipped with, volume storage units, disk space that can be attached to VM instances as volume storage, and RAM (in total and per virtual machine). If more resources are needed, resource requests can be made to CANFAR, and if granted, are allocated within three days. These persistent VMs are where testing of the Dragonfly Pipeline is carried out. However, to do efficient and fast data reduction that requires a large amount of resources at once, batch processing is used. 

Batch processing allows queuing of large number of computationally intensive tasks. This means that many users can share the same disk space and processing resources. For the Dragonfly Pipeline Software, it also means different night's data on the same target can be processed in parallel by launching many batch processing jobs at the same time. This drastically speeds up the time it takes to reduce data, and all issues with the data are quickly  identified in parallel and dealt with on time lines of a few days instead of several weeks. Batch jobs are deployed and monitored using the CANFAR batch login node. Once an account is granted on the node, easy access is possible by using SSH via the use of SSH key pairs. Information for registering for an account on the CANFAR batch login node and user documentation can be found at \url{http://www.canfar.net/en/docs/batch_processing/}.

\chapter{THE DRAGONFLY NEARBY GALAXIES SURVEY. IV. A Giant Stellar Disk in NGC 2841}
\chaptermark{THE DRAGONFLY NEARBY GALAXIES SURVEY. IV.}

This chapter was published in the Astrophysical Journal~\citep{Zhang2018}. The introduction to this paper has been re-worked minimally and printed in the introduction to this thesis, and so the reviewer who has read the introduction may wish to skip Section~\ref{sec:diskintro}. However, the paper is included in this chapter as published, for the sake of completeness. 

\section{Abstract}
Neutral gas is commonly believed to dominate over stars in the outskirts of galaxies, and investigations of the disk-halo interface are generally considered to be in the domain of radio astronomy. This may simply be a consequence of the fact that deep HI observations typically probe to a lower mass surface density than visible wavelength data. This paper presents low surface brightness optimized visible wavelength observations of the extreme outskirts of the nearby spiral galaxy NGC 2841. We report the discovery of an enormous low-surface brightness stellar disk in this object. When azimuthally averaged, the stellar disk can be traced out to a radius of $\sim$70 kpc (5 $R_{25}$ or 23 inner disk scale lengths). The structure in the stellar disk traces the morphology of HI emission and extended UV emission. Contrary to expectations, the stellar mass surface density does not fall below that of the gas mass surface density at any radius. In fact, at all radii greater than $\sim$20 kpc, the ratio of the stellar to gas mass surface density is a constant 3:1. Beyond $\sim$30 kpc, the low surface brightness stellar disk begins to warp, which may be an indication of a physical connection between the outskirts of the galaxy and infall from the circumgalactic medium. A combination of stellar migration, accretion and in-situ star formation might be responsible for building up the outer stellar disk, but whatever mechanisms formed the outer disk must also explain the constant ratio between stellar and gas mass in the outskirts of this galaxy.

\section{Introduction}
The sizes of galaxy disks and the extent to which they have well-defined edges remain poorly understood. Galaxy sizes are often quantified using $R_{25}$, the isophotal radius corresponding to $B=25$ mag arcsec$^{-2}$, but this is an arbitrary choice. In fact, the literature over the last three decades has produced conflicting views regarding whether there is a true physical edge to galactic stellar disks. Early studies seemed to show a truncation in the surface brightness profiles of disks at radii where star formation is no longer possible due to low gas density~\citep{vanderKruitSearle1982}, but more recent investigations have found examples of galaxy disks where the visible wavelength profile is exponential all the way down to the detection threshold~\citep{Bland-Hawthorn2005,vanDokkum2014,vlajic2011}. There is considerable confusion in the literature regarding the relationship between the profile shape and the size of the disk. Most disks fall in one of three classes of surface brightness profile types~\citep{PohlenTrujillo2006,Erwin2008}: Type I (up-bending), Type II (down-bending) and Type III (purely exponential). The existence of Type II disks has been pointed to as evidence for physical truncation in disks. However, while the position of the inflection in the profile can certainly be used to define a physical scale for the disk, this scale may not have any relationship to the ultimate edge of the disk~\citep{Bland-Hawthorn2005,PohlenTrujillo2006}.

The common view in the literature is that the HI disks of galaxies are considerably larger than their stellar disks. This arises from the observation that HI emission extends much further in radius than the starlight detected in deep images~\citep{vanderKruitFreeman2011,Elmegreen2016}. A rationale for this is the possible existence of a minimum gas density threshold for star formation~\citep{FallEfstathiou1980,Kennicutt1989}, although this idea is challenged by the fact that extended UV (XUV) emission is seen in many disks at radii where the disks are known to be globally stable~\citep{Leroy2008}. Studies suggest that large scale instability is decoupled from local instability, and the latter may be all that is required to trigger star formation. For example, a study by~\cite{Dong2008} analyzed the Toomre stability of individual UV clumps in the outer disk of M83. They found that even though the outer disk is globally Toomre stable, individual UV clumps are consistent with being Toomre unstable. These authors also found that the relationship between gas density and the star-formation rate of the clumps follows a local Kennicutt-Schmidt law. In a related investigation,~\cite{Bigiel2010} carried out a combined analysis of the HI and XUV disks of 22 galaxies and found no obvious gas surface density threshold below which star formation is cut off, suggesting that the Kennicutt-Schmidt law extends to arbitrarily low gas surface densities, but with a shallower slope. 

On the basis of these considerations, it is far from clear that we have established the true sizes of galactic disks at any wavelength. Absent clear evidence for a physical truncation, the `size' of a given disk depends mainly on the sensitivity of the observations. This basic fact applies to both the radio and the visible wavelength observations, and relative size comparisons which do not account for the sensitivity of the observations can be rather misleading. For example, it is commonly seen that the gas in galaxies extends much further in single dish observations than it does in interferometric observations, because single dish observations probe down to lower column densities~\citep{Koribalski2016}. At visible wavelengths, the faintest surface brightness probed by observations has been stalled at $\sim 29.5$ mag arcsec$^{-2}$ for several decades~\citep{Abraham2016book}, with this surface brightness `floor' set by systematic errors~\citep{Slater2009}. 

The Dragonfly Telephoto Array (Dragonfly for short) addresses some of these systematic errors and is optimized for low surface brightness observations; see~\cite{Abraham2014} for more details. Dragonfly has demonstrated the capability to  routinely reach $\sim$32 mag arcsec$^{-2}$ in azimuthally averaged profiles~\citep{vanDokkum2014,Merritt2016a}.

In this paper, we present ultra-deep visible wavelength observations taken with Dragonfly of the spiral galaxy NGC 2841. This galaxy is a particularly clean example of XUV emission in an isolated environment~\citep{Afanasiev1999}. It is notable for being the archetype for the flocculent class of spiral galaxies. The disk is globally Toomre-stable~\citep{Leroy2008} and it shows no evidence for grand design structure, although near-infrared observations do show some long dark spiral features in its interior~\citep{Block1996}. Our aim is to determine if the stellar disk, as traced by visible wavelength light, extends at least as far as the neutral gas mapped by the THINGS survey~\citep{Walter2008}, and the XUV emission mapped by GALEX~\citep{Thilker2007}. Instead of just comparing sizes of disks in different wavelengths, we will compare mass surface densities up to the sensitivity limit of the respective data sets. Throughout the paper, we assume the distance to NGC 2841 is 14.1 Mpc~\citep{Leroy2008}.

In \S\ref{sec:data} we describe our observations and the specialized reduction techniques we have adopted in order to obtain deep profiles with careful control of systematic errors. Our results are presented in \S\ref{sec:results}, and our findings are discussed in \S\ref{sec:discussion}. 

\section{Observations and Data Reduction}
\label{sec:data}

Broadband images of NGC 2841 were obtained between 2013 and 2016 using Dragonfly as part of the Dragonfly Nearby Galaxy Survey~\citep{Merritt2016a}. Dragonfly is comprised of multiple lenses with their pointing offset from one another by a few arcminutes. Between 2013 and 2016, the number of lens and camera subsystems on the Array increased from 8 to 24. A total of 3351 ten-minute exposure images of NGC 2841 were obtained in Sloan \textit{g} and \textit{r} bands, distributed over the multiple cameras. Sky flats were taken daily at twilight and dawn. Data reduction was carried out using the Dragonfly Pipeline, full details for which can be found in~\cite{Zhang2018}. The full-width at half maximum (FWHM) of the final combined NGC 2841 image is 7 arcseconds.  

The ultimate limiting factor in low surface brightness observations of nearby galaxies is the wide-angle point-spread function (PSF;~\citealt{Slater2009, Abraham2014, Sandin2015}). The largest-scale component of the PSF is the so-called `aureole'~\citep{Racine1996, King1971}. In conventional telescopes, the aureole is dominated by scattered light from internal optical components ~\citep{Bernstein2007}. An important point emphasized in~\cite{Zhang2018} is that this stellar aureole varies on a timescale of minutes, and so its origin is most likely atmospheric.~\cite{AtmIceCrystals2013} suggest that high-atmosphere aerosols (mainly ice crystals) are the culprit. An efficient way to detect the existence of atmospheric conditions which result in prominent stellar aureoles is to monitor the photometric zeropoints of individual exposures and identify those with deviations from the nominal zeropoint for a given camera at a given air mass. In our analysis of data from NGC 2841, exposures with a photometric zeropoint deviant from the nominal zeropoint by more than $\sim$0.1 mag were excluded from the final combined image. Out of the 3351 exposures obtained, 1034 were used. Most of the exposures excluded were taken in obviously marginal weather conditions (e.g., thin clouds). However, $\sim$25\% of the exposures were identified as having wider-than-normal wide-angle PSFs only by using the procedure of monitoring the zeropoint values of the exposures.

Sky subtraction was done in two passes. In the first pass, a sky model was fit to the \textsc{SExtractor}~\citep{SExtractor} background map for each image and subtracted. Sky-subtracted frames were then used to create an average combined image (including both \textit{g} and \textit{r}-band data). \textsc{SExtractor} was run on this average combined image to produce a segmentation map. A mask was created by growing the segmentation map with settings to capture sources all the way out to their low surface brightness outer edges. In the second pass, sky models were fit to the \textsc{SExtractor} background map of non-sky-subtracted images again, but this time the mask was input into \textsc{SExtractor} for the creation of the background map. This ensures we do not over subtract the sky by fitting a sky model to ultra-faint galaxy light. After this careful sky subtraction procedure, there were no residual large-scale gradients visible to the eye in an image where non-sky pixels were masked. In order to measure any residual large-scale gradients not obvious by visual inspection, a third order polynomial was fit to a masked image. The peak to peak range of sky background model values are ~0.05\% the sky value for both the g and r-band images. This was measured on a 57 by 70 arcminute image of NGC 2841, with the long edge aligned in the north-south direction. It is important to note, however, that the regions responsible for the 0.05\% variation in sky are all on the edges of the image, and do not overlap with NGC 2841. 

The sky background and its error was determined by measuring the flux in elliptical annuli placed around NGC 2841 in each of the \textit{g} and \textit{r}-band Dragonfly images. Randomly placed elliptical annuli were used to sample the sky because the error in the sky value depends on the scale over which the sky is measured. For example, the variance in a sample of 10x10 pixel sky boxes will be different to a sample of 100x100 pixel sky boxes. The outer-most data points in the surface brightness profile are the most sensitive to the accuracy of the sky background measurement, therefore, that is the scale on which it is critical to know the sky background error. The shape of the sky sampling “box” was chosen to be an elliptical annulus because that most resembles the shape within which we have to determine the sky for the surface brightness profile. 1000 elliptical annuli were placed randomly in a 30 by 30 arcminute region around NGC 2841. The ellipticity and position angle of the elliptical annuli was fixed to be that of the largest surface brightness profile isophote. The annuli sizes were allowed to randomly vary but not below the size of the largest surface brightness profile isophote. In order to sample the local sky background values, the elliptical annuli were not allowed outside of a 30 by 30 arcminute region around NGC 2841. A mask was created so that no light from any sources was included in the determination of the sky level. First, the segmentation map produced by \textsc{SExtractor}~\citep{SExtractor} was grown to include the faint outer extents of the sources. For the brightest stars, as well as NGC 2841, the mask was then grown further until no light was visible from these sources using the histogram stretch option in SAOImage DS9. The average and standard deviation of all the sky value measurements in the 1000 elliptical annuli were used to define the sky value and the error on the sky value, respectively. The mean sky surface brightnesses in g and r band were 21.1 mag arcsec$^{-2}$ in g-band and 20.2 mag arcsec$^{-2}$ in r band. The percentage errors on these sky levels were 0.01\% and 0.007\%, corresponding to limiting surface brightness levels of 30.9 and 30.6 mag arcsec$^{-2}$ for the g and r-band images, respectively. The error in the sky value is the dominant source of uncertainty in the surface brightness profile at large radii. 

The HI map used in comparisons below was taken from The HI Nearby Galaxy Survey (THINGS), made using data from the NRAO Very Large Array (VLA)~\citep{Walter2008}. We obtained a far UV (FUV) map of NGC 2841  from the Galaxy Evolution Explorer (GALEX) Nearby Galaxies Survey~\citep{GildePaz2007} using the Detailed Anatomy of Galaxies (DAGAL) image repository~\citep{DAGAL2015}.

\section{Analysis and Results}
\label{sec:results}
\subsection{Size of the stellar disk}

\begin{figure*}[!htbp]
\centering
\includegraphics[width=0.8\textwidth]{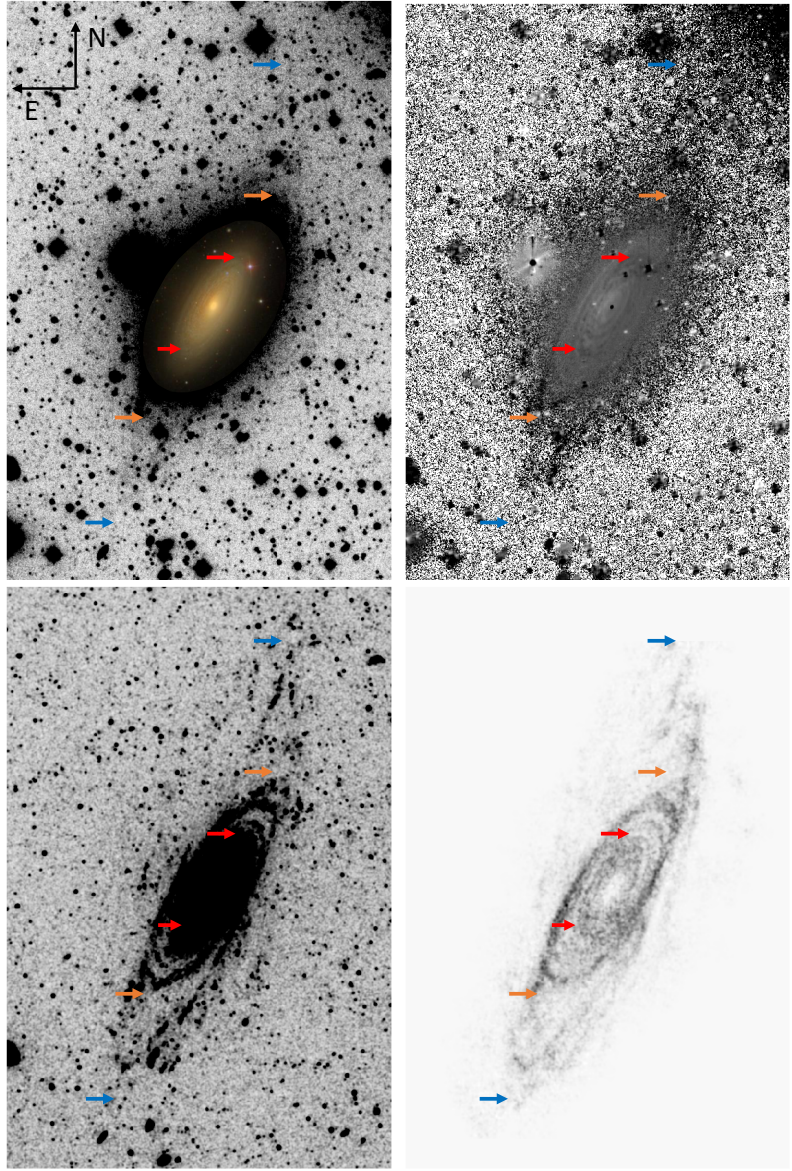}
\caption{A panchromatic view of NGC 2841. Top left: Dragonfly g-band image. The color image embedded in the center was taken from the Sloan Digital Sky Survey data release 14 skyserver website~\citep{SDSSDR14}. Top right: Dragonfly (\textit{g}-\textit{r}) color image. Bottom left: GALEX FUV image. Bottom right: THINGS HI image. The arrows (red, orange, blue) in each image mark the following radii along the galaxy: $R_{25}$ (14.2 kpc), 30 kpc, 60 kpc. The UV, HI gas and visible wavelength emission are traced out to similar radii in the disk of NGC 2841, with the peaks of the visible wavelength disk corresponding to the peaks of the HI and UV emission.}
\label{fig:image}
\end{figure*}

Our images of NGC 2841 in \textit{g}-band and (\textit{g}-\textit{r}) color are shown in Figure~\ref{fig:image}, together with maps of FUV and HI emission. Figure~\ref{fig:image}'s \textit{g}-band image shows a giant disk extending as far as the HI and XUV emission. This disk is visible out to $\sim$60 kpc radius, which is $\sim$4 $R_{25}$ ($R_{25}=14.2$ kpc). To guide the eye, three pairs of arrows are marked in Figure~\ref{fig:image}, colored red, orange, and blue, corresponding to radii of 14.2 kpc ($R_{25}$), 30 kpc, 60 kpc respectively. The orange arrows (30 kpc) mark the edge of the well-studied inner disk of this commonly-observed galaxy~\citep{Block1996,Leroy2008,Silchenko2000,Afanasiev1999}. The disk beyond $\sim$30 kpc appears warped in all three wavelengths. Interestingly, there may be two distinct warped disks, which is most obvious in the (\textit{g}-\textit{r}) color and HI images. The peaks of the HI disk correspond to the peaks seen in the XUV and the visible wavelength data, which suggests that, at large radii, stars in this galaxy are mainly in a disk and are not part of a stellar halo.

\begin{figure}[!htbp]
\centering
\includegraphics[width=0.52\textwidth]{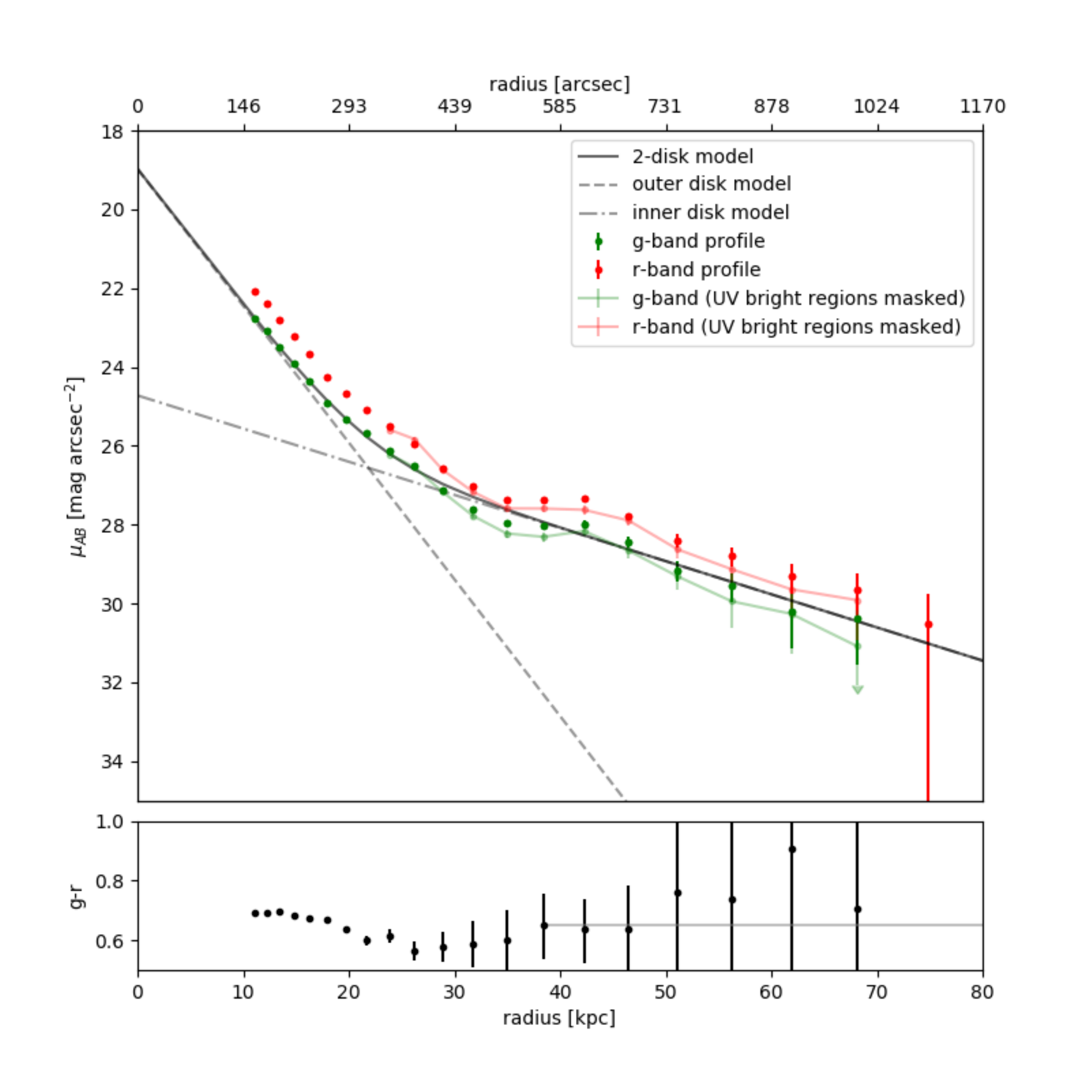}
\caption{Top: Surface brightness profile of NGC 2841 in Sloan \textit{g} and \textit{r}-band. A 2-disk model is fit to the surface brightness profile. The inner and outer disks have disk scale lengths of 3.1$\pm$0.1 kpc and 13$\pm$3 kpc respectively. Plotted in a lighter shade of green and red with lines connecting the data points is the derived surface brightness profile after masking the UV bright regions. Bottom: The (\textit{g}-\textit{r}) color profile. The error bars in both plots include RMS and sky errors. } 
\label{fig:profile}
\end{figure}

In order to make \textit{g} and \textit{r}-band surface brightness profiles, stars and other sources were masked so that they did not contribute to the low surface brightness outer disk. The method used to create the mask for making the surface brightness profile was similar to that for measuring the sky level. The only difference is that NGC 2841 was removed from the segmentation map used to create the mask and the step to mask NGC 2841 by visual inspection in SAOImage DS9 was not applied. The mask for the bright star to the east of the galaxy (see Figure~\ref{fig:image}) overlaps with the central bulge of NGC 2841. Based on previous surface brightness profiles created of the inner disk, the bulge is no longer dominant beyond 60 arcseconds~\citep{Boroson1981}, so we can safely fit a pure exponential disk to the surface brightness profile beyond 200 arcseconds, and this is where the surface brightness profile in this paper begins. To create the profile, isophotes were fitted using the {\tt iraf.stsdas.isophote.ellipse} routine~\citep{Jedrzejewski1987} in PyRAF\footnote{PyRAF are products of the Space Telescope Science Institute, which is operated by AURA for NASA}. An initial x and y coordinate for the center of the galaxy was input, but the center, position angle and ellipticity of isophotes were allowed to vary. However, beyond a radius of $\sim$35 kpc, there is no longer enough signal to noise for the ellipse-fitting routine to allow these parameters to vary, and so the ellipse shape is fixed beyond that radius. The routine also extracts the average unmasked pixel value in each isophote. The \textit{g} and \textit{r}-band surface brightness profiles of NGC 2841 are shown in Figure~\ref{fig:profile}, where the error bars include random as well as systematic sky-errors.   

The visible light in the outskirts of NGC 2841 is clearly part of an extended disk, because the visible wavelength morphology of the galaxy traces the HI and UV disks. A 21-cm kinematic study of the galaxy shows there is a warp in the gas disk~\citep{Bosma1978}. Visually, \textit{g} and \textit{r}-band light beyond $\sim$30 kpc is dominated by a warped outer disk, aligned with the HI and the UV star forming disk. The warp is most obvious in the \textit{g}-\textit{r} color, and the HI images. As a further indicator of a stellar disk warp, the position angle of the fitted elliptical isophote jumps from -30 $\pm$1.5 degrees clockwise from the y-axis within 30 kpc to -26.5 at 30 kpc. Note, however, that the position angle is not allowed to vary in the ellipse fitting routine beyond $\sim$35 kpc due to low signal to noise. Because the disk is warped, and because the surface brightness profile appears to up-bend, a two-exponential disk model is most appropriate and this was fit to the galaxy surface brightness profile. The inner and outer disks have scale lengths of 3.1$\pm$0.1 kpc and 13$\pm$3 kpc, respectively. The two-disk model is shown in Figure~\ref{fig:profile}, with the solid grey line being the sum of the two components. Our measured surface brightness profile extends to $\sim$70 kpc, which is $\sim$23 inner disk scale lengths. 

\subsection{Is the outer disk light contaminated by scatter from the wide-angle PSF?}

As described in the Introduction, the wide-angle point-spread function can play an important role in the measurement of profiles at very low surface brightness levels. Since the visibility of the stellar aureole varies as a function of atmospheric conditions, part of the data reduction pipeline for Dragonfly data rejects science exposures with zeropoints that deviate from a nominal zeropoint by more than $\sim$0.1 mag. This procedure removes the science exposures most contaminated by scattered light from the PSF. To test the significance of remaining contamination, we convolved a measured PSF with a one-dimensional model galaxy profile similar to NGC 2841 to observe the change in the surface brightness profile at large radii. A bulge central surface brightness of $\mu_0=20.1$ and bulge effective radius of $R_e=0.94$ kpc was used~\citep{Boroson1981} together with the two-disk model found in this paper. 

The measured PSF has a radius of 10 arcmin and spans 18 magnitudes in surface brightness. The inner part of the PSF was measured using the brightest unsaturated star in the field. The outer part of the PSF was measured using the brightest saturated star in the field. The IRAF\footnote{IRAF is distributed by the National Optical Astronomy Observatory, which is operated by the Association of Universities for Research in Astronomy (AURA) under a cooperative agreement with the National Science Foundation.} routine {\tt pradprof} was used to compute a radial profile around each star, which was then median binned in the radial direction to remove contamination by other sources. We note that this can only overestimate the PSF compared to the PSF that would be obtained without contamination from other sources. 

The outcome of this exercise was that, because of the careful control of systematics in our experimental setup, the surface brightness profile of NGC 2841 remains unaffected by the wide-angle PSF down to at least $\mu=32$ mag arcsec$^{-2}$. 

\subsection{How is the extended light distributed?}

Is the extended visible wavelength emission from NGC 2841 simply the visible wavelength counterpart of the UV knots identified by GALEX? Or is this light truly distributed at all azimuthal angles around the disk? 

To explore whether the visible wavelength light in the outer regions of NGC 2841 is entirely the visible wavelength emission from the UV knots identified by GALEX, we measured the \textit{g} and \textit{r}-band surface brightness profiles again after masking out the UV bright regions to see how much signal is left outside of the star forming regions. This surface brightness profile is shown also shown in Figure~\ref{fig:profile} as line-connected green and red data points for the \textit{g} and \textit{r}-band profiles, respectively. While the surface brightness profiles in both \textit{g} and \textit{r}-band have dimmed (by $\sim20\%$ beyond 30 kpc) as a result of this masking, the overall shape and extent of the light profiles remain similar. We therefore conclude that the outer disk light is not simply the visible wavelength counterpart of the UV knots identified by GALEX. 

\begin{figure}[!htbp]
\centering
\includegraphics[width=0.52\textwidth]{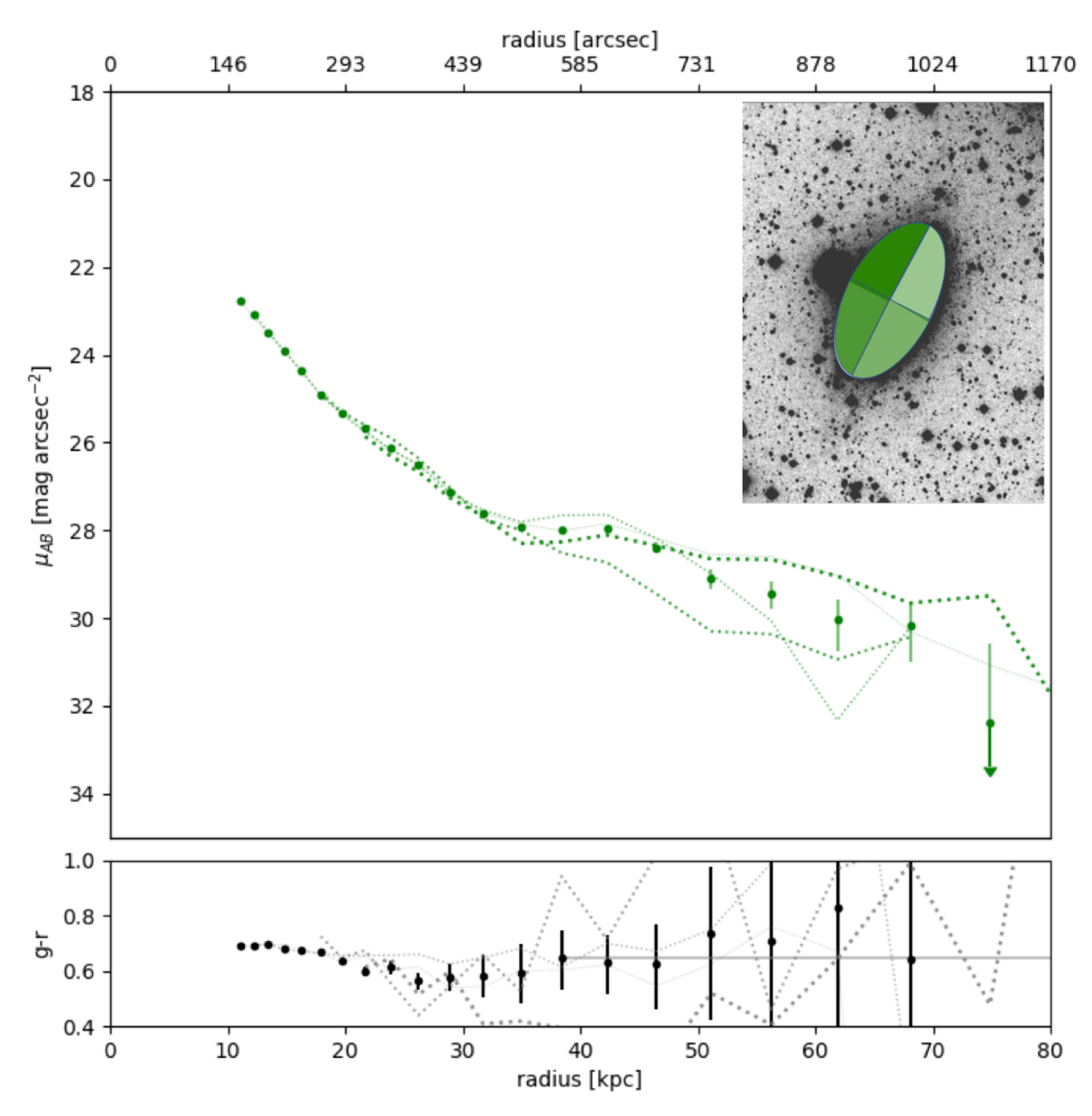}
\caption{Top: The surface brightness profile of NGC 2841 in the Sloan \textit{g} filter is shown again in green data points with error bars. Dotted green line profiles are azimuthal wedge \textit{g}-band profiles. The darkest dotted line corresponds to the darkest green wedge shown in the top right-hand side image, with subsequent lighter shades of green corresponding to other lighter green wedges. Bottom: The (\textit{g}-\textit{r}) color profile is shown again in black. Dotted grey lines are the azimuthal wedge color profiles. The darkest grey dotted line corresponds to the darkest green wedge shown in the top right-hand side image, with subsequent lighter shades of grey corresponding to other lighter green wedges. The error bars in both plots include RMS and sky errors and are indicative of error bar size for all azimuthal wedge profiles.} 
\label{fig:profile_wedge}
\end{figure}

A surface brightness profile averages light from all angles. This means in the low surface brightness galaxy outskirts the profile shape might be dominated by features in a small azimuthal wedge of the galaxy. To see if there is extended galaxy emission at all azimuthal angles, surface brightness profiles for azimuthal wedges were measured and plotted in Figure~\ref{fig:profile_wedge}. While there is scatter in the profile in different azimuthal wedges, there is consistently light at all angles, lending further evidence to a smooth underlying disk that is the continuation of disk visible in Figure~\ref{fig:image}.

\subsection{Mass of stellar disk and comparison to the gas disk}
\begin{figure}[!htbp]
\centering
\includegraphics[width=0.5\textwidth]{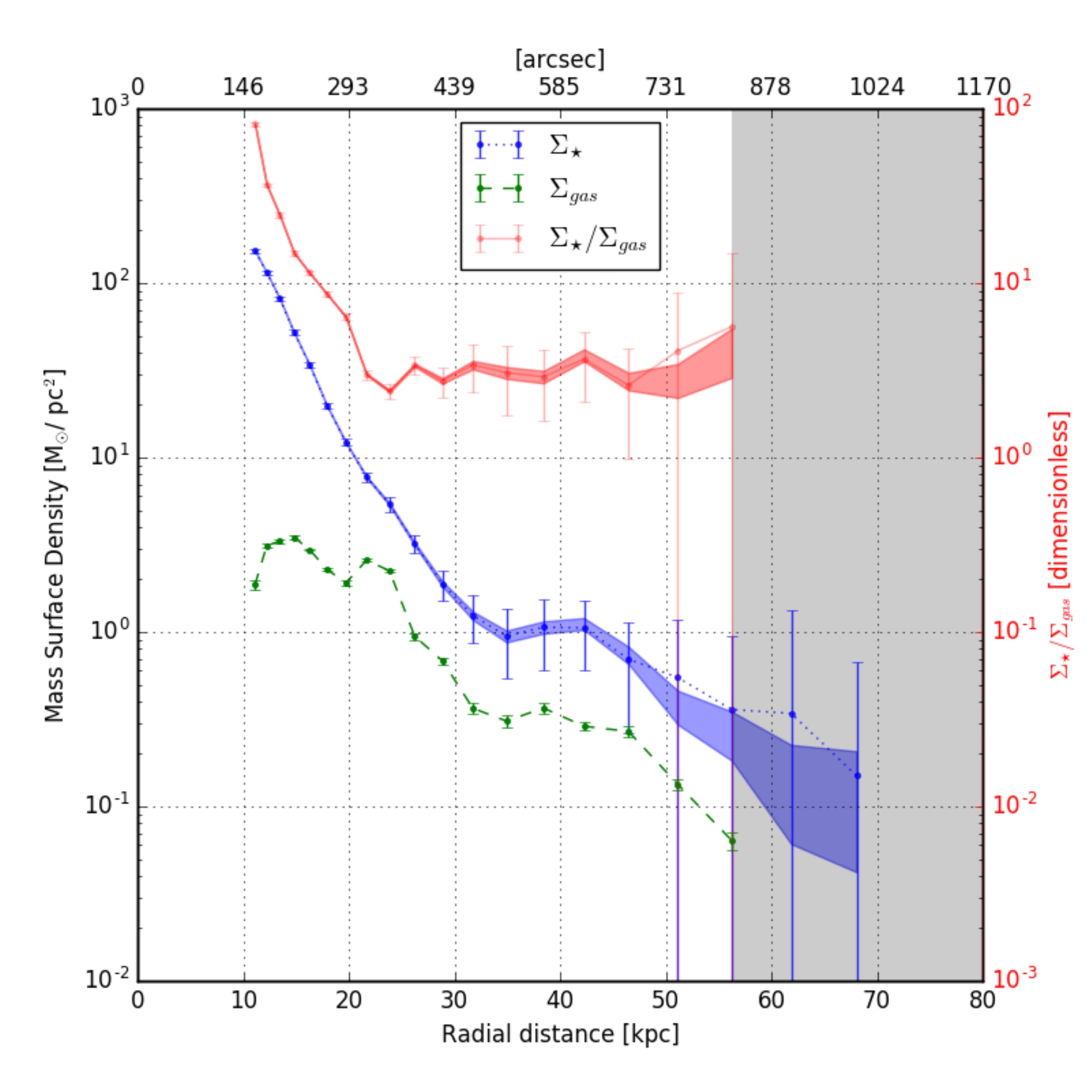}
\caption{The mass surface density of stars and gas ($\Sigma_{\star},\Sigma_{gas}$ $[\text{M}_{\odot} \text{ pc}^{-2}]$) and the ratio of the two ($\Sigma_{\star}/\Sigma_{gas}$) are plotted as a function of radius for NGC 2841. The error bars include the RMS uncertainty and the systematic error due to sky uncertainty in the g-band image and the (\textit{g}-\textit{r}) color used to calculate the stellar mass surface density. The shaded profiles assume that the color in the outer disk beyond 40 kpc is a constant. See the text for more details.}  
\label{fig:profile_masses}
\end{figure}

The stellar and the gas mass surface density profiles of NGC 2841 are shown in Figure~\ref{fig:profile_masses}. The stellar mass surface density was calculated in the same way as described in~\cite{vanDokkum2014}, using relations given in~\cite{Bell2001}:
\begin{equation}
\log_{10}(\rho [\text{M pc}^{-2}])=-0.4(\mu_g [\text{mag arcsec}^{-2}] -\text{DM}) + 1.49(g-r) + 1.64 + \log_{10}(1/C^2)
\label{eqn:stellarmass}
\end{equation}
Using a distance of 14.1 Mpc~\citep{Leroy2008}, $\text{DM}=30.7$ is the distance modulus and $\text{C}=0.0684$ is the conversion factor from arcsecond to kiloparsecs. 

Since the mass density depends on color, we explored the impact of color on the inferred mass density using two different approaches. Densities obtained using both approaches are shown in Figure~\ref{fig:profile_masses}, as error bars and as shaded regions. In the first approach, we used the measured (\textit{g}-\textit{r}) color, and its uncertainty, which includes RMS and sky errors in both filters. The uncertainties in the mass measurements using this approach are displayed as error bars and include color and \textit{g}-band RMS and systematic sky errors. In the outskirts, sky errors dominate. This is the most conservative indication of how well the stellar mass in the outskirts of NGC 2841 can be measured. The second approach, which is plotted as the shaded region in Figure~\ref{fig:profile_masses}, was to use the measured (\textit{g}-\textit{r}) color within 40 kpc, and a constant value of 0.65 beyond that. This choice of color is indicated by the grey horizontal line in the lower panel of Figure~\ref{fig:profile}. Our rationale for adopting this constant color is because beyond 50 kpc the measurements of (\textit{g}-\textit{r}) color have very large uncertainties due to uncertainties in the sky level in both \textit{g} and \textit{r}-band. One can view the shaded region as a potential stellar mass surface density profile if the color in the outer disk remains at a constant 0.65 beyond 40 kpc. The errors indicated by the shaded region in the stellar mass surface density in Figure~\ref{fig:profile_masses} include both RMS errors and systematic sky-errors in the \textit{g}-band data.

The gas mass surface density was calculated using equation A1 from~\cite{Leroy2008}:
\begin{equation}
\Sigma_{\text{gas}} [\text{M}_\odot \text{pc}^{−2}] = 0.020 \cos i \text{ } I_{21 \text{cm }}[\text{K km s}^{-1}]
\label{eqn:gasmass}
\end{equation}
where $i=1.29$ is the inclination in radians~\citep{Leroy2008}. $I_{21 \text{cm}}$ is the 21 cm flux from the THINGS HI map. This gas mass equation includes a factor 1.36 to reflect the presence of helium. The gas mass surface density is plotted in green in Figure~\ref{fig:profile_masses}. Error bars shown account for the RMS scatter in each isophotal annulus. The shaded grey region is where the THINGS HI map and the GALEX FUV map has no detection at their respective sensitivity limits.

To better illustrate the relationship between stellar and gas mass, the ratio of the two is plotted in Figure~\ref{fig:profile_masses} on the right-hand axis, in red. The ratio of stellar mass to gas mass remains remarkably constant (at 3:1) from just beyond $\sim$20 kpc, to the limit of the THINGS data. Remarkably, at the sensitivity limit of the THINGS survey and at the current depth of Dragonfly's observations of NGC 2841, there is no radius at which the mass surface density of HI gas begins to dominate over that of the stars. This has been measured out to 50 kpc, or 16 inner disk scale lengths. At radii greater than 50 kpc, the uncertainty in the sky levels of both the \textit{g} and \textit{r} Dragonfly images means the stellar mass is not measured with enough precision to conclude it is greater than the gas mass. However, the color in the disk beyond 50 kpc would have to be bluer than (\textit{g}-\textit{r}) = 0.3 in order for the stellar mass to drop below that of the gas mass. 

\subsection{Timescales}

\begin{figure}[!htbp]
\centering
\includegraphics[width=0.5\textwidth]{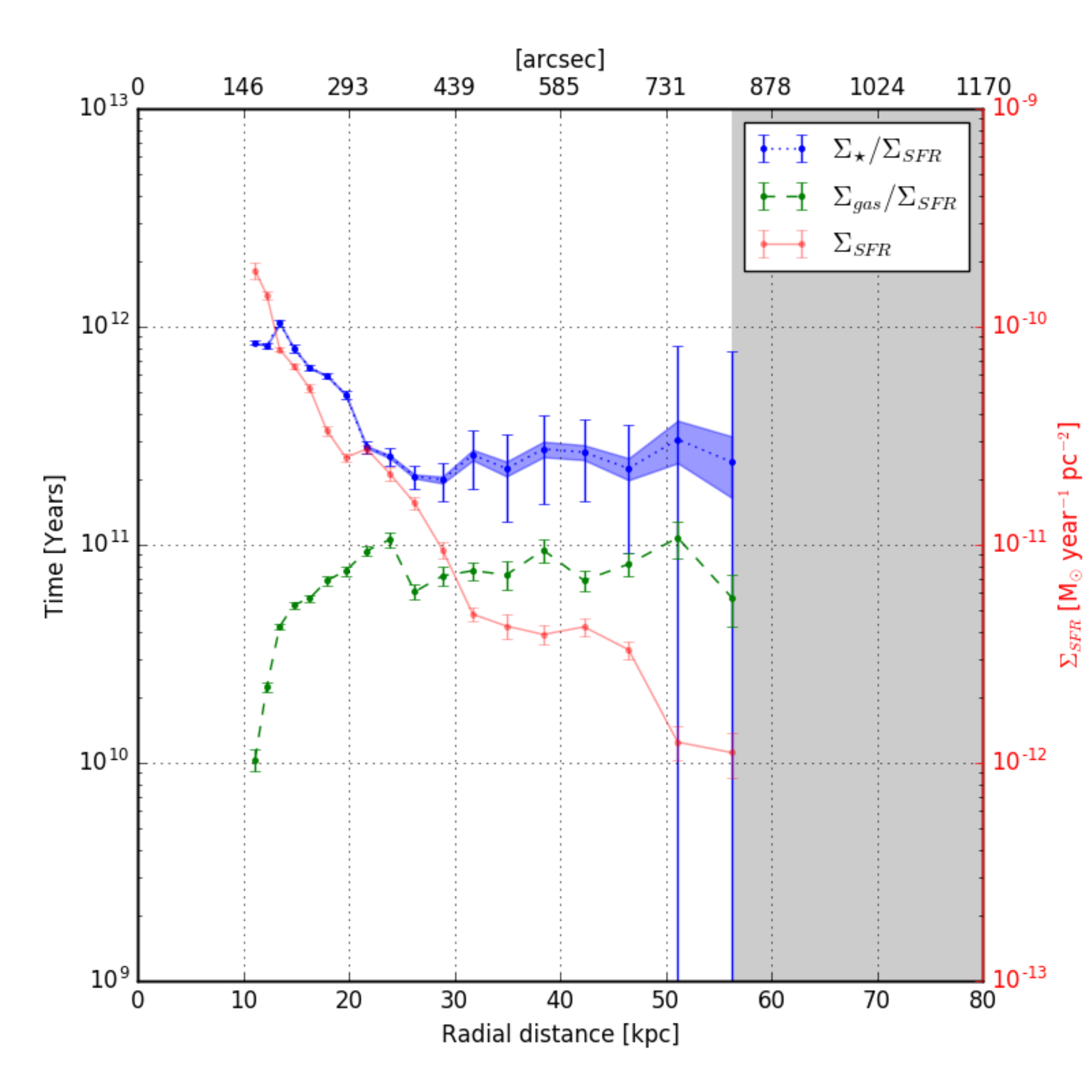}
\caption{The stellar mass buildup time ($\Sigma_{\star} / \Sigma_{\text{SFR}}$, blue) and gas depletion time ($\Sigma_{\text{gas}} / \Sigma_{\text{SFR}}$, green) and SFR surface density (black) are shown as a function of radius. The error bars include the RMS uncertainty in the HI, UV, \textit{g}-band and \textit{r}-band measurements, as well as the systematic error due to sky uncertainty in the g-band image and the (\textit{g}-\textit{r}) color used to calculate the stellar mass surface density. }
\label{fig:profile_timescales}
\end{figure}

The gas depletion time ($\Sigma_{\text{gas}} / \Sigma_{\text{SFR}}$) and the stellar mass buildup time ($\Sigma_{\star} / \Sigma_{\text{SFR}}$) for NGC 2841 are plotted as a function of radius in Figure~\ref{fig:profile_timescales}. The star formation rate (SFR) surface density used was obtained from the GALEX FUV image using equation 1 taken from~\cite{Wilkins2012}:
\begin{equation}
\Sigma_{\text{SFR}} [\text{M}_{\odot} \text{ year}^{-1} \text{ pc}^{-2}] = 10^{-34} \text{ L}_{\text{FUV}} [\text{ergs} \text{ s}^{-1} \text{ Hz}_{-1} \text{ kpc}^{-2}] \text{ B}_{\text{FUV}} 
\label{eqn:SFR}
\end{equation}
where $\text{B}_{\text{FUV}}$ varies with the slope of the initial mass function (IMF). The value used here is $0.9 \pm 0.18$, based on the Kennicutt 1983 IMF~\citep{Kennicutt1983}. The SFR surface density is also plotted in Figure~\ref{fig:profile_timescales}, on the right-hand axis, in red. The SFR and HI mass surface density error bars only includes the RMS scatter in each isophotal annulus. The error bars in  $\Sigma_{\star} / \Sigma_{\text{SFR}}$ include fractional errors in $\Sigma_{\star}$ and $\Sigma_{\text{SFR}}$ added in quadrature. For $\Sigma_{\star}$, the error bars were calculated using two methods in the same way as in Figure~\ref{fig:profile_masses}: the error bars include the \textit{g} and (\textit{g}-\textit{r}) errors while the shaded profile uses a (\textit{g}-\textit{r}) color model and only include the \textit{g}-band errors.

The stellar mass buildup and gas depletion timescales remain constant from beyond $\sim$20 kpc, the same region with a constant 3:1 ratio of stellar to gas mass surface density. The stellar mass buildup time in this region is ~250 Gyr and the gas depletion time is on the order of 70 Gyr. Both of these timescales are much longer than the age of the Universe. 

\section{Discussion}
\label{sec:discussion}

\subsection{The origin of outer disk stars}

The underlying stellar disk in NGC 2841 discovered using Dragonfly is gigantic, reaching beyond the size of the most sensitive HI and UV disk observations to $\sim$70 kpc ($\sim$5 $R_{25}$ or $\sim$23 inner disk scale lengths). The surface brightness profile of the galaxy shows an upward bending break at 30 kpc, when the \textit{g}-band surface brightness is $\sim$28 mag arcsec$^{-2}$. Similar upward bending (Type I) surface brightness profiles are common at large radii, measured using a combination of photometry and star counts~\citep{Barker2009,Barker2012,Watkins2016}. Of note is a multi-object spectroscopy study of stars in M31 by ~\cite{Ibata2005}, which traced disk stars out to $\sim$70 kpc. In NGC 2841, the position of the upbend corresponds to the start of a low surface brightness warp in the outer disk. This warp is visible in both \textit{g} and \textit{r}-band, THINGS HI and GALEX UV images. One common assumption is that in the outer disks of galaxies, neutral gas is the dominant baryonic component based on the observation that in general HI disks extend much further than stellar disks~\citep{vanderKruitFreeman2011,Elmegreen2016}. A comparison of the stellar mass to gas mass surface densities shows that for NGC 2841, there is no radius at which the mass surface density of gas begins to dominate over that of the stars. Beyond $\sim$20 kpc, NGC 2841 also has the interesting property that the stellar to gas mass surface density ratio is a constant 3:1.

A central question is: how did this giant stellar disk form? There are three main ways to populate the outer disk with stars: (1) stellar migration, (2) accretion of stars, and (3) in-situ star formation. Note that one mechanism does not preclude the others. We discuss the merits and weaknesses of each of these possibilities below.

(1) \textit{Stellar Migration:} \cite{SellwoodBinney2002} showed that transient spiral arms in galaxy disks can scatter stars into orbits at different radii, but allow the stellar orbits to remain circular. Subsequent simulations by Ro{\v s}kar~\citep{Roskar2008a, Roskar2008b} showed that a downward bending (type II) surface brightness profile can be explained by a star formation threshold combined with stellar migration, which can move stars beyond their formation radius. There are several issues with appealing to stellar migration to populate the outer disk of NGC 2841. Firstly, it is unclear whether stellar migration can move so much mass to such a large range of radii beyond 30 kpc.~\cite{Watkins2016} has similar concerns with appealing to stellar migration as a means of populating the outer disk of three nearby galaxies, where stars have to be moved several disk scale lengths beyond the extent of spiral arms. Secondly, stellar migration should not create an upwards bending surface brightness profile. Thirdly, the stars, as well as the gas, beyond 30 kpc are in a warped disk, and it is unclear how the migrated stars would end up in such a warped orbit. Perhaps the stars were displaced to large radius long ago and the warp was induced by a later interaction that warped both the stars and the gas.

(2) \textit{Accretion:} In order for accreted stars to build up a co-planar disk, incoming stars need to have a narrow range in angular momenta that matches the existing disk, otherwise simulations show that the accreted stars tend to end up in a bulge or stellar halo~\citep{Toomre1977,Schweizer1990}. In between these extremes lies infall with a slight mismatch in angular momentum, the result of which is a warped disk~\citep{Binney1992}. The warp in the disk of NGC 2841 beyond 30 kpc may hint at past accretion onto the disk with a slightly different angular momentum than that of the underlying stellar population. A quarter of a century ago,~\cite{Binney1992} noted presciently that: ``...it is by no means inconceivable that warps are in direct physical contact with material that is only now joining the galactic system''. Since our data for NGC241 extends out to $\sim$60 kpc, it is tempting to associate the outer warped disk in this system with material that has only recently infallen. If this interpretation is correct, we have reached far enough into the outskirts of NGC 2841 to be probing its circumgalactic environment, and may be witnessing a slightly more evolved part of the cool-flow driven galactic component referred to by ~\cite{Bland-Hawthorn2017} as the `proto-disk'. 

The total stellar mass in the disk beyond $\sim$30 kpc (where the warped disk starts to dominate) is $1.4\pm0.2\times10^8\text{ M}_{\odot}$. For comparison, the Small Magellanic Cloud has a stellar mass of $\sim$7$\times10^8\text{ M}_{\odot}$~\citep{Dooley2017}. The outer disk could potentially be formed by one or several small accretion events with the angular momenta of incoming dwarf galaxies almost aligned with that of NGC 2841. 

The similarity in angular momenta required in this scenario may not be particularly improbable, since it seems that satellites of nearby galaxies can map out organized structures. Well-known examples include the Great Plane of Satellites around M31~\citep{Ibata2013} and the Vast Polar Orbital structure around the Milky Way Galaxy~\citep{Pawlowski2015}. One hypothesis for the existence of these planar structures is accretion along large scale filamentary structures~\citep{Ibata2013}. If this is the case, the accreted dwarf galaxies could deposit their gas and stars onto the outer disk, possibly creating a constant stellar mass to gas mass ratio after a few rotations.  

(3) \textit{In-Situ Star Formation:} Stars in the outer disk of NGC 2841 clearly trace the distribution of the HI gas. This observation tends to favor a model in which the stars were formed in-situ. However, at the current SFR, it would take $\sim$200 Gyr to build up the outer disk stellar mass. On the other hand, the global star formation in the Universe was much higher in the past than today, peaking at redshifts of 2-3, with a SFR an order of magnitude greater than today~\citep{Madau2014}. At a star-formation rate that is 10 times that of the current SFR in NGC 2841, it would take $\sim$15 Gyr to build up the outer disk stellar mass. This is only slightly longer than a Hubble time, so such a scenario could be made to work. If in-situ star formation is responsible, then it begs the question of how the outer disk achieved such high levels of star formation in the past. The entire disk is currently Toomre stable and star formation is occurring in UV knots. It is possible the gas density in the outer disk was higher in the past, but this may not be enough to stimulate sufficient star formation. \cite{Cormier2016} compared a sample of HI-rich galaxies to a control sample and showed that there is no increase in molecular gas mass or SFR in the outer disk in the HI rich sample. The difference was that the HI rich sample was able to able to sustain star formation in the outer disk for a longer period of time. Other studies support this by showing that gas depletion takes longer than a Hubble time and that SFRs and the Toomre stability parameter are not correlated~\citep{Bigiel2010}. \cite{Semenov2017} carried out simulations that suggest global gas depletion times are long because only a small fraction of gas is converted into stars before star-forming regions are disrupted. Therefore, gas has to cycle in and out of the star-forming state many times before being turned into stars.

None of the mechanisms described above strike us as unreasonable, so perhaps the most likely scenario is that the outer disk of NGC 2841 is being built up by a combination of multiple mechanisms. For example, cool flow infall of gas which somehow triggers in-situ star-formation. However, one then wonders how these mechanisms conspired together to result in a constant ratio between stellar and gas mass surface density beyond $\sim$20 kpc. 

The present paper reports results for a single galaxy, NGC 2841, the results from which indicate that with experimental setups optimized for low surface brightness imaging, stellar disks can be probed out to radii where the disk starts to warp. This may be an indication we are seeing parts of the disk that encroach upon the circumgalactic medium. Future papers will carry out similar analyses on other galaxies in the Dragonfly Nearby Galaxies survey, four of which have accompanying THINGS HI data. If NGC 2841 is any guide, the key questions for understanding galactic outskirts must now include: what fraction of massive spirals contain enormous underlying stellar disks? Are these disks always more massive than gaseous disks revealed by HI imaging? Is the mass ratio of stars to gas a constant in the outer disk, as seen in NGC 2841? At ultra-low surface brightness levels, do stellar disks always trace HI, and are these disks always warped in a manner consistent with infall? Is the geometry of the warps correlated with the positions of companion galaxies, as would be expected if the warped disk is built up by infall, and companion galaxies trace dark matter filaments? This long list of questions befits the richness of the phenomena being revealed at low surface brightness levels in the outskirts of galaxies. In any case, it seems to us that the key to answering these questions is to approach them in the appropriate panchromatic context, focusing on comparisons of surface mass densities, and not just on arbitrary definitions of the `sizes' of disks at various wavelengths.

\section{Acknowledgements}

We thank the anonymous referee for a thoughtful and constructive report, which improved the paper. Support from NSERC, NSF grants AST-1312376 and AST-1613582, and from the Dunlap Institute (funded by the David Dunlap Family) is gratefully acknowledged. We thank the staff at New Mexico Skies Observatory for their dedication and support. JZ thanks Rhea-Silvia Remus for useful discussions. 
Funding for the Sloan Digital Sky Survey IV has been provided by the Alfred P. Sloan Foundation, the U.S. Department of Energy Office of Science, and the Participating Institutions. SDSS-IV acknowledges support and resources from the Center for High-Performance Computing at the University of Utah. The SDSS web site is www.sdss.org.

\chapter{The Dust Content of the Spider HI Cloud}
This chapter is temporarily withheld.

\chapter{Summary and Future Work}

This thesis presents the development of the data reduction pipeline for the Dragonfly Telephoto Array, as well as observations of the outer disk of NGC 2841 and the Spider HI cloud in the Milky Way Galaxy. In this chapter we present a summary of this thesis and the future direction of this work. 

\section{Thesis Summary}

\subsection{Chapter 2: The Dragonfly Pipeline}
The Dragonfly Telephoto Array is optimized for low surface brightness visible wavelength imaging. It is able to routinely image at a level of 30 mag arcsec$^{-2}$. This is achieved via hardware as well as software. Key issues limiting low surface brightness observations that are addressed via software include careful sky modeling and subtraction, and the consideration of the effects of the wide-angle PSF. In the Dragonfly Pipeline, sky modeling and subtraction is done in two stages to ensure that the sky is not over or under subtracted in the vicinity of galaxies being studied. Images where there is considerable power in the wide-angle PSF are rejected from the final combined images. Chapter 2 describes in detail how the Dragonfly Pipeline addresses these key systematic errors.

The Dragonfly Software can be run without human intervention or inspection of images. This is important because extragalactic observations typically contain many thousands of dark, flat and science exposures for each target field of view.  Chapter 2 also presents the Dragonfly Pipeline's software architecture, data flow, methods of automatic rejection of problematic frames, data storage and retrieval solutions, and cloud-orchestration. 

\subsection{Chapter 3: A Giant Stellar Disk in NGC 2841}
Neutral gas is commonly believed to dominate over stars in the outskirts of galaxy disks. This may simply be a consequence of the fact that deep HI observations typically probe to a lower mass surface density than visible wavelength data. The Dragonfly Telephoto Array was used to observe a nearby spiral galaxy, NGC 2841. Comparisons of the stellar disk was made to the THINGS HI and GALEX UV disks of this galaxy. 

The underlying stellar disk in NGC 2841 discovered using Dragonfly is gigantic, reaching beyond the size of the most sensitive HI and UV disk observations to $\sim$70 kpc (∼5 R$_{25}$). Contrary to expectations, the stellar mass surface density does not fall below that of the gas mass surface density at any radius. In fact, at all radii greater than $\sim$20 kpc, the ratio of the stellar to gas mass surface density is a constant 3:1. Visually, the bright peaks of the galaxy disk in visible wavelength are aligned with the peaks in the HI and UV images. The stellar disk, just like the HI and UV disks, begins to warp beyond $\sim$30 kpc. At current star formation rates, the stellar mass in this outer disk would take $\sim$200 Gyr to build up. The disk warp may be an indication of a physical connection between the outskirts of the galaxy and infall from the circumgalactic medium.

\subsection{Chapter 4: The Spider HI Cloud}
Correlating the scattered and thermally emitted light from dust in the Spider HI cloud has enabled a measurement of the ratio of the intensity of the two radiative processes in the presence of the same interstellar radiation field (ISRF). The scattered light observations were made with the Dragonfly Telephoto Array, in Sloan \textit{g} and \textit{r}-bands. The thermal emission data was obtained by SPIRE on the Herschel Space Observatory in the 250$\mu$m channel. The challenges overcome to achieve the measurement of the ratio of scattered to thermally emitted light in Spider include (1) sky subtraction and (2) careful accounting of noise sources in both Dragonfly and Herschel data. It is the topic of future work to use this ratio to test dust models. 

\section{Future Work}

\subsection{The Dragonfly Pipeline}
The Dragonfly Pipeline is currently in working order. However, there are always improvements that can be made. I will briefly describe a number of ideas for enhancing Dragonfly's capability for observing low surface brightness phenomena. 

\subsubsection{Improved Sky Subtraction}
The current method of sky subtraction in the Dragonfly Pipeline relies on the existence of pixels in each image which can be assumed to contain only sky signal. These pixels need to be distributed evenly across the whole field of view of the image in order for a sky model to be fit to these pixels and the model subtracted. As shown in Chapter 4, where I analyze the Spider HI cloud, the existence of empty sky regions may not be a good assumption in some cases of interest for Dragonfly. One such application currently being explored using Dragonfly is imaging of the cosmic web in H-alpha. New sky modeling and subtraction techniques need to be developed and incorporated into the Dragonfly Pipeline for these purposes. 

\subsubsection{Better Understanding of Confusion from Compact Sources}
The current limiting factor for the detection of low surface brightness objects in areas of low foreground cirrus is likely to be the confusion from multiple sources in our large pixels. As a reference, in Chapter 4, in order to identify pixels with negligible signal from stars and galaxies, where there is only sky and cirrus signal, up to 70\% of pixels in a Dragonfly image could be masked for images convolved to 36", in order to compare with Herschel. Confusion would not be as severe at the native Dragonfly resolution, where the PSF has a FWHM of $\sim$5", but it is still a limiting factor in affected pixels. It is difficult to distinguish a collection of faint stars and galaxies from a low surface brightness object. In this case, PSF modeling and subtraction could make a big difference, as would an upgrade of the instrument to use cameras with smaller pixels. 

\subsubsection{Better Testing Framework}
The scripts for the Dragonfly Database were written at the same time as unit tests for these scripts. Unit tests allow small units, or components, of code to be tested. The implementation of unit tests for software associated with the Dragonfly Database means that every time an update to a script is made, it is quick to determine whether an unintentional bug has been introduced into the code. 

Unit tests were not originally used when developing the Dragonfly Pipeline. The current testing framework for the Dragonfly Pipeline (excluding the Dragonfly Database scripts) is ad hoc and cumbersome. For example, if one updates a small part of a single script, there may not be a need to run the entirety of the script to test it, as is currently being done. This means time is wasted when making small updates to the code for testing purposes. Furthermore, the current testing framework is such that tests are essentially running sub-steps in the pipeline. This means that one has to make sure one has the testing data in the right directories, calling the script in the particular ways, with hard-coded path names. Testing things in this way is more time consuming and more fragile than using a unit test framework.    

\subsubsection{Compression of Raw Dragonfly Data}
The current data flow from New Mexico Skies (NMS), the site where Dragonfly is hosted, does not compress data before file transfer over the internet. Occasionally, when there are network issues at NMS, automatic file transfer cannot finish the next day. The backlog of data then requires human intervention to sort out. One way of overcoming this issue is to compress the FITS files before file transfer to the data storage RAID machines in Toronto. The files can remain compressed as they are uploaded onto the CANFAR VOSpace cloud storage facilities too. Automatic scripts that compress and decompress fits files when uploading and downloading from the VOSpace would make this procedure seamless. Preliminary tests of compression and decompression were done using the NASA's High Energy Astrophysics Science Archive Research Center (HEASARC) software \textit{fpack} and \textit{funpack}, and these showed that lossless compression would reduce fits file sizes from $\sim$74 MB to $\sim$53 MB, which is a 30\% reduction. 

\subsubsection{Dragonfly Telescope monitoring system}
\begin{figure*}[!htbp]
\centering
\includegraphics[width=0.85\textwidth]{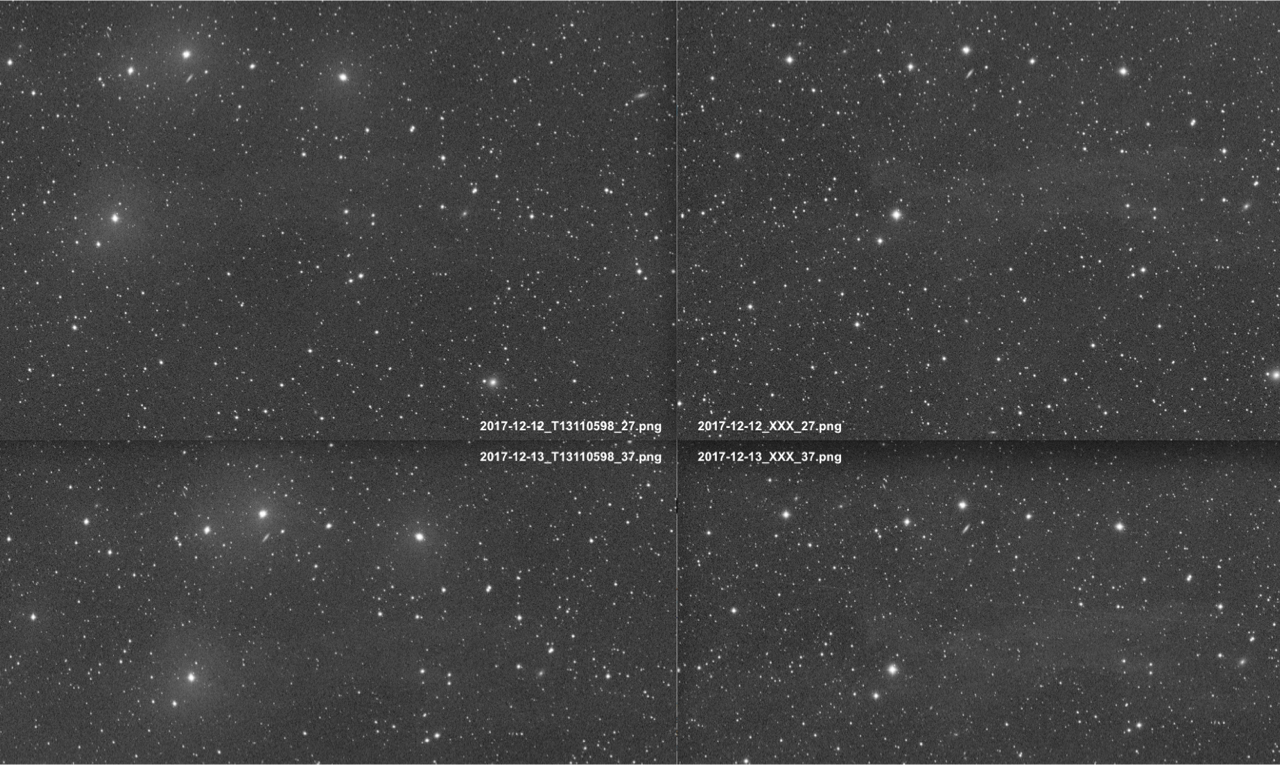}
\caption{The right hand side shows two images taken by the camera with a serial number of T13110598, on two different nights: 2017-12-12 and 2017-12-13. The right hand side shows two images taken at the same time as their left hand side counterparts, but by another lens-camera subsystem. Notice that the images on the left have wide-angle PSFs, whereas the images on the right do not. Images taken by all other lens-camera subsystems at the same time as these do not have significant wide-angle PSFs, and look similar to the images on the right.}
\label{fig:lensmoisture_aureole}
\end{figure*}

\begin{figure*}[!htbp]
\centering
\includegraphics[width=0.75\textwidth]{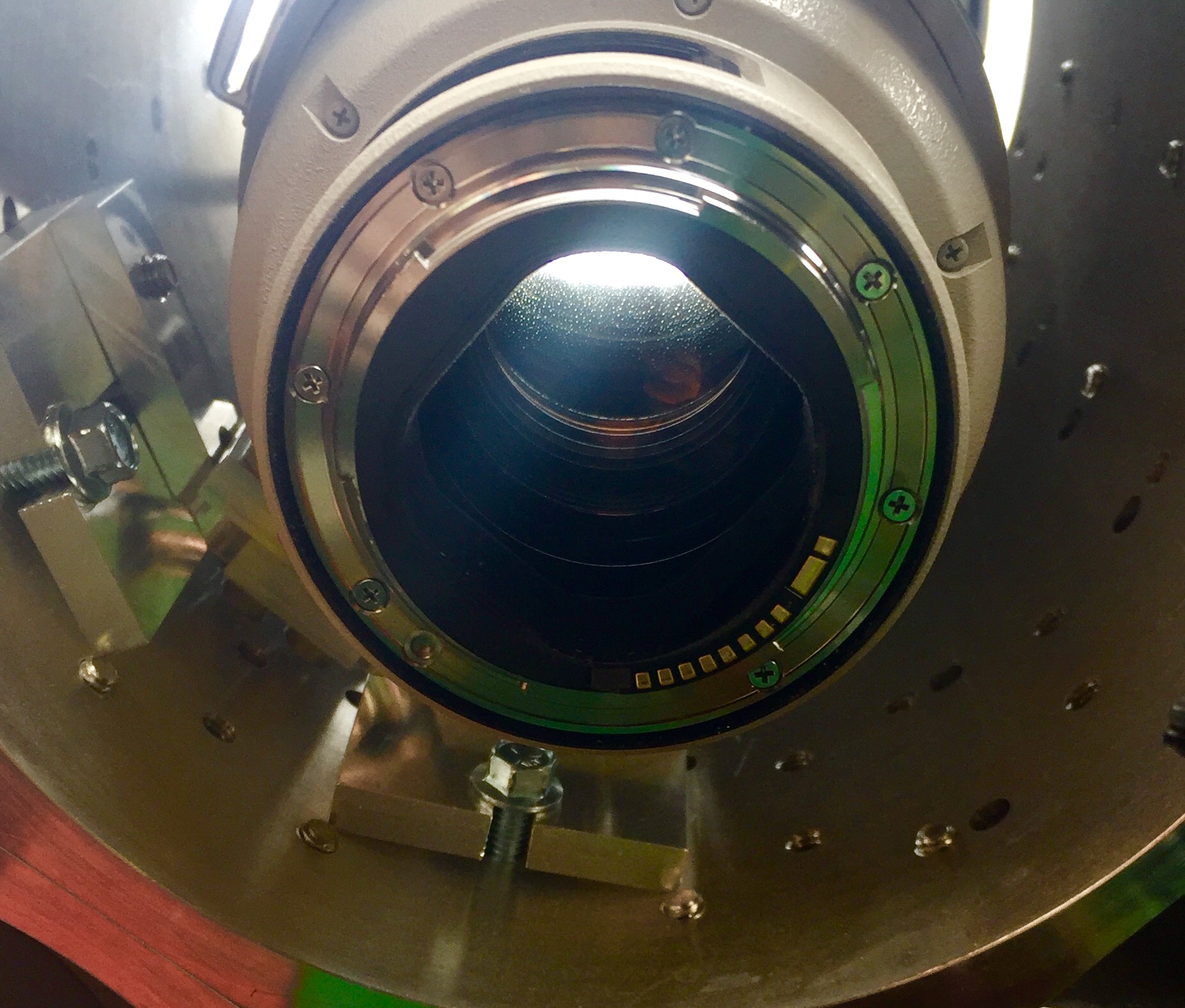}
\caption{The lens attached to the camera with a serial number of T13110598 in March 2018. Notice the moisture on the inside of the lens.}
\label{fig:lensmoisture_lens}
\end{figure*}

The health and performance of the lens-camera subsystems on Dragonfly and the observing conditions at the NMS observatory site are not currently systematically monitored. For example, in March, 2018, it was discovered serendipitously that one of the lenses had moisture on an interior surface. What alerted the team to this issue was noticing that the zeropoints of this lens-camera subsystem were consistently lower than what they normally are. Upon inspecting the images, it was obvious that stars in images taken by this subsystem always had stellar aureoles, even when images taken at the same time by all the other subsystems did not. Usually we associate stellar aureoles with atmospheric conditions that scatter light to large angles, however, in this case, the origin was clearly instrumental. Example images taken with this lens-camera subsystem are shown on the left hand side of Figure~\ref{fig:lensmoisture_aureole}. The right hand side of this Figure shows a representative image taken at the same time from another camera-lens subsystem. An image showing the moisture on the lens of the problematic subsystem is shown in Figure~\ref{fig:lensmoisture_lens}. 

This lens moisture is but one of the many problems that can occur with the hardware on Dragonfly. Other problems (such as broken fans on the cameras, significant numbers of dead pixels on the CCD) can also be diagnosed with raw or processed Dragonfly images. As the Dragonfly Database becomes populated with information regarding both raw and processed images, a Dragonfly Telescope monitoring system could be developed. This would mean that instead of the moisture in the lens being diagnosed at least 3 months after the fact, it would be identified and a notification could be sent to the team the next day.  

\subsection{The Growth of Galaxy Disks}
\begin{figure*}[!htbp]
\centering
\includegraphics[width=0.75\textwidth]{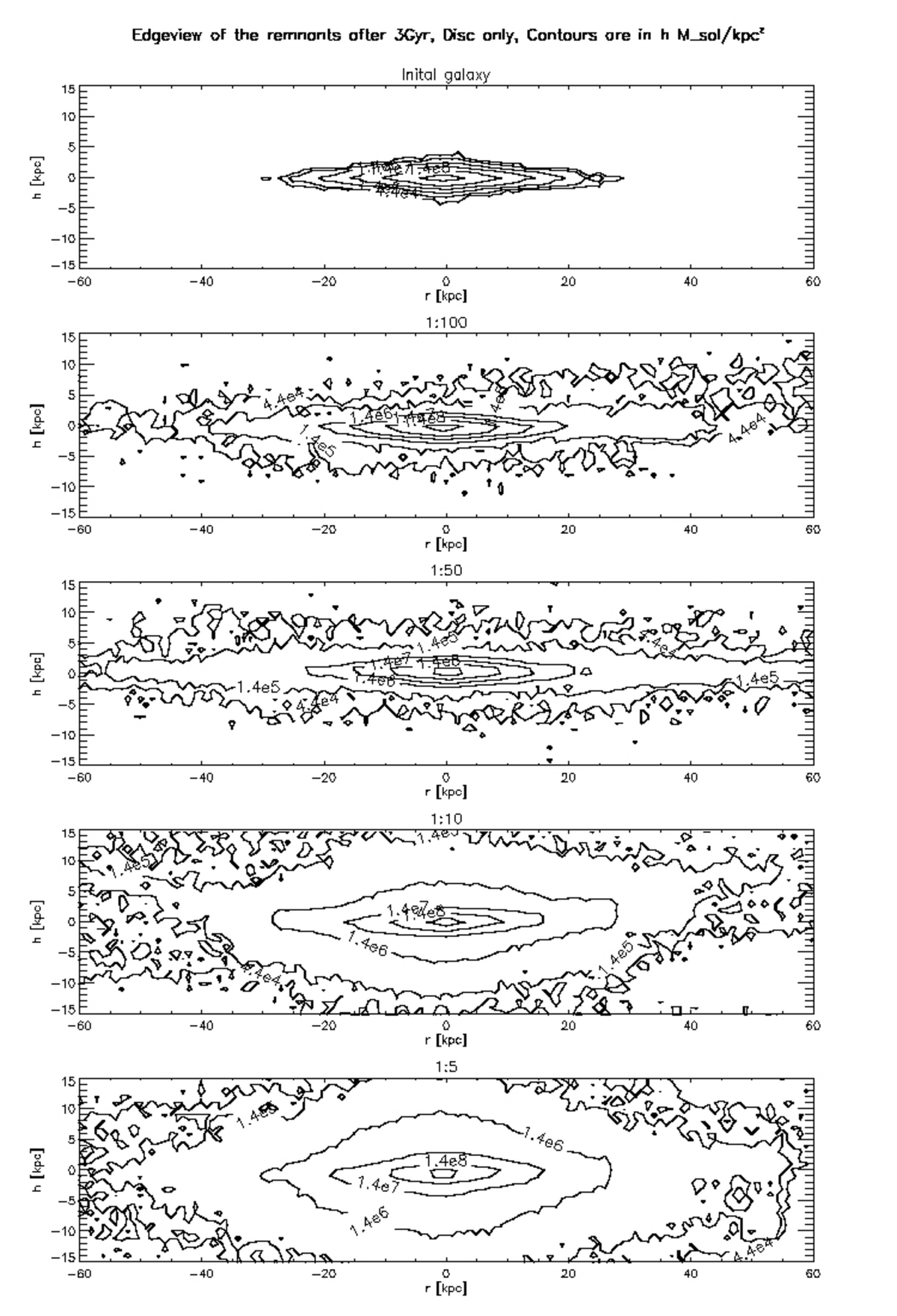}
\caption{This Figure has not been published yet and is provided by Dr. Rhea-Silvia Remus from the University Observatory Munich. The top panel is an edge-on view of a galaxy simulated with a bulge and disk model taken from NGC 2841. Subsequent panels show what this simulated galaxy would look after co-planar accretion of satellite galaxies with different mass ratios. These simulations show that if the mass ratio of the satellite to accreting galaxy is less than 1:10, the deposited stars grow the disk and not a bulge or stellar halo.}
\label{fig:accretionsim}
\end{figure*}

The Dragonfly Telephoto Array has demonstrated spectacular performance in studying the outer disk of NGC 2841. The stellar disk of this galaxy was found to extend further than the HI gas disk at the sensitivity of the THINGS survey, to five times R$_{25}$. Both the stellar and HI disks of this galaxy are warped in the outskirts. Comparing the current star formation rates to the stellar mass in this outer stellar disk, in-situ star formation seems like an unlikely formation mechanism for the outskirts of this newly discovered enormous stellar disk. The most likely formation scenario is accretion of satellite galaxy (or galaxies) co-planar to the existing NGC 2841 disk. It is unclear whether satellite accretion in this way can grow the stellar disk radius without growing the stellar bulge or stellar halo. In collaboration with Dr. Rhea-Silvia Remus and her student Geray Karademir, we started to explore this question. They simulated co-planar satellite accretion onto an NGC 2841 like galaxy, with its mass, and bulge and disk model. Preliminary results indicate that if the satellite galaxy has a non-zero impact parameter and has a mass ratio of less than 1:10, the accretion event grows the galaxy disk and not a bulge or stellar halo. Images of the relaxed galaxy post accretion of satellite galaxies of different masses is presented in Figure~\ref{fig:accretionsim}. 

These preliminary results are promising, but further checks are warranted. NGC 2841 has an XUV disk, does the accretion event stimulate an XUV disk that continues after the accreted stars have already settled into the disk? If there is gas in the satellite galaxy being accreted, will it accrete and settle in such a way as to create a constant stellar to gas mass ratio in the galaxy outskirts? 

Accretion is currently the best explanation for the old extended stellar disk in NGC 2841. However, this doesn't mean that all outskirts of galaxy disks are formed via this mechanism. About 30\% of nearby disk galaxies exhibit XUV disks. Are these disks stimulated via accretion events, as might be for NGC 2841, or do they have other origins? Are the star formation rates in some XUV disks able to explain the extended stellar disks in their host galaxies if an extended disk exists? In which subset of galaxies do stars dominate over gas in the outskirts of their disks? In order to answer these sorts of questions, a larger sample of galaxies needs to be studied. Further studies of the outskirts of other galaxies in the Dragonfly Nearby Galaxies Survey will be carried out in the future. 

\subsection{Dragonfly and Cirrus}
A measurement of the ratio of scattered to thermally emitted light from the Spider HI cloud was made in Chapter 4. The largest remaining uncertainty of this ratio comes from incomplete knowledge of whether the dispersion in the correlation plot of scattered to thermally emitted light was due to noise in Dragonfly or noise in Herschel. The fact that the $\chi^2$ of the best fit linear model to the correlation plot was not unity (e.g. for the fits to correlations done on images convolved to 18", $\chi^2$ was about two) indicates that we have not yet accounted for all sources of error in our noise models. Further work will be carried out to include in our noise models the fact that some light from stars and galaxies have not been masked (because they fell outside the central bright masked regions of the sources). In order to do this accurately, an accurate PSF needs to be used to create the simulated image. Currently, the PSF used to simulate the stars and galaxies in Dragonfly data were created using the internal PSF of SkyMaker. This is a simple model PSF that does not necessarily reflect the real PSF in the Dragonfly images. A more realistic PSF can be measured by directly analyzing the Dragonfly images, which will enable a proper measurement of the contribution to noise from stars and galaxies outside of the masks.  

An illustration of how this ratio can be used to test dust models was presented at the end of Chapter 4. The dust model from~\cite{Compiegne2011}, together with an isotropic scattering phase function and/or isotropic angular distribution of the interstellar radiation field (ISRF) over-predicted scattered light in \textit{g} and \textit{r} bands by a factor of 4.5. However, the scattering phase function is quite anisotropic, and the ISRF should have a dominant contribution from the inner galaxy, and so should not be isotropic either. Future work will involve using more realistic models of the scattering phase function and ISRF angular distribution. 

When the slope of the linear portion of the correlation plot of scattered to thermally emitted light from dust has been thoroughly investigated, the non-linear portion of the correlation can be investigated. Possible causes of this non-linearity are optical depth effects, or changes in emissivity of the dust. Further insight might be gained by the correlation of scattered light and thermal emission with the Hydrogen column density, as derived using the Dominion Radio Astrophysical Observatory (DRAO) HI Intermediate Galactic Latitude Survey (DHIGLS) 21 cm data~\citep{Blagrave2017}. 

\section{Conclusions}
The sky at surface brightness levels of 29 mag arcsec$^{-2}$ or fainter is uncharted and predicted to be full of astrophysical phenomena on a huge range of physical scales: from dust in the interstellar medium (ISM) to the cosmic web. The Dragonfly Telephoto Array and the Dragonfly Pipeline represents a disruptive telescope and software design that addresses key sources of systematic errors that has stalled our ability do low surface brightness observations at 29 mag arcsec$^{-2}$ for the last 40 years. This thesis documents the Dragonfly Pipeline and how it addresses the systematics of scattered light and sky subtraction. Dragonfly and its software was then used to study phenomena on two vastly different physical scales: galaxies and dust in the ISM. This thesis presents the discovery that, counter to common understanding, the stellar disk of a galaxy can be larger and more massive than its HI gas disk. This discovery is a key piece of the puzzle in understanding galaxy disk growth. From here, it is critical to study a larger sample of galaxies to see how common this type of large stellar disk is. If it is very common, we need to rethink our assumptions about how large galaxy stellar disks are, and how they grow. It is also demonstrated that Dragonfly data, in combination with Herschel data, can be used to study dust and its radiative properties. By using scattered light observations together with thermal emission from dust grains, the imperfect knowledge about the intensity of the interstellar radiation field (ISRF) is no longer an issue. The ratio of scattering to absorption cross sections of dust can be derived without knowing the absolute value of the ISRF. This is an important alternative method of measuring dust physical and radiative properties, which is a notoriously difficult branch of astronomy. If this technique can be developed to use all sky surveys, then it will provide unprecedented information about the evolution of dust in the Milky Way. 

The Dragonfly Telephoto Array has opened up a new parameter space of exploration for astronomy. It is still unclear what the surface brightness limits achievable by Dragonfly are. The Dragonfly team is currently taking an ultra deep exposure of the edge on, nearby galaxy, NGC 4565 (the Needle Galaxy). We hope to characterize what the limiting factors are for Dragonfly when it comes to deep imaging. One contender is faint and unresolved point sources that blur together to look like low surface brightness galaxies, cirrus and structures in galaxy outskirts. We will address this with the use of higher resolution data. The ultimate limiting factor is likely to be foreground cirrus. One undertaking we will work on is cirrus foreground subtraction using images of thermal emission from dust. It is a subject of great interest to continue to understand what limits our ability to do better and try to overcome it. It is plausible that at some stage, our ability to do sky subtraction will be the limiting factor, at which point, a space telescope could provide the answer.

\addcontentsline{toc}{chapter}{Bibliography}
\bibliographystyle{aasjournal}
\bibliography{zhang,Planck_bib,extra_arXivEdit_bib}


\end{document}